\documentclass[showpacs,pre,onecolumn]{revtex4-1}
\usepackage{amsmath}
\usepackage{graphicx}
\usepackage{subfigure}
\usepackage{color}

\begin{document}

\title{Symmetric and asymmetric solitons and vortices in linearly coupled
two-dimensional waveguides with the cubic-quintic nonlinearity}
\author{Nir Dror and Boris A. Malomed }
\affiliation{Department of Physical Electronics, School of Electrical Engineering,
Faculty of Engineering, Tel Aviv University, Tel Aviv 69978, Israel}

\begin{abstract}
It is well known that the two-dimensional (2D) nonlinear Schr\"{o}dinger
equation (NLSE) with the cubic-quintic (CQ) nonlinearity supports a family
of stable fundamental solitons, as well as solitary vortices (alias vortex
rings), which are stable for sufficiently large values of the norm. We study
stationary localized modes in a symmetric linearly coupled system of two
such equations, focusing on \textit{asymmetric} states. The model may
describe \textquotedblleft optical bullets" in dual-core nonlinear optical
waveguides (including \textit{spatiotemporal vortices} that were not
discussed before), or a Bose-Einstein condensate (BEC) loaded into a
\textquotedblleft dual-pancake" trap. Each family of solutions in the
single-component model has two different counterparts in the coupled system,
one symmetric and one asymmetric. Similarly to the earlier studied coupled
1D system with the CQ nonlinearity, the present model features \textit{%
bifurcation loops}, for fundamental and vortex solitons alike: with the
increase of the total energy (norm), the symmetric solitons become unstable
at a point of the direct bifurcation, which is followed, at larger values of
the energy, by the \textit{reverse bifurcation} restabilizing the symmetric
solitons. However, on the contrary to the 1D system, both the direct and
reverse bifurcation may be of the subcritical type, at sufficiently small
values of the coupling constant, $\lambda $. Thus, the system demonstrates a
\textit{double bistability }for the fundamental solitons. The stability of
the solitons is investigated via the computation of instability growth rates
for small perturbations. Vortex rings, which we study for two values of the
\textquotedblleft spin", $s=1$ and $2$, may be subject to the azimuthal
instability, like in the single-component model. In particular, complete
destabilization of asymmetric vortices is demonstrated for a sufficiently
strong linear coupling. With the decrease of $\lambda $, a region of stable
asymmetric vortices appears, and a single region of bistability for the
vortices is found. We also develop a quasi-analytical approach to the
description of the bifurcations diagrams, based on the variational
approximation. Splitting of asymmetric vortices, induced by the azimuthal
instability, is studied by means of direct simulations. Interactions between
initially quiescent solitons of different types are studied too. In
particular, we confirm the prediction of the reversal of the sign of the
interaction (attractive/repulsive for in-phase/out-of-phase pairs) for the
solitons with the odd spin, $s=1$, in comparison with the even values, $s=0$
and $2$.
\end{abstract}

\maketitle

\section{Introduction}

\label{sec:Introduction} In the last two decades, much interest has been
drawn to the studies of spatiotemporal solitons (STSs, alias
\textquotedblleft light bullets") in nonlinear optics \cite{Agrawal,review}.
They are supported through the balance between the temporal dispersion,
spatial diffraction, and nonlinearity. It is well known that two-dimensional
(2D) and three-dimensional (3D) STSs cannot be stable in uniform media with
the cubic (Kerr) nonlinearity, due to possibility of the collapse in the
same setting \cite{review,Barge}. To avoid the collapse, other types of the
nonlinearity were proposed. In particular, the stability of multidimensional
solitons is readily secured by saturable \cite{Alfano,Enns}, quadratic ($%
\chi ^{(2)}$) \cite{Blagoeva,Mihalache}, and cubic-quintic (CQ)
nonlinearities \cite{Teixeiro,Malomed}. While the creation of STSs in 3D has
not been reported so far, quasi-2D STSs were made in $\chi ^{(2)}$ crystals
\cite{Liu}.

While the fundamental (zero-vorticity) STSs supported by the above-mentioned
non-Kerr nonlinearities are stable, the stability of solitary vortices
(alias vortex rings, or spinning solitons, with integer \textquotedblleft
spin" $s$ referring to the corresponding topological charge) are vulnerable
to the destabilization by azimuthal perturbations. In particular, spinning
solitons in 2D models with $\chi ^{(2)}$ or saturable nonlinearity are
unstable, as demonstrated by simulations \cite{Torner} and the experiment
\cite{Petrov}. The azimuthal instability breaks the solitary vortices with $%
s=1$ into two or three fragments, each re-trapping into a moving fundamental
soliton, so that the intrinsic spin moment of the unstable mode is
transformed into the orbital momentum of the set of separating fragments.

Vortex solitons may be stable in media with \textit{competing nonlinearities}%
. In particular, stable 2D vortices, with $s=1$ and $2$, were found in a
model combining the $\chi ^{(2)}$ and self-defocusing cubic nonlinearities
\cite{Towers}. Another model which is known to support stable 2D spinning
solitons includes self-focusing cubic and self-defocusing quintic
nonlinearities. Quiroga-Teixeiro and Michinel \cite{Michinel} were the first
to demonstrate in direct simulations that 2D solitons with $s=1$ may be
stable in the CQ model, provided that their power (norm) is sufficiently
large. The detailed analysis \cite{Crasovan,Stability}, which made use of
the linearized version of the model for computing growth rates of
perturbation eigenmodes, had demonstrated that, for spins $s=1$ and $2$, the
CQ model exhibits relatively large stability regions, which cover,
respectively, $\approx 9\%$ and $\approx 8\%$ of the respective existence
regions, in terms of soliton's norm (total power). It has also been
demonstrated that the spinning solitons with the vorticity up to $s=5$ may
also be stable, but in extremely narrow regions \cite{Warchall}.

The subject of the present work are 2D solitons, both the fundamental and
spinning ones, in a symmetric system of two linearly coupled nonlinear Schr%
\"{o}dinger equations (NLSEs) with the CQ nonlinear terms. The main
objectives of the analysis are the \textit{symmetry-breaking bifurcations}
and asymmetric solitons generated by them. As explained below, the model may
have realizations in both nonlinear optics and BECs (Bose-Einstein
condensates, where the NLSE is known as the Gross-Pitaevskii equation, GPE
\cite{Pit}). The 1D version of the model was studied in Ref.~\cite{Albuch},
where it was found that a bifurcation of the supercritical type destabilizes
symmetric two-component solitons, generating a pair of asymmetric ones. At
larger values of the norm (power), the branches of symmetric and asymmetric
solitons merge back, restabilizing the symmetric ones, which gives rise to a
\textit{bifurcation loop}. At relatively small values of the linear-coupling
constant, $\lambda $, the reverse bifurcation is of the subcritical type,
which may feature a large region of the bistability between the symmetric
and asymmetric 1D solitons. With the increase of $\lambda $, the bistability
region gradually disappears and the bifurcation loop shrinks, vanishing and
leaving only the symmetric solitons in the system at still larger values of $%
\lambda $.

Spontaneous-symmetry-breaking bifurcations of 2D solitons and vortices in
linearly-coupled systems were studied in Ref.~\cite{Gubeskys}, in terms of
two parallel pancake-shaped BECs linked by the tunneling of atoms across the
potential barrier separating them. The model was based on two GPEs with the
linear coupling and cubic nonlinearity. The stability of the solitons and
vortices against the collapse and (as concerns the vortices) against the
splitting was provided by a 2D periodic potential (an optical lattice)
present in both equations. In agreement with the results known from other
models, the symmetric solitons and vortices underwent the symmetry-breaking
bifurcations in the system with the self-attractive nonlinearity. Unlike
that setting, in the present work we consider the uniform space, the
stability of the single-component solitons and vortices being provided by
the CQ nonlinearity.

The paper is organized as follows. The model is introduced, and its physical
realizations are discussed, in section~\ref{sec:model}. The set of
stationary solutions to the coupled equations is parameterized by the
linear-coupling constant, $\lambda $, and the propagation coefficient, $k$.
In section~\ref{sec:loops} we present asymmetric solutions, and produce
bifurcation diagrams for the fundamental and spinning solitons, with
vorticities $s=0,1,2$. Section~\ref{sec:stability} reports a detailed
stability analysis, performed by dint of numerical calculations of the
growth rates for eigenmodes of small azimuthal perturbations. In section~\ref%
{sec:VA}, we develop an explicit analytical approximation for the solutions,
based on the variational method. Results of direct numerical simulations,
which demonstrate the splitting of azimuthally unstable asymmetric spinning
solitons, are displayed in section~\ref{sec:splitting}. In section~\ref%
{sec:Interactions} we report numerical results for interactions between
solitons. The paper is concluded by section~\ref{sec:Conclusion}.

\section{The model}

\label{sec:model} The system of 2D linearly coupled equations with the CQ
nonlinearity is taken in the scaled form (cf. the 1D system introduced in
Ref. \cite{Albuch}):
\begin{eqnarray}
i\psi _{z}+\psi _{xx}+\psi _{yy}+|\psi |^{2}\psi -|\psi |^{4}\psi
&=&-\lambda \phi ,  \notag \\
i\phi _{z}+\phi _{xx}+\phi _{yy}+|\phi |^{2}\phi -|\phi |^{4}\phi
&=&-\lambda \psi .  \label{ModelPsiPhi}
\end{eqnarray}%
In the terms of the BEC, this is the system of GPEs for wave functions of
the condensate in parallel tunnel-coupled pancake-shaped traps, with the
negative scattering length accounting for the cubic self-attraction, the
scattering length itself being eliminated by the rescaling \cite{Gubeskys}.
In this context, evolution variable $z$ is actually time, the quintic terms
account for repulsive three-body collisions, provided that collision-induced
losses may be neglected \cite{Abdullaev}.

Actually, Eqs. (\ref{ModelPsiPhi}) may find a more relevant interpretation
in the application to optics, where the equations appear as normalized NLSEs
for the transmission of spatiotemporal light signals in a dual-core planar
waveguides. In this context, $z$ is the propagation distance, $x$ is the
transverse coordinate, and $y$ is the temporal variable (reduced time),
provided that the sign of the group-velocity dispersion in the waveguide is
anomalous \cite{Agrawal}. Accordingly, terms $\psi _{xx},\phi _{xx}$ and $%
\psi _{yy},\phi _{yy}$ account for the paraxial diffraction and dispersion
of light, respectively, while $1/\lambda $ determines the coupling length in
the dual-core waveguide. The equations are made symmetric with respect to $x$
and $y$ by means of rescaling of the spatial and temporal variables. Also, $%
\lambda $ is fixed to be real and positive, which can also be achieved in
the general case by means of an obvious transformation. As for the
self-focusing-defocusing CQ nonlinearity, it was theoretically predicted
\cite{Dutta} and observed \cite{Smektala} in diverse optical media. Some of
them admit the fabrication of dual-core planar waveguides.

The fundamental STS in a planar waveguide, although it has never been
reported in an experiment, is a well-known concept \cite{review}. On the
other hand, the \textit{spatiotemporal vortex}, whose snapshot would seem as
an elliptic ring running at the speed of light in the plane of the
waveguide, is a novel object. In the experiment, it may be coupled into the
planar waveguide by an oblique vortical laser beam shone onto the waveguide
under an appropriate angle. In that sense, the physical purport of the
spatiotemporal vortices is essentially different from that of $(2+1)$%
-dimensional spatial solitons with the embedded vorticity, which are
understood as hollow cylindrical beams of light propagating in a bulk medium
\cite{Torner,Petrov,Towers,Michinel,Crasovan,Stability}. In the experiment,
vortical spatial solitons, built as multi-beam complexes with the phase
distribution carrying the effective vorticity (rather than cylindrical
beams), were created in photorefractive crystals with the saturable
nonlinearity, their stability against splitting being maintained by a
photoinduced lattice potential \cite{Neshev}).

We aim to find stationary axisymmetric solutions to Eqs.~(\ref{ModelPsiPhi})
as
\begin{eqnarray}
\psi  &=&U(r)\exp (is\theta )\exp (ikz),  \notag \\
\phi  &=&V(r)\exp (is\theta )\exp (ikz),  \label{psiphi}
\end{eqnarray}%
where $r$ and $\theta $ are the polar coordinates in the $(x,y)$ plane, $k$
is the propagation constant, and integer $s$ is the above-mentioned spin.
Substituting expressions~(\ref{psiphi}) into Eqs.~(\ref{ModelPsiPhi}), we
arrive at equations for real functions $U$ and $V$:
\begin{eqnarray}
-kU+\frac{d^{2}U}{dr^{2}}+\frac{1}{r}\frac{dU}{dr}-\frac{s^{2}}{r^{2}}%
U+U^{3}-U^{5} &=&-\lambda V,  \notag \\
-kV+\frac{d^{2}V}{dr^{2}}+\frac{1}{r}\frac{dV}{dr}-\frac{s^{2}}{r^{2}}%
V+V^{3}-V^{5} &=&-\lambda U,  \label{ModelUV}
\end{eqnarray}%
with the boundary conditions demanding that the solution must feature
asymptotic forms $r^{|s|}$ at $r\rightarrow 0$, and $\exp (-\sqrt{k}r)$ at $%
r\rightarrow \infty $ (hence $k$ must be positive). The energies (norms) of
the two components of the soliton (alias their norms) are defined as usual,
\begin{equation}
E_{U,V}=2\pi \int_{-\infty }^{+\infty }r(U^{2},V^{2})dr,  \label{Nuv}
\end{equation}%
the total energy being $E_{\mathrm{total}}=E_{U}+E_{V}$. The asymmetry of
the two-component soliton is characterized by ratio
\begin{equation}
\Theta =\frac{E_{U}-E_{V}}{E_{U}+E_{V}}~.  \label{Theta}
\end{equation}

\section{Asymmetric solitons and bifurcation loops}

\label{sec:loops} Stationary symmetric and asymmetric soliton solutions for $%
s=0,1$ and $2$ were generated in a numerical form, applying the
Newton-Raphson method to Eqs.~(\ref{ModelUV}). We have also examined the
possibility of the existence of elliptic solitons (i.e., anisotropic
localized solutions to Eqs.~(\ref{ModelUV})), using several numerical
algorithms adjusted for the 2D setting, such as a generalized Petviashvili
iteration method, and a modification of the squared-operator method
presented in Ref.~\cite{Lakoba}. No stationary elliptic solutions have been
found.

Symmetric soliton solutions in the present model, with $\psi =\phi $ and $%
\Theta =0$, can be obtained from their counterparts previously found in the
single-component CQ model, i.e., one equation from system~(\ref{ModelPsiPhi}%
), with $\lambda =0$, for a single wave function, $\varphi $: $\psi
(x,y,z;k)=\phi (x,y,z;k)\equiv \varphi (x,y,z;k-\lambda )$. Accordingly, the
symmetric solutions emerge at $k=\lambda $, with the known minimum
(threshold) values of the energy in the single component \cite{Stability}: $%
E_{\mathrm{thr}}=11.73,~48.38,$ and $88.34$, for $s=0,1,$ and $2$,
respectively. Within the family of the symmetric solitons, $k$ varies from $%
k_{\min }\equiv \lambda $, which corresponds to $E=E_{\mathrm{thr}}$, up to $%
k_{\max }=\lambda +3/16$, corresponding to $E\rightarrow \infty $ ($k=3/16$
is the value of the propagation constant in the single-component 2D model at
which the energy diverges, along with the soliton's radius, for any $s$ \cite%
{Crasovan}).

\begin{figure}[tbp]
\subfigure[]{\includegraphics[width=2.1in]{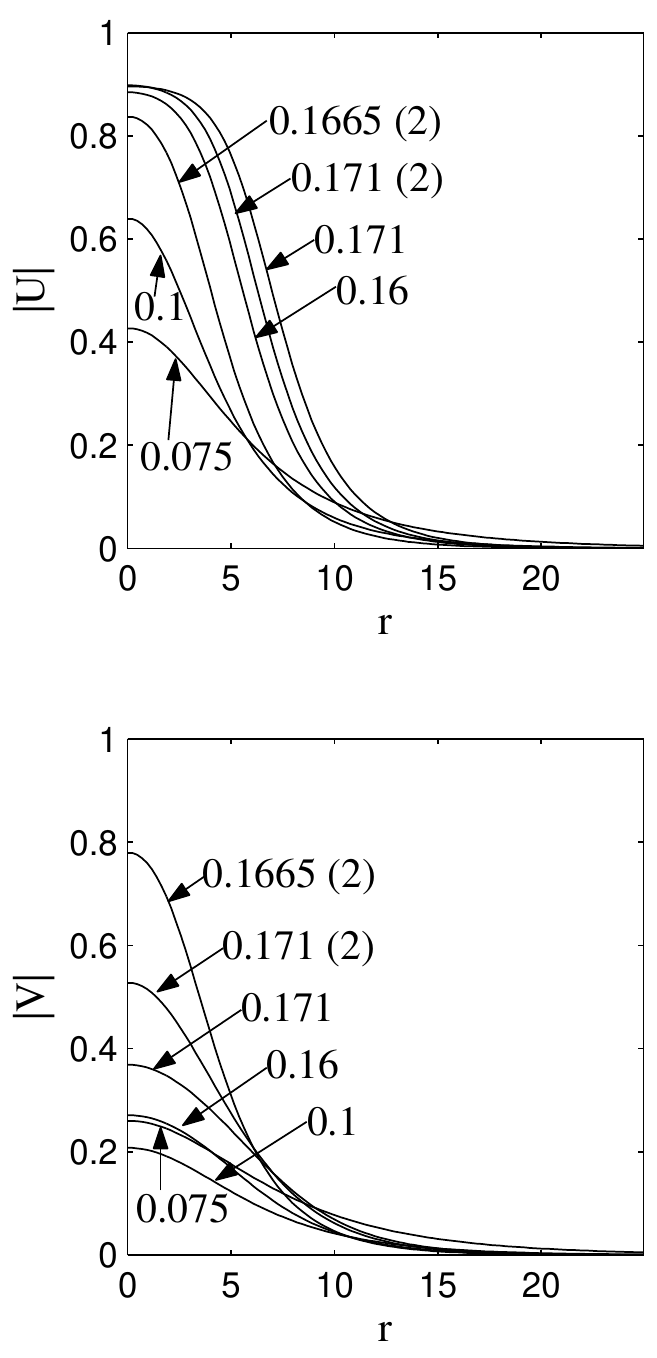}} \quad %
\subfigure[]{\includegraphics[width=2.1in]{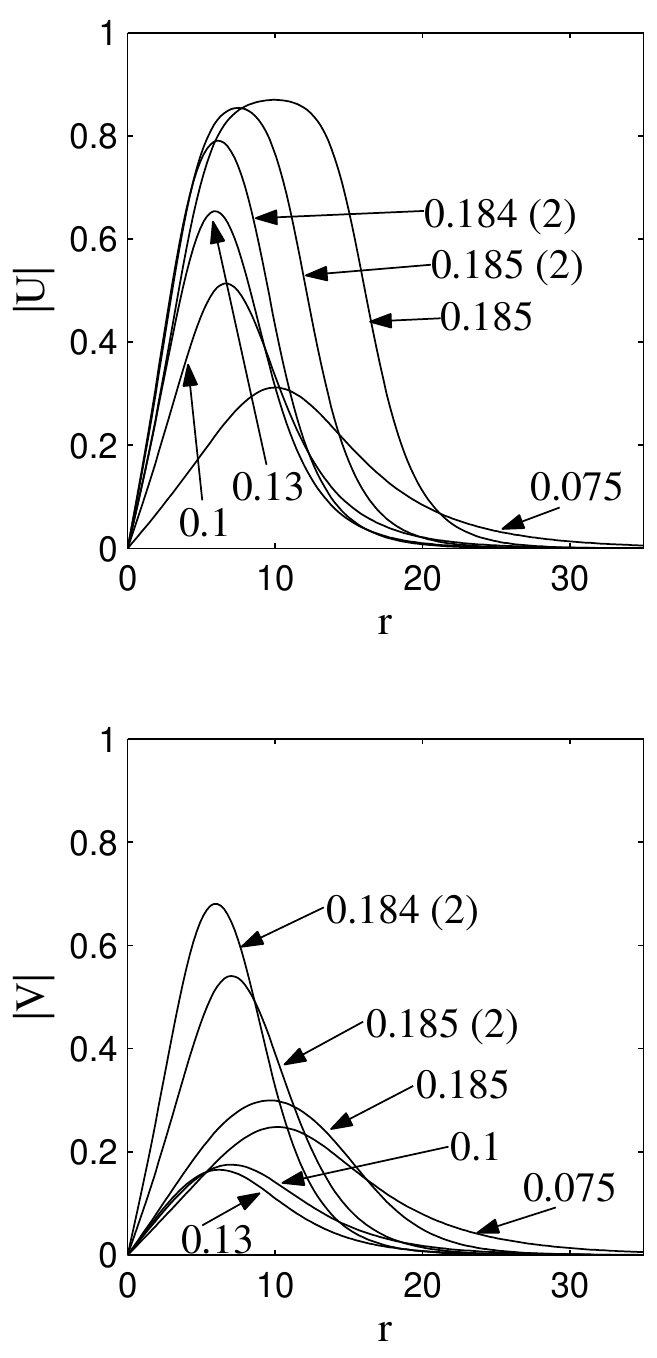}} \quad %
\subfigure[]{\includegraphics[width=2.1in]{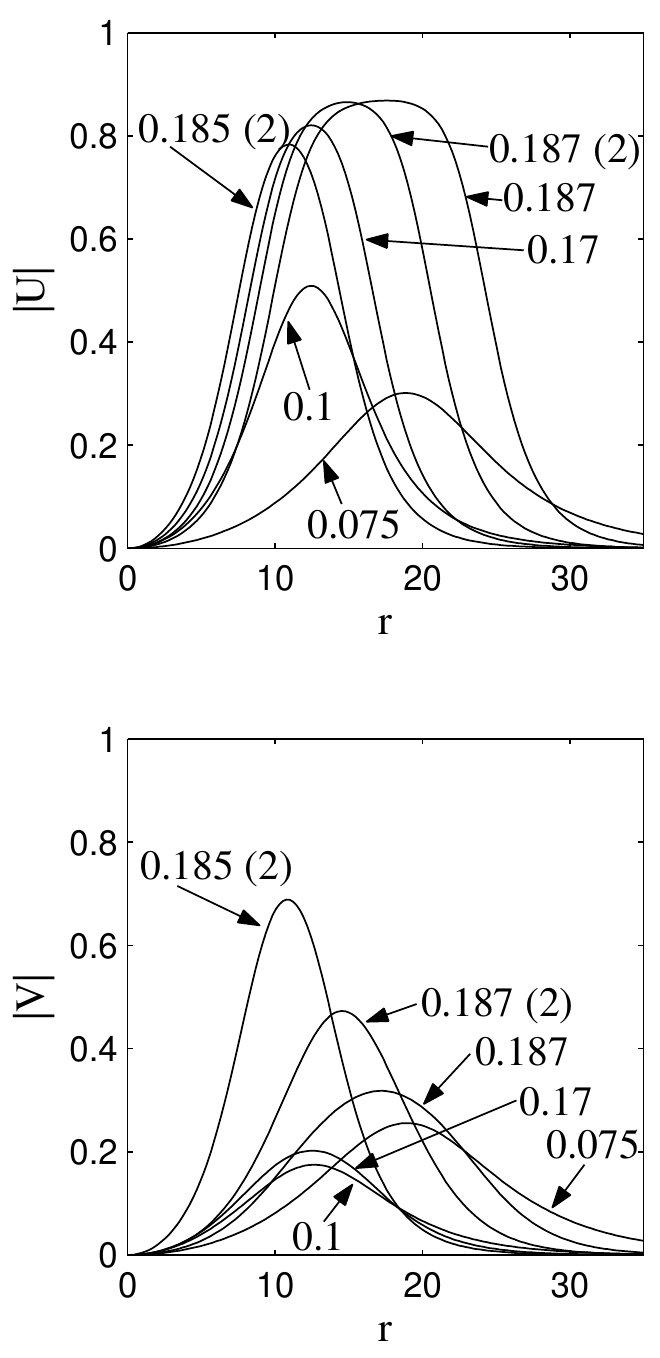}}
\caption{Examples of asymmetric solitons with vorticities $s=0$ (a), $s=1$
(b), and $s=2$ (c), for coupling constant $\protect\lambda =0.05$. Both
components are shown, with values of propagation constant $k$ indicated near
the corresponding curves. Notice that in the $(k,\Theta )$ plane ($\Theta $
is the asymmetry measure defined as per Eq.~(\protect\ref{Theta})), families
of the solutions form closed loops, which include the reverse bifurcation of
the subcritical type, see Figs. \protect\ref{BifLoopS0}-\protect\ref%
{BifLoopS2} below. Therefore, asymmetric solutions are found for $k$
increasing up to a certain maximum value, and then turning back and
decreasing until hitting the reverse-bifurcation point. In the present
panels, the solitons pertaining to the \textquotedblleft backward" subfamily
are labeled by $(2)$, to distinguish them from their counterparts with the
same values of $k$ belonging to the \textquotedblleft forward" part of the
family.}
\label{AsyProfiles}
\end{figure}

For each value of the spin, a unique family of asymmetric solitons, with $%
U\neq V$ and $\Theta \neq 0$, can be found. Typical radial profiles of
asymmetric solitons with $s=0,1$ and $2$, for $\lambda =0.05$ and several
values of $k$, are shown in Fig.~\ref{AsyProfiles}.

\begin{figure}[tbp]
\subfigure[]{\includegraphics[width=2.2in]{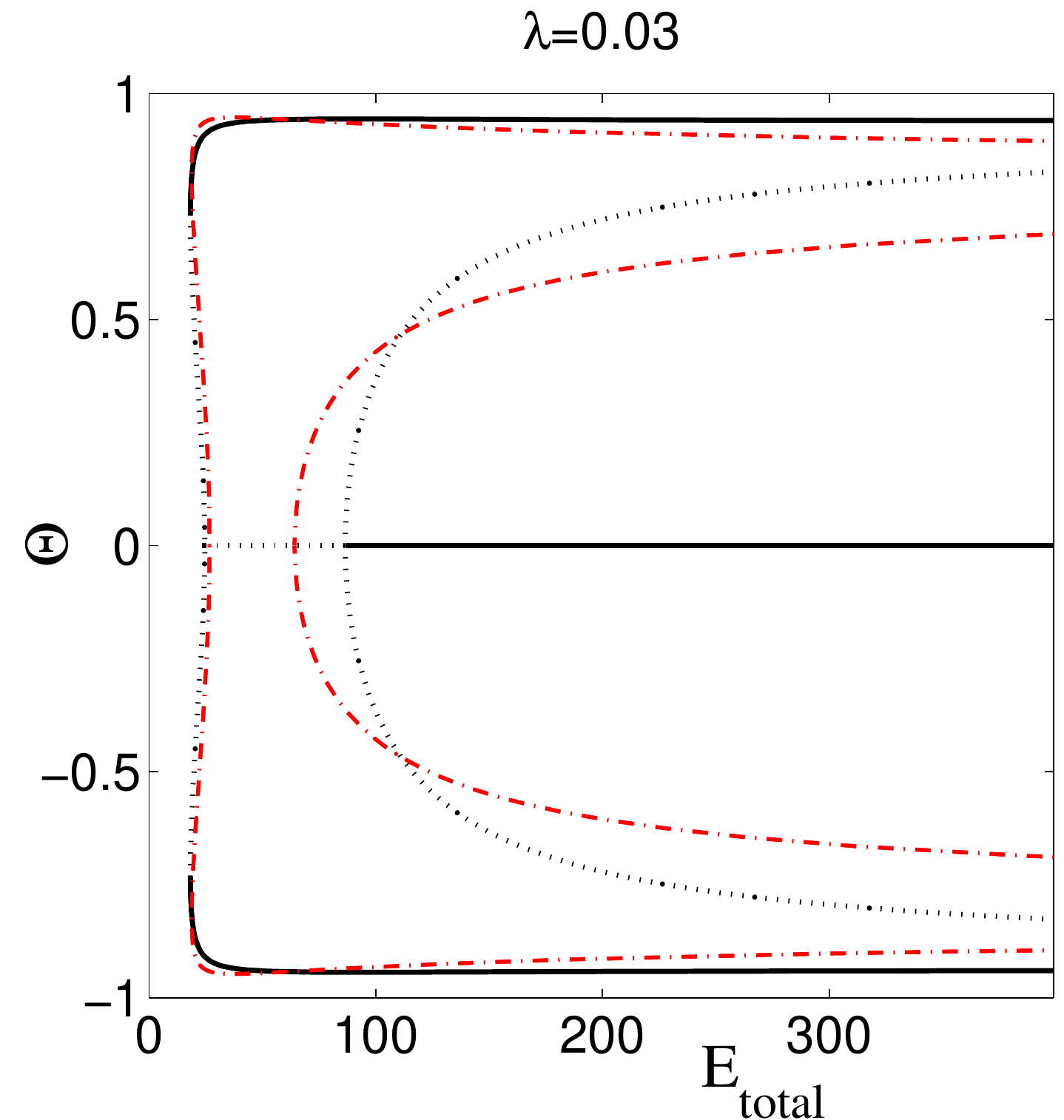}} %
\subfigure[]{\includegraphics[width=2.2in]{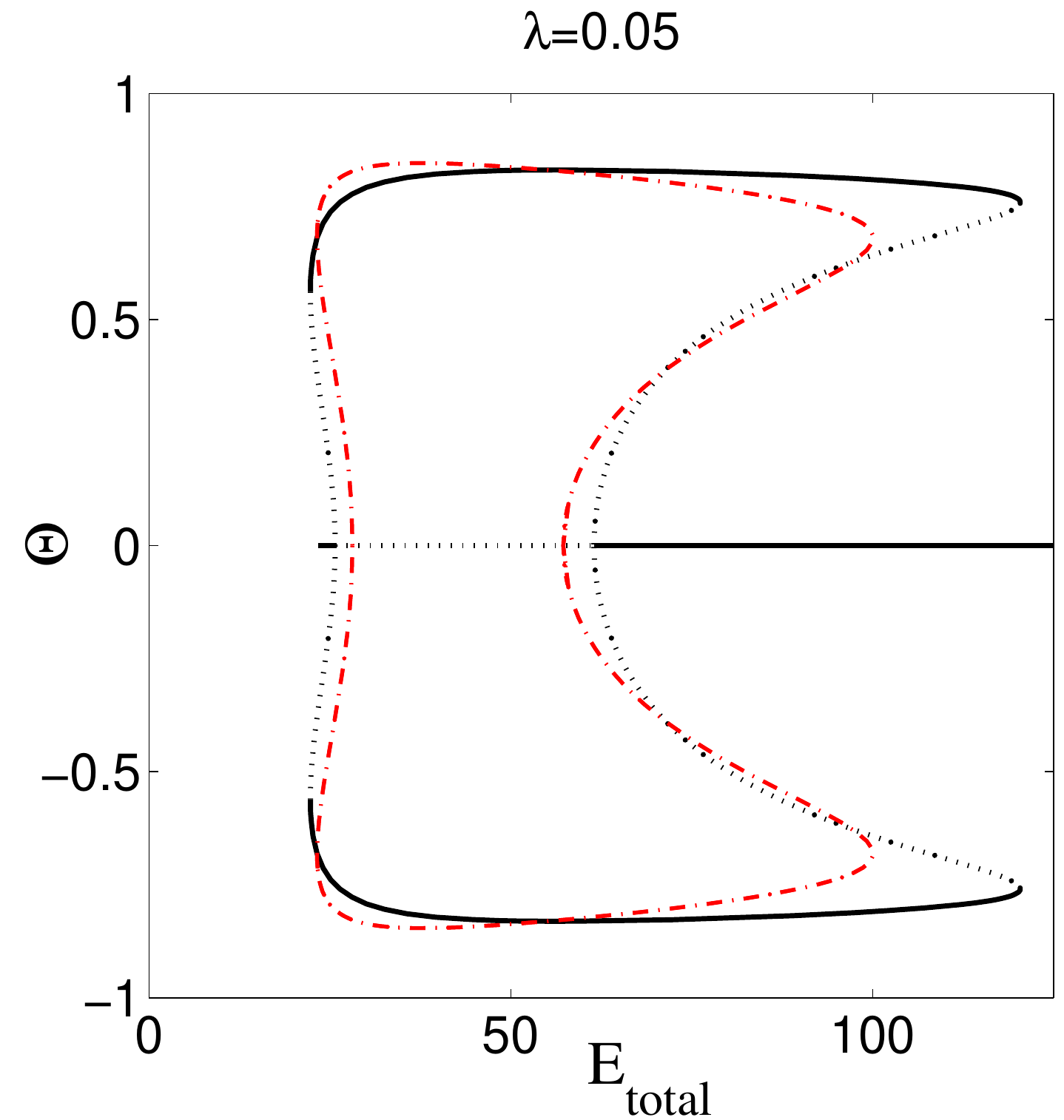}} %
\subfigure[]{\includegraphics[width=2.2in]{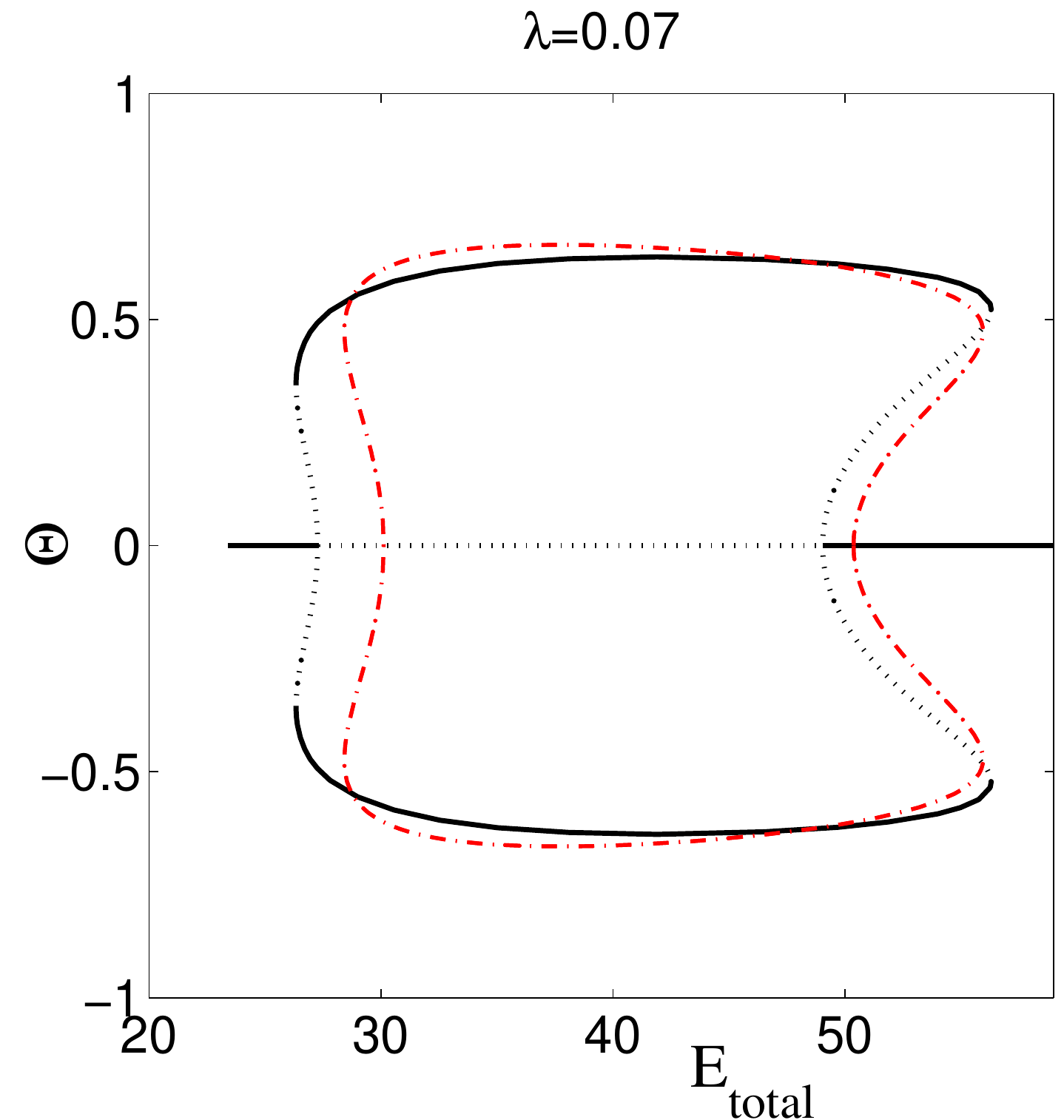}} \newline
\subfigure[]{\includegraphics[width=2.2in]{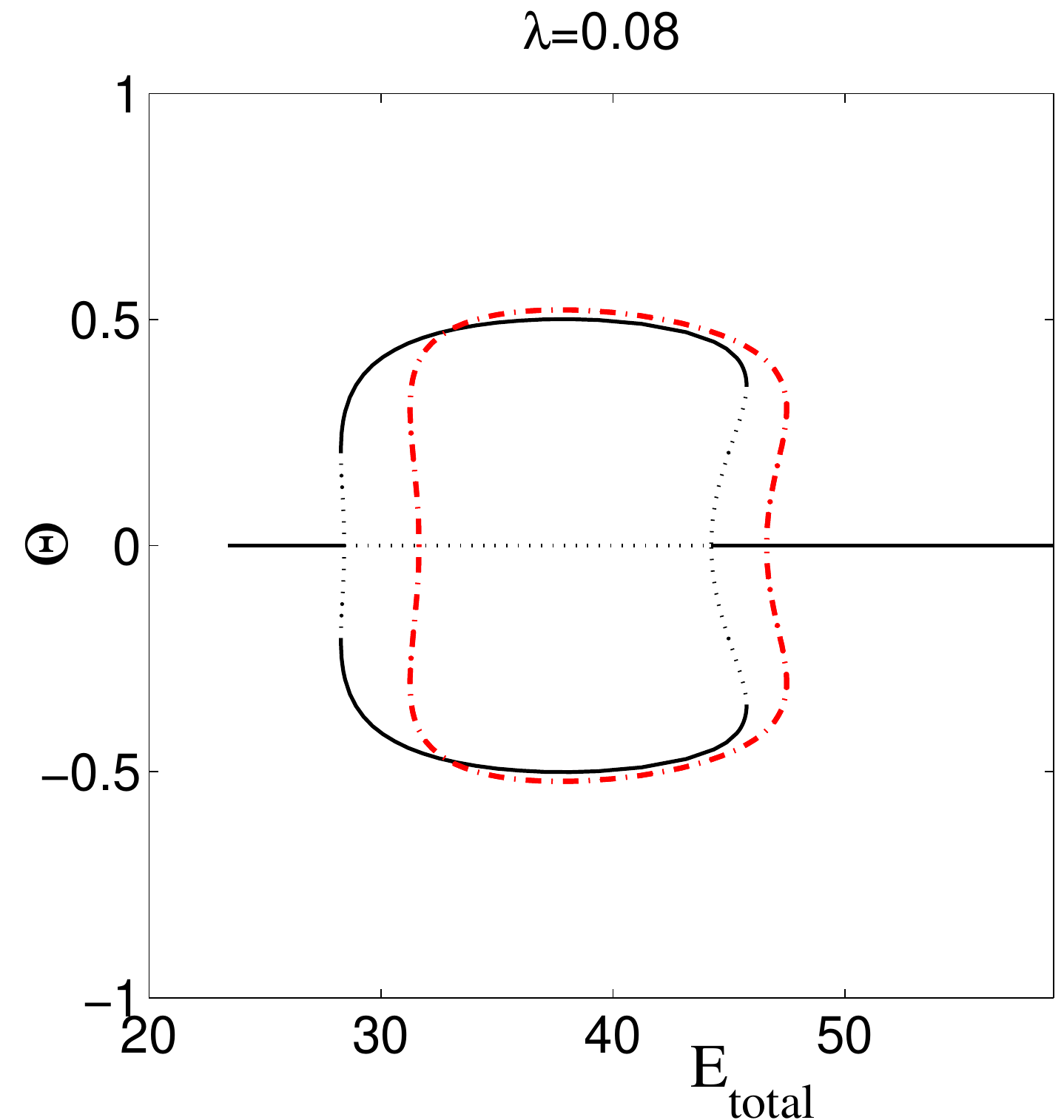}} %
\subfigure[]{\includegraphics[width=2.2in]{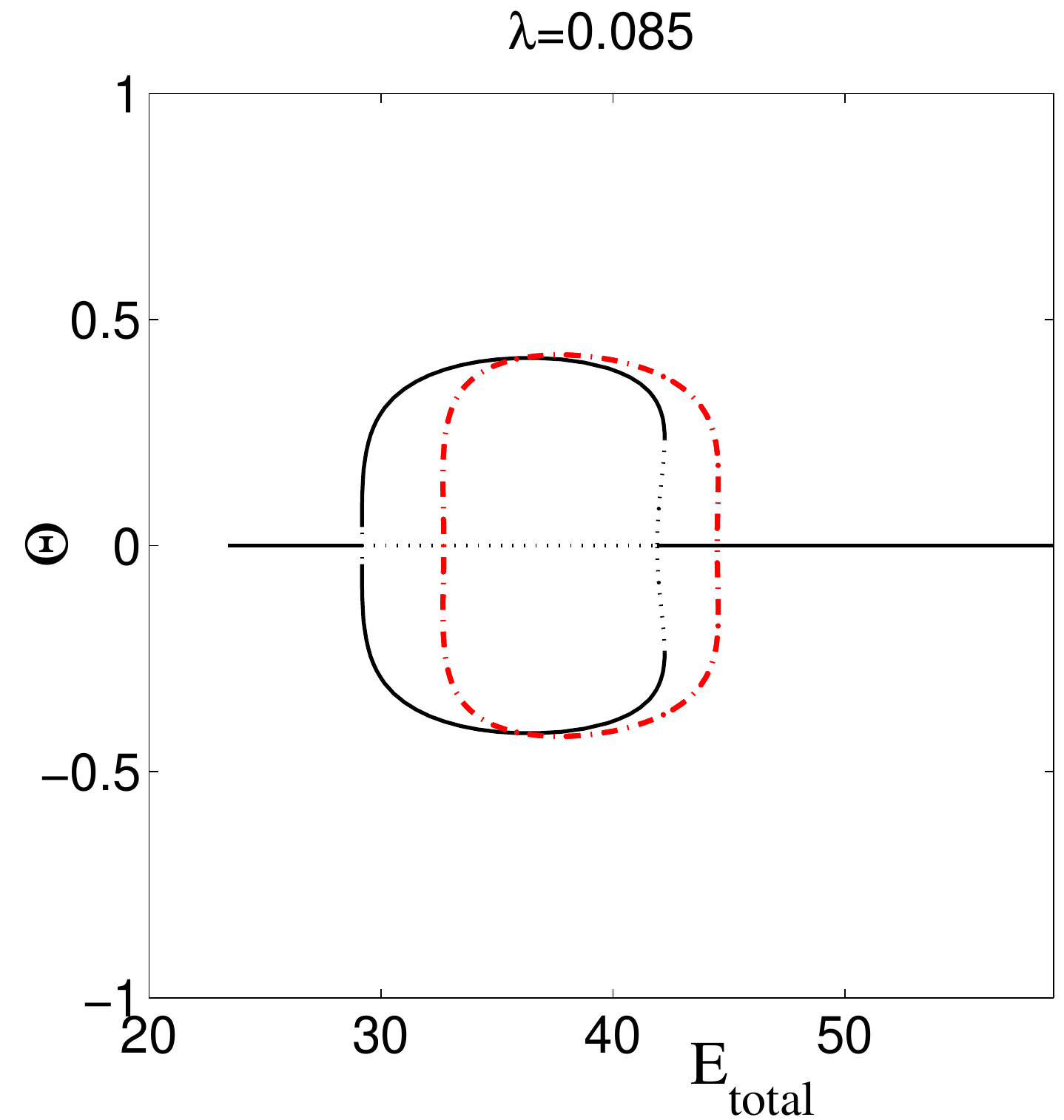}} %
\subfigure[]{\includegraphics[width=2.2in]{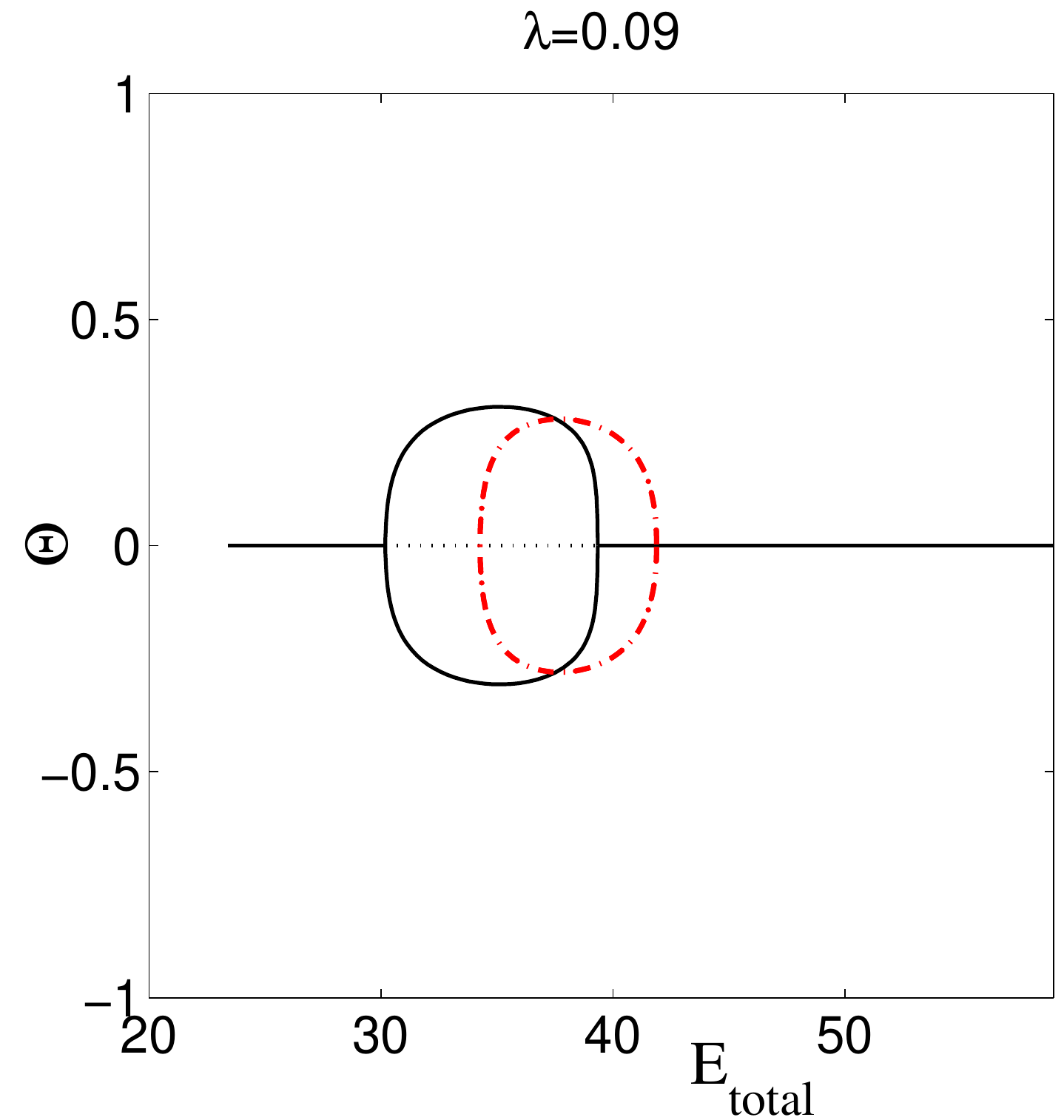}} \newline
\subfigure[]{\includegraphics[width=2.2in]{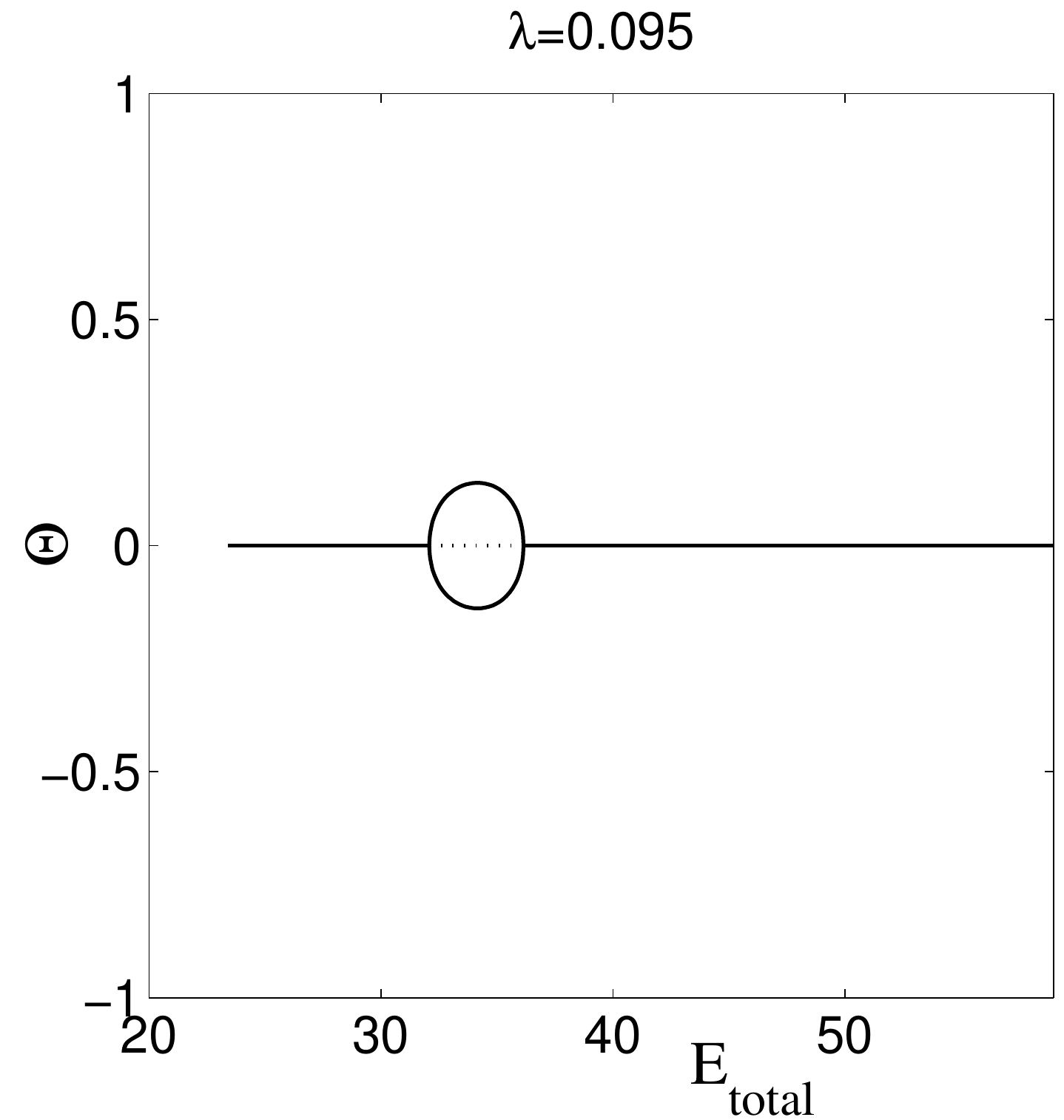}} %
\subfigure[]{\includegraphics[width=2.2in]{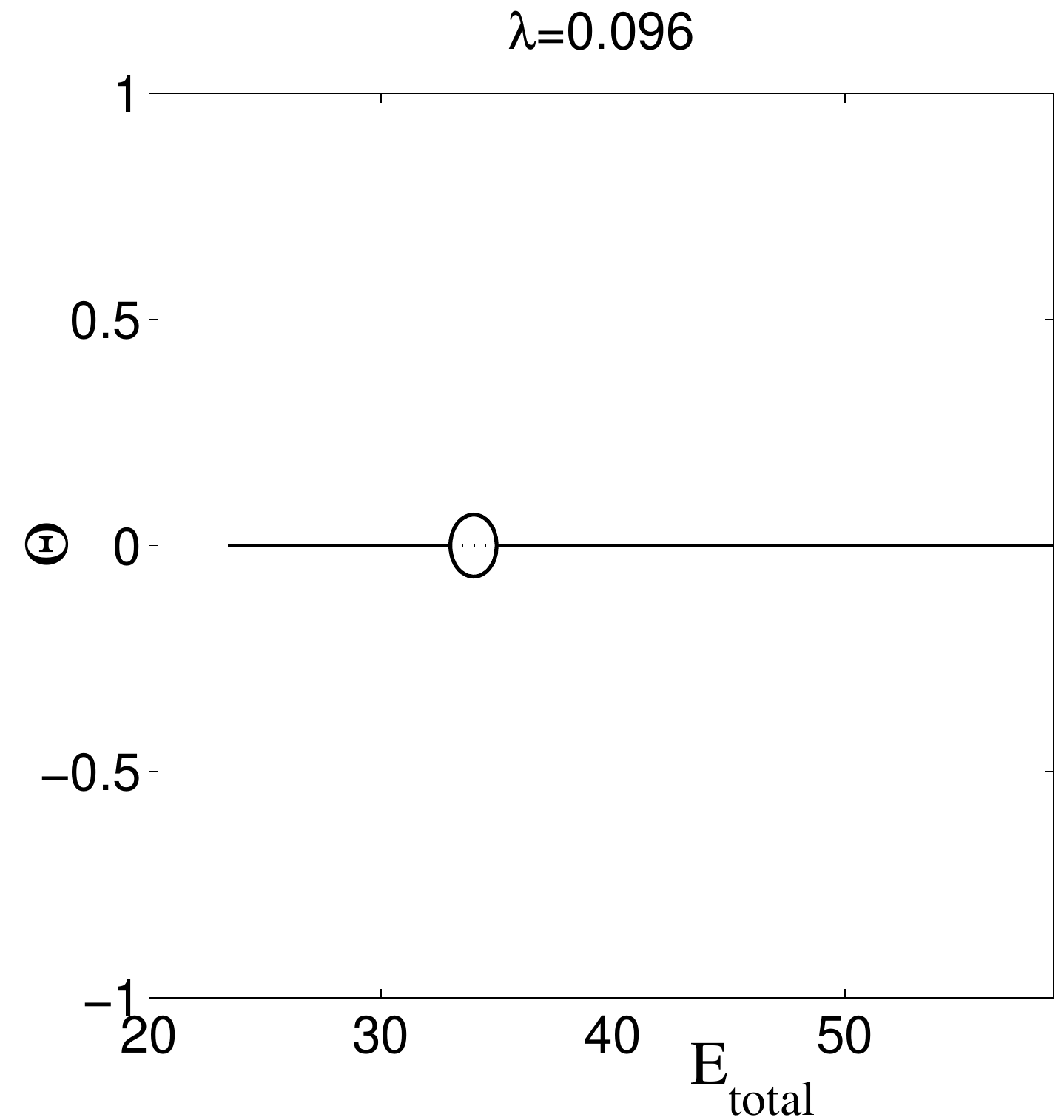}} %
\subfigure[]{\includegraphics[width=2.2in]{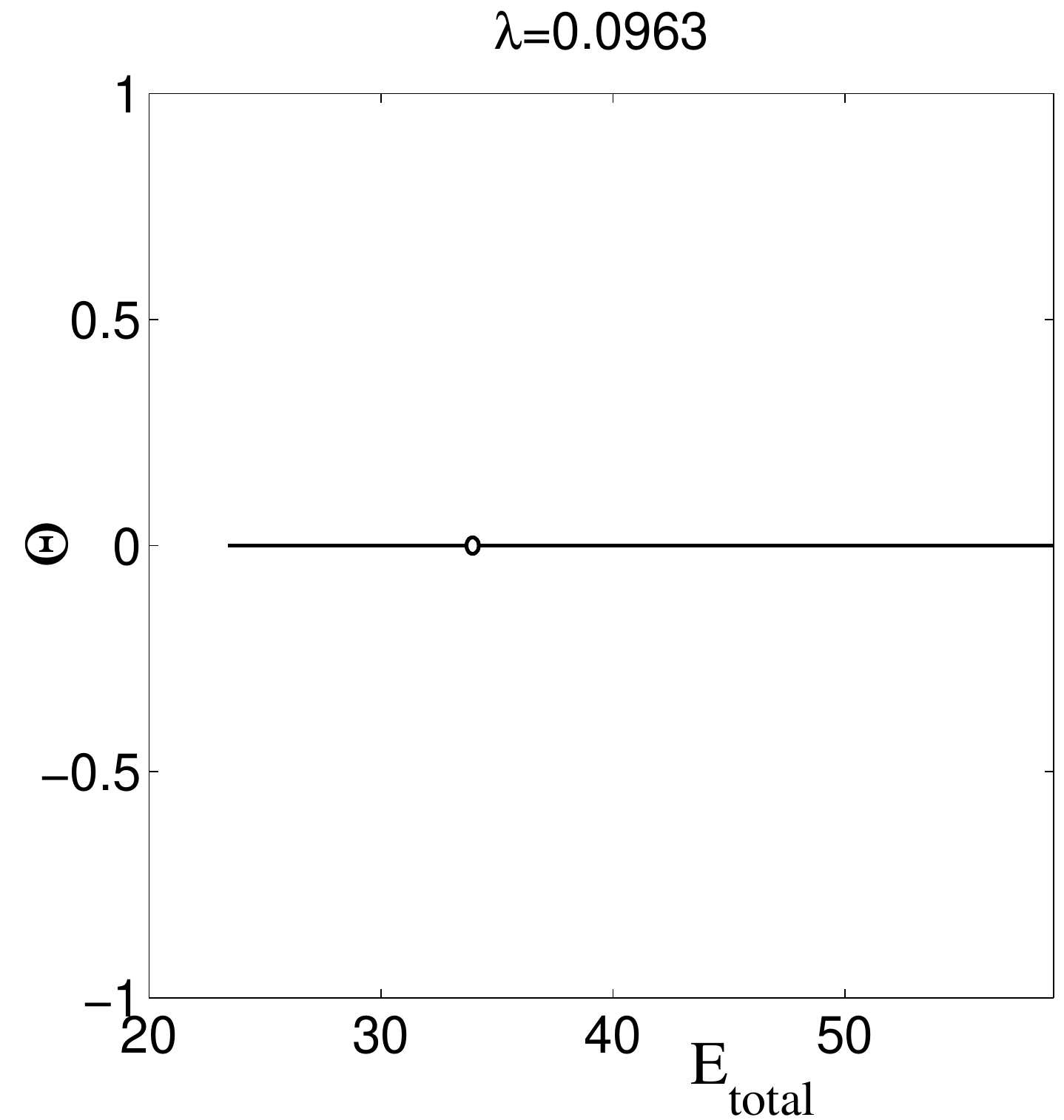}}
\caption{(Color online) The bifurcation diagrams, in the $(E_{\mathrm{total}%
},\Theta )$ plane, for fundamental solitons ($s=0$), at different values of
the linear-coupling constant, $\protect\lambda $. Here and in other
bifurcation diagrams, stable and unstable branches are shown by solid and
dotted lines, respectively (see also Fig.~\protect\ref%
{Stability_Lambda005_s0-a} below, for details of the stability of solution
branches displayed in this figure). The bifurcation loops produced by the
variational approximation (section~\protect\ref{sec:VA}) are shown too, by
dashed-dotted red curves.}
\label{BifLoopS0}
\end{figure}

\begin{figure}[tbp]
\subfigure[]{\includegraphics[width=2.2in]{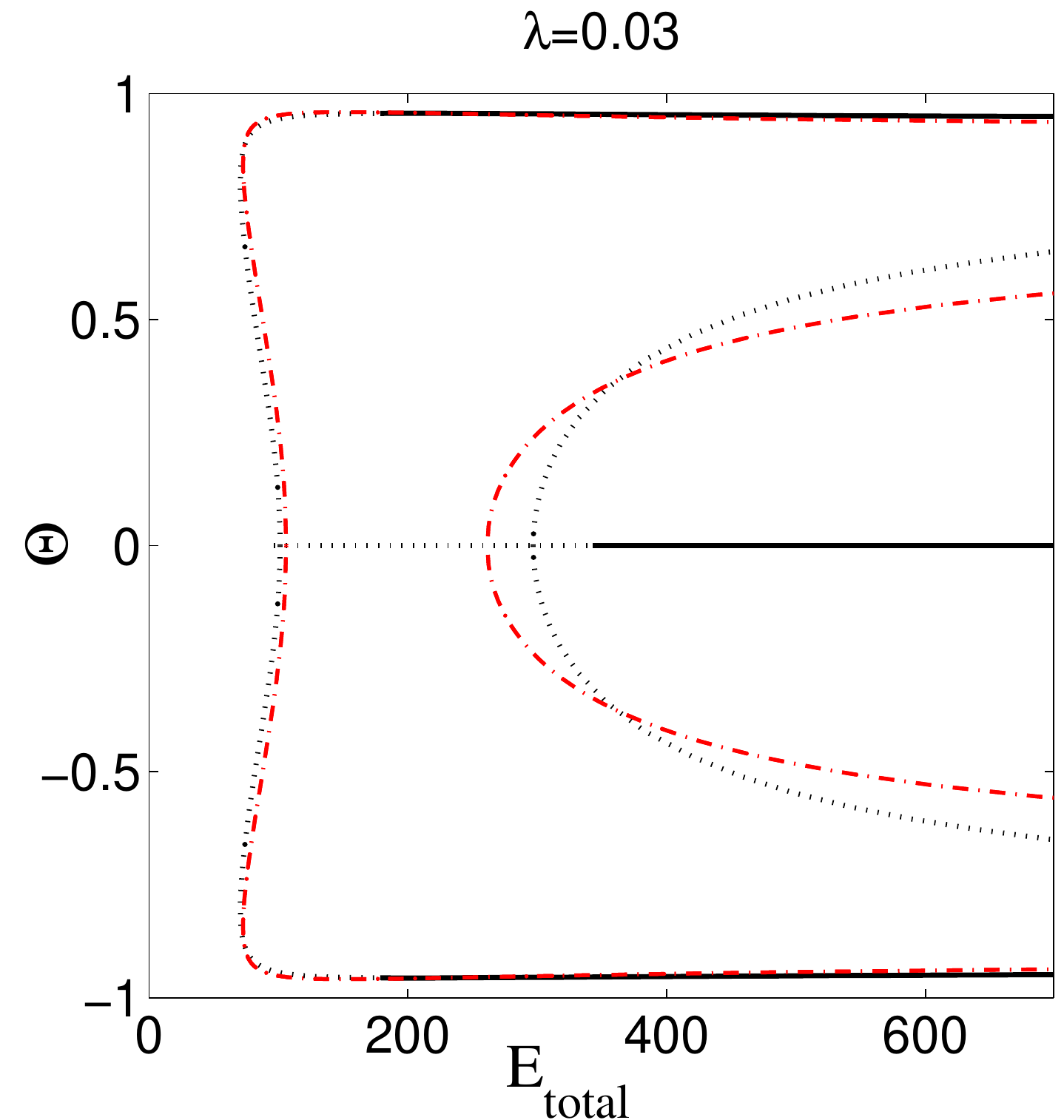}} %
\subfigure[]{\includegraphics[width=2.2in]{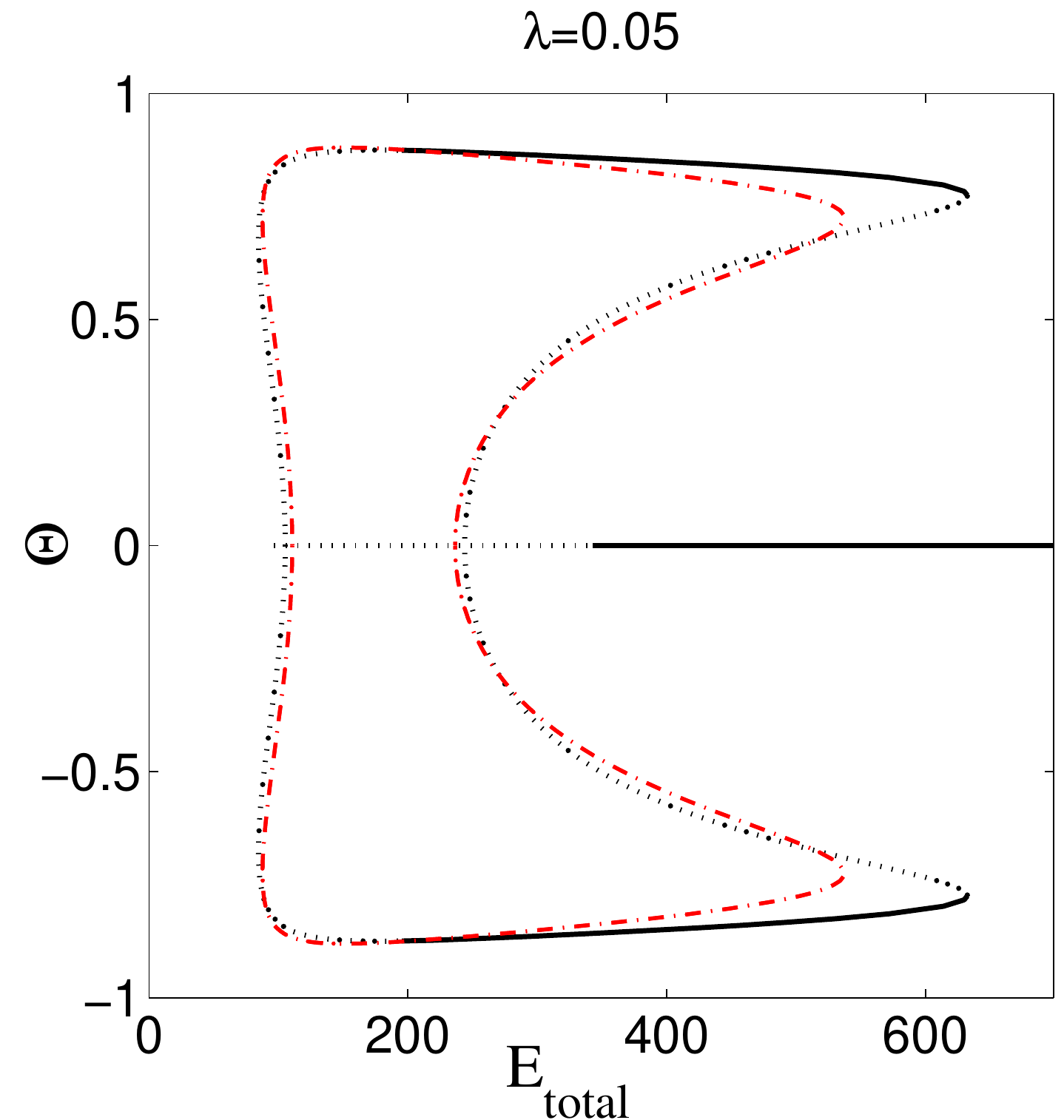}} %
\subfigure[]{\includegraphics[width=2.2in]{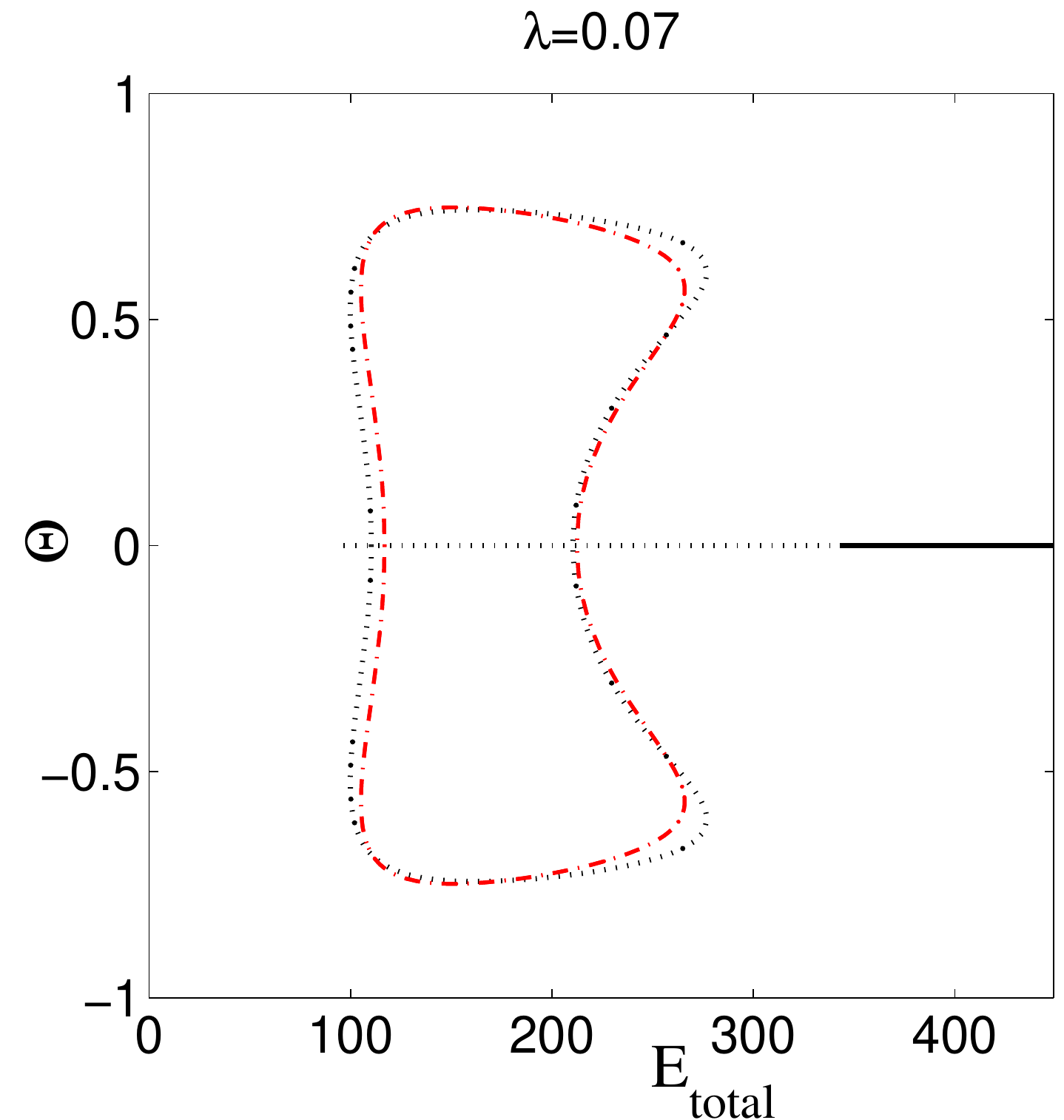}} \newline
\subfigure[]{\includegraphics[width=2.2in]{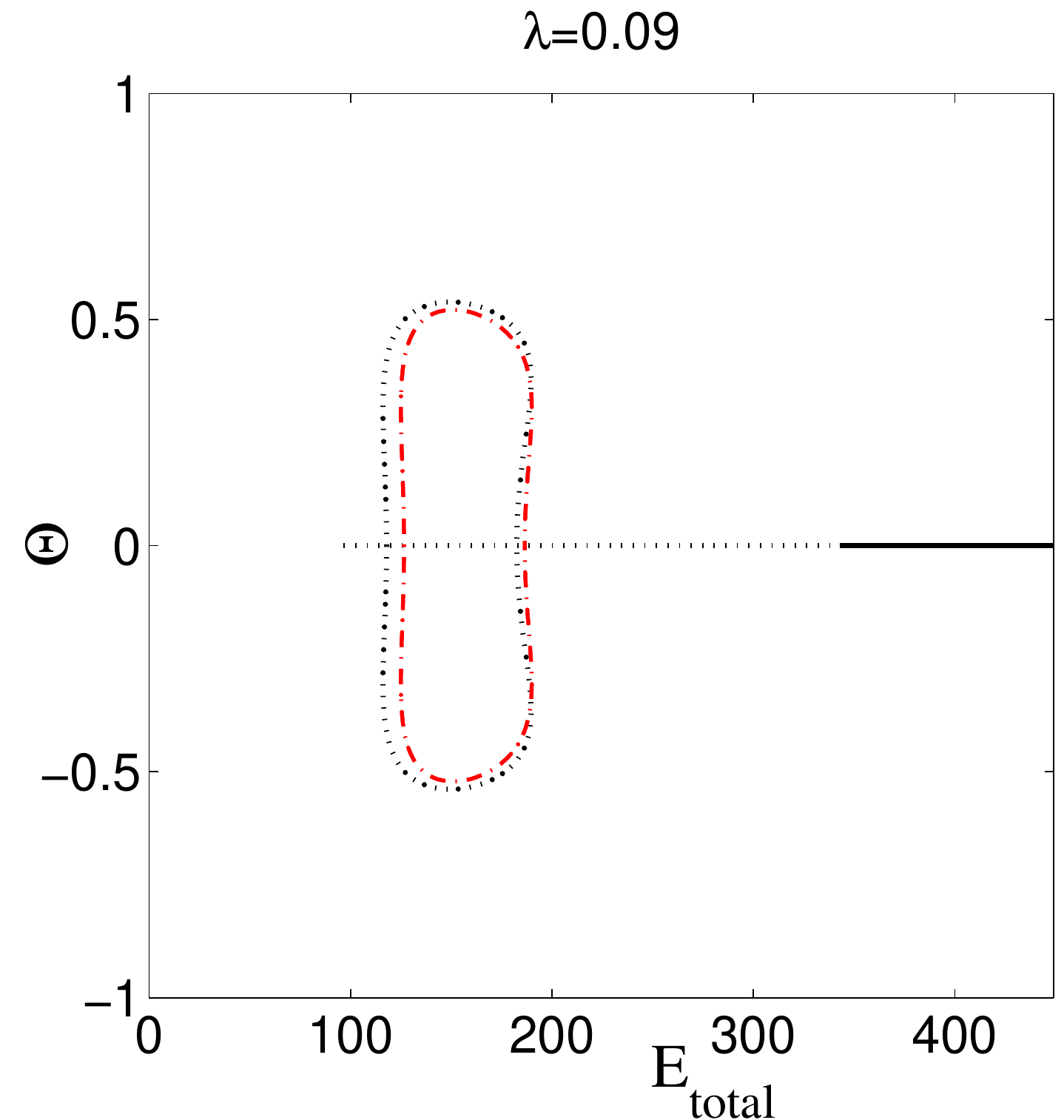}} %
\subfigure[]{\includegraphics[width=2.2in]{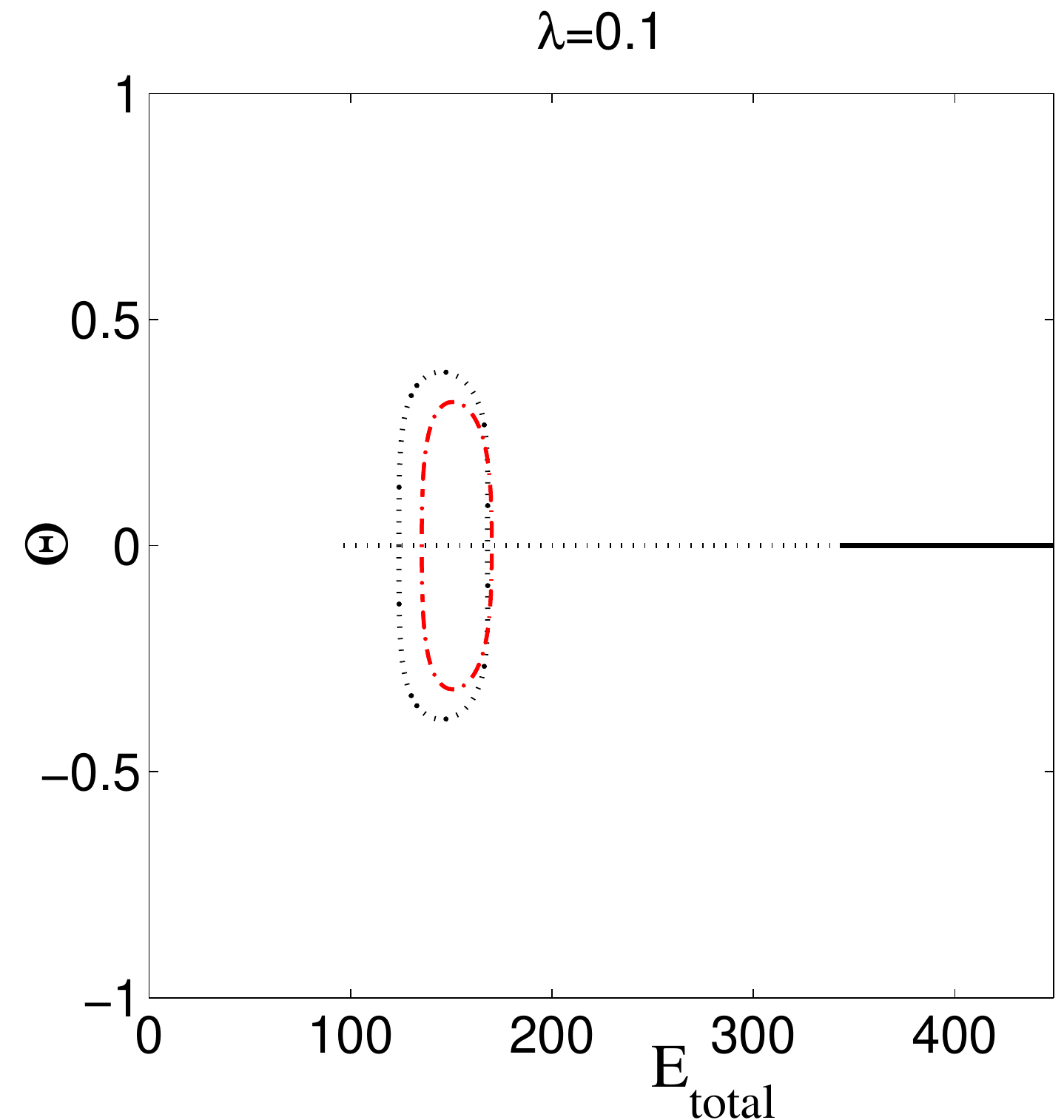}} %
\subfigure[]{\includegraphics[width=2.2in]{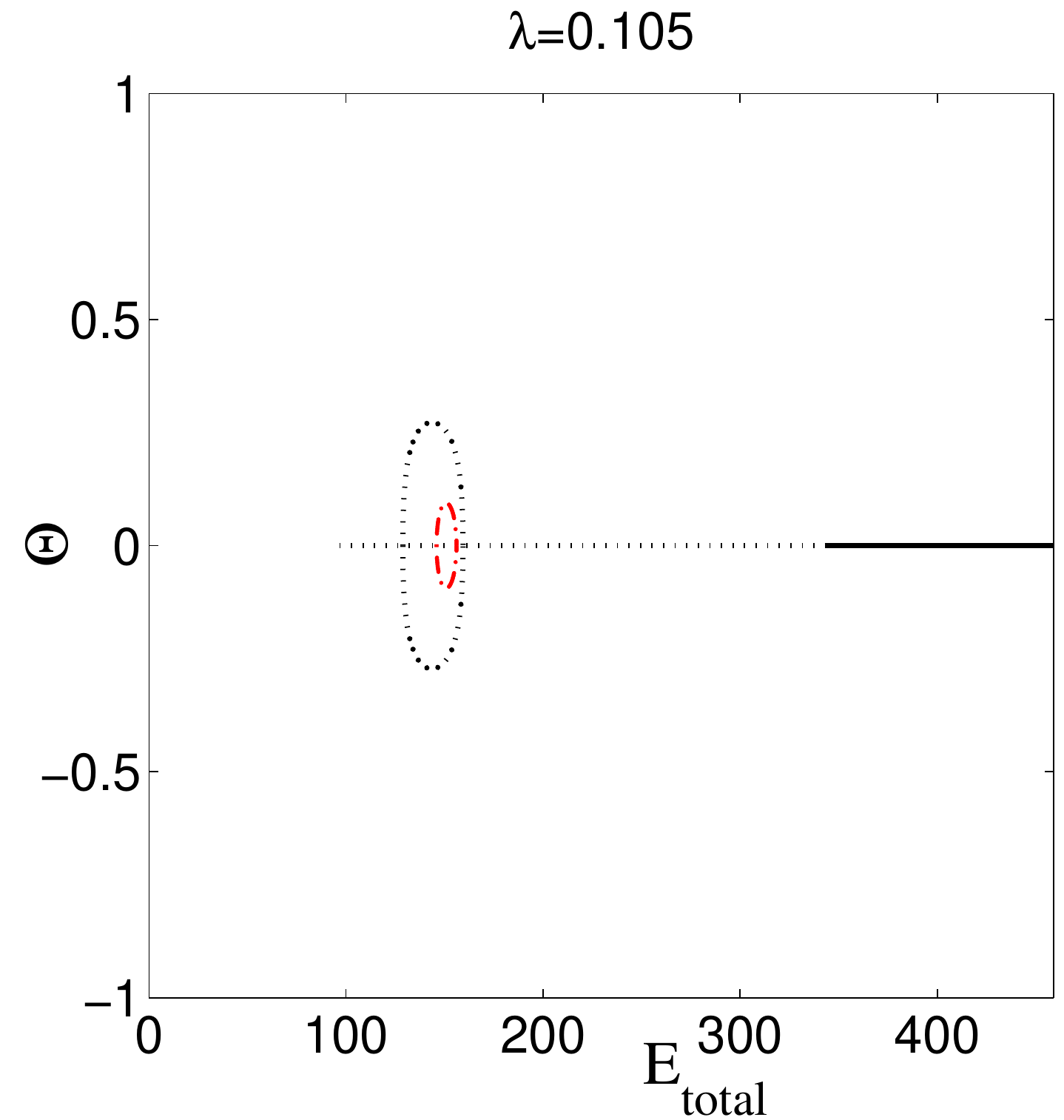}}
\caption{(Color online) The same as in Fig.~\protect\ref{BifLoopS0}, but for
vortex solitons with $s=1$. See also Fig.~\protect\ref%
{Stability_Lambda005_s1-a} below for details of the stability.}
\label{BifLoopS1}
\end{figure}

\begin{figure}[tbp]
\subfigure[]{\includegraphics[width=2.2in]{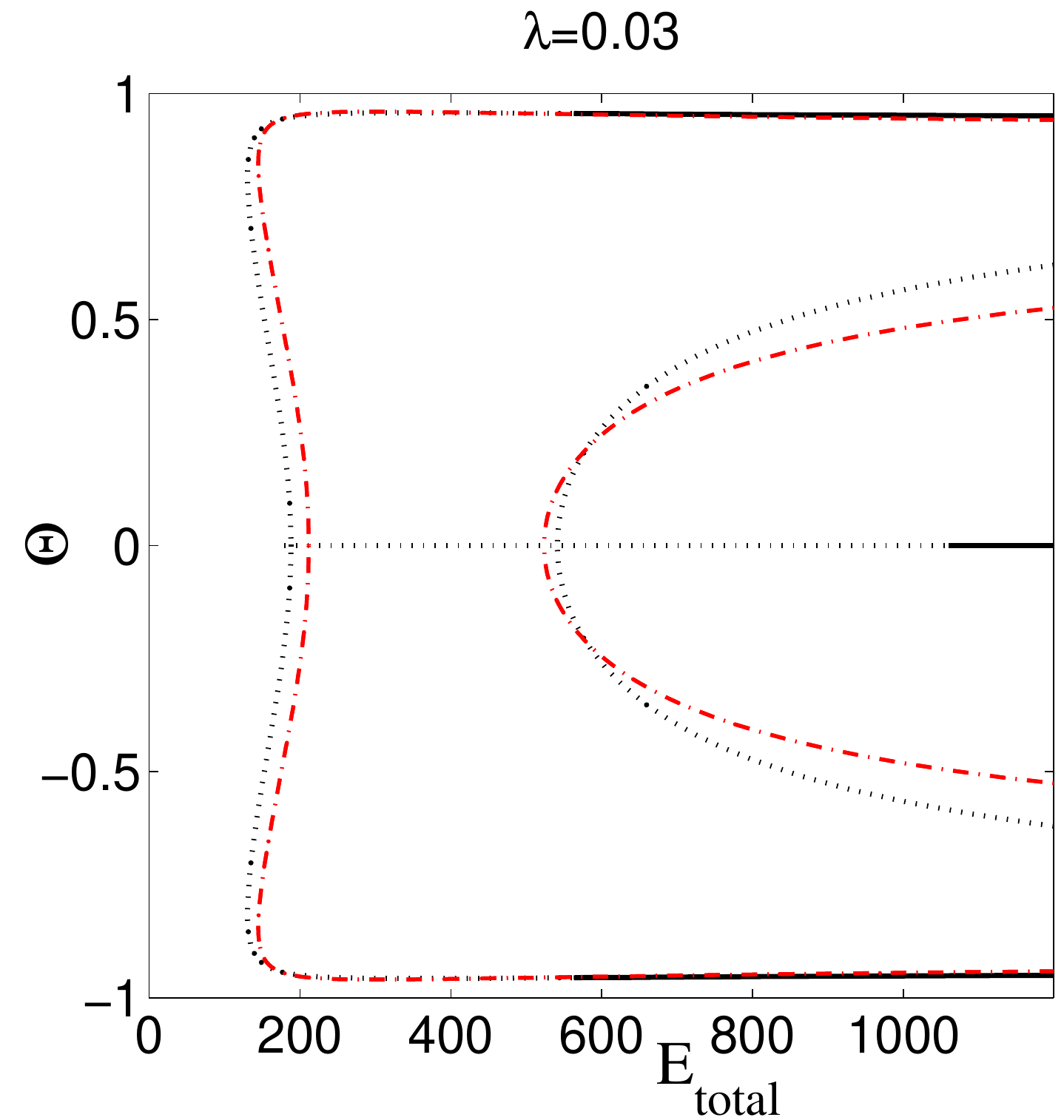}} %
\subfigure[]{\includegraphics[width=2.2in]{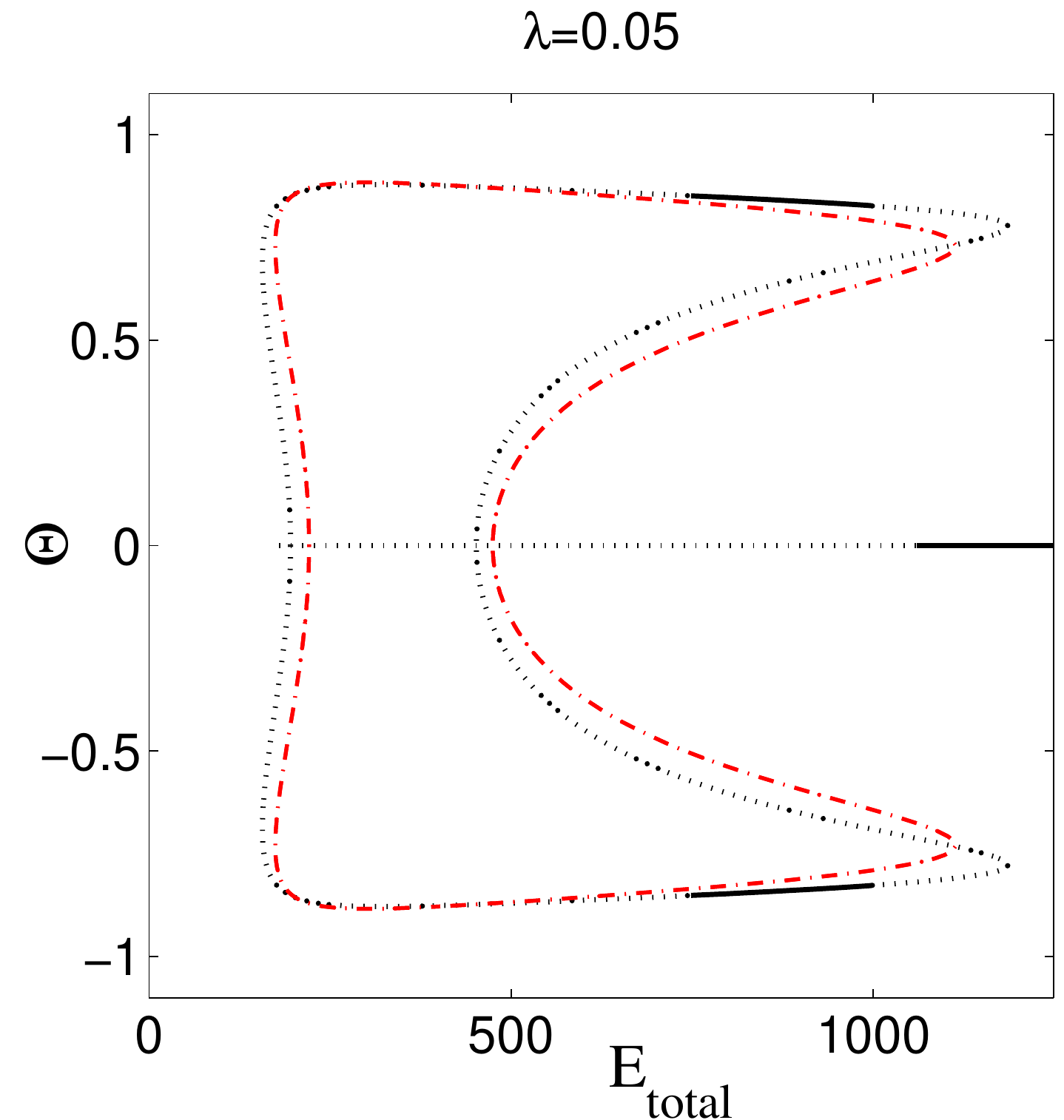}} %
\subfigure[]{\includegraphics[width=2.2in]{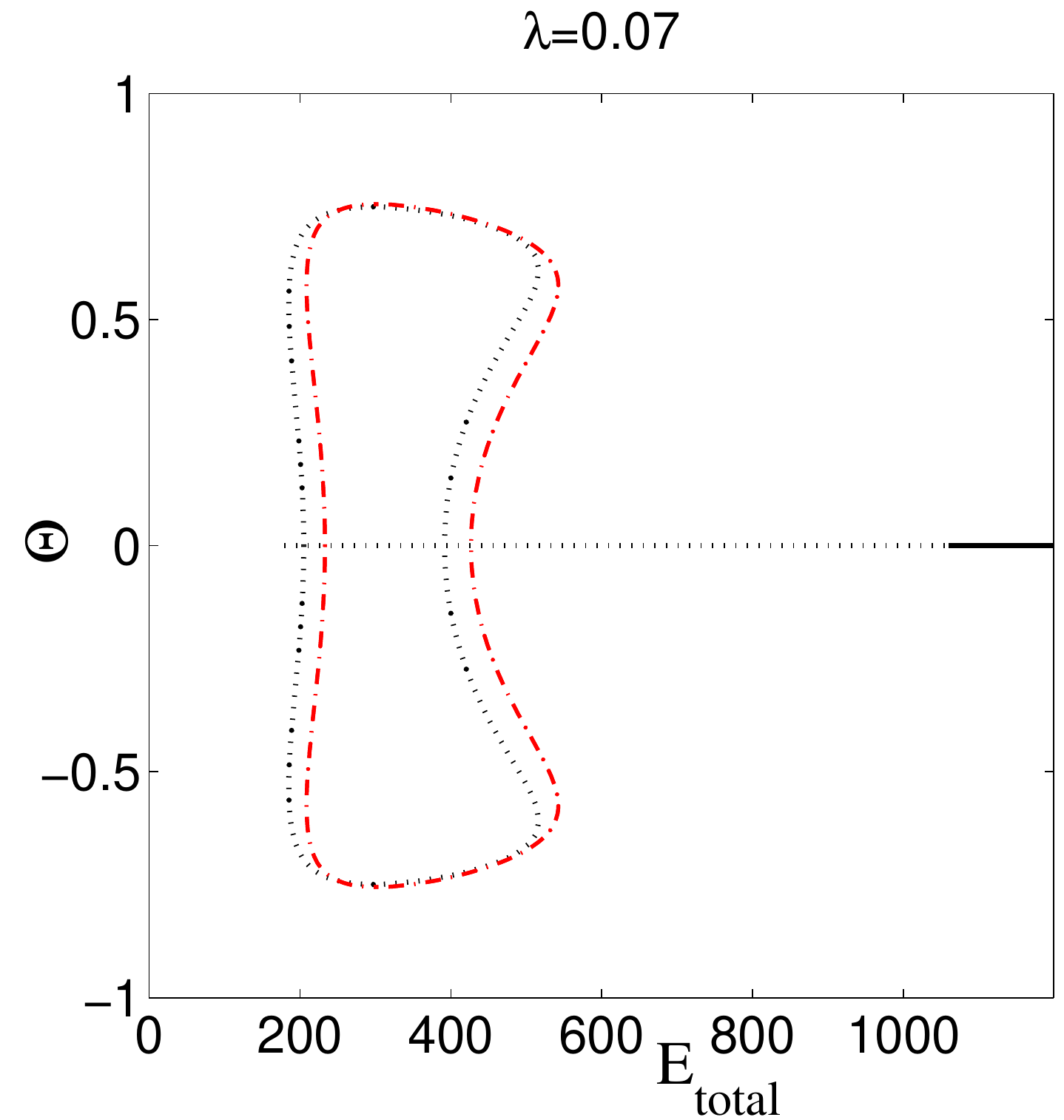}} \newline
\subfigure[]{\includegraphics[width=2.2in]{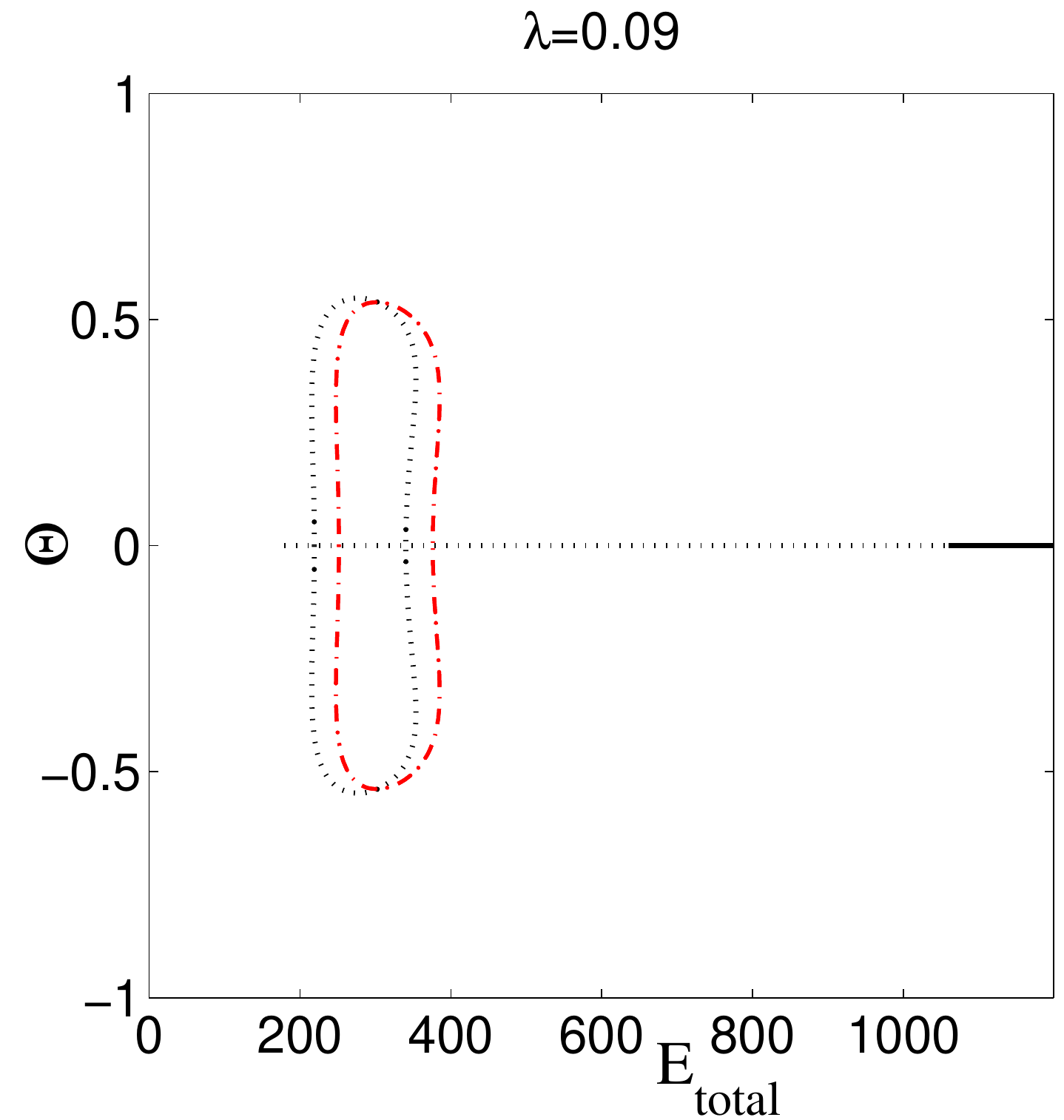}} %
\subfigure[]{\includegraphics[width=2.2in]{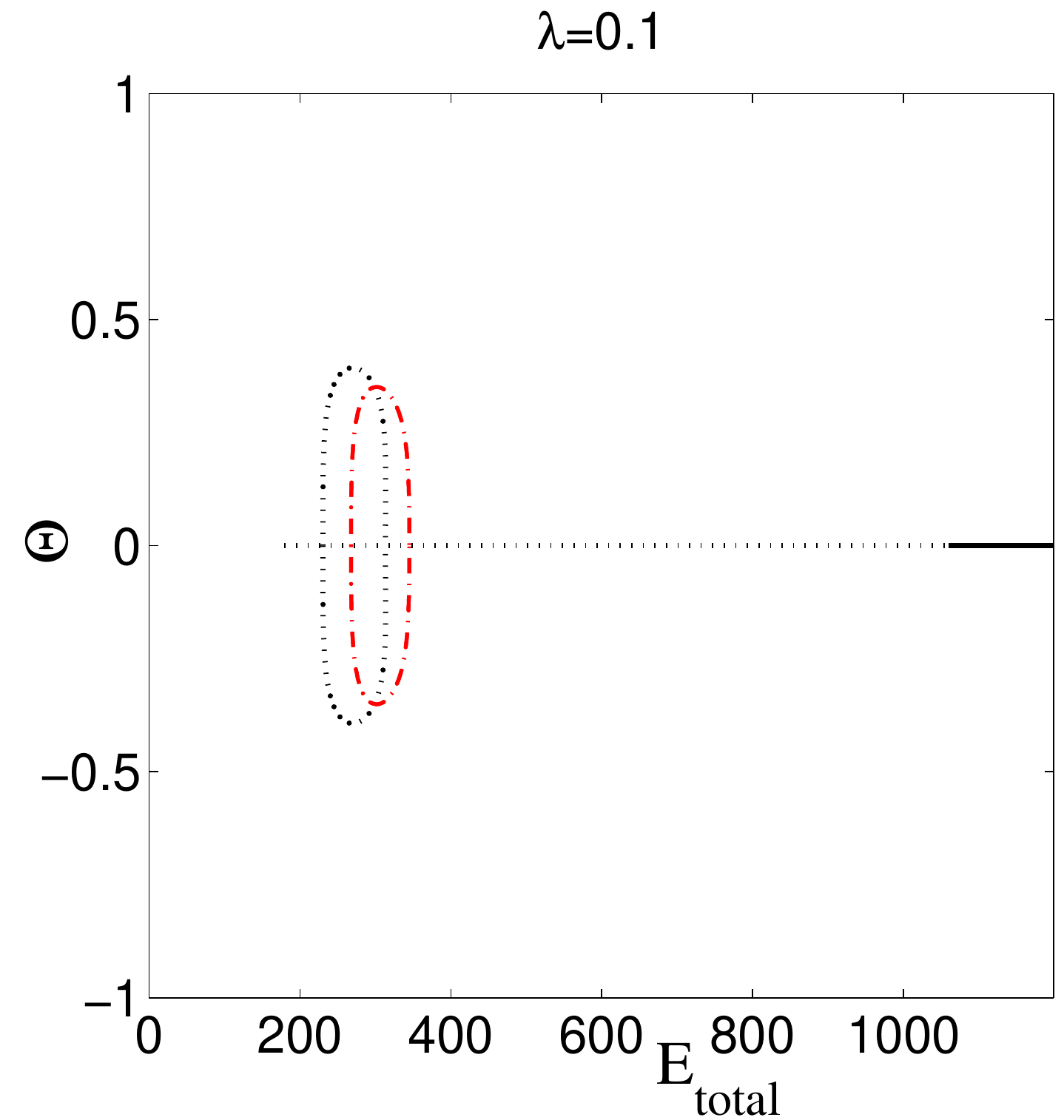}} %
\subfigure[]{\includegraphics[width=2.2in]{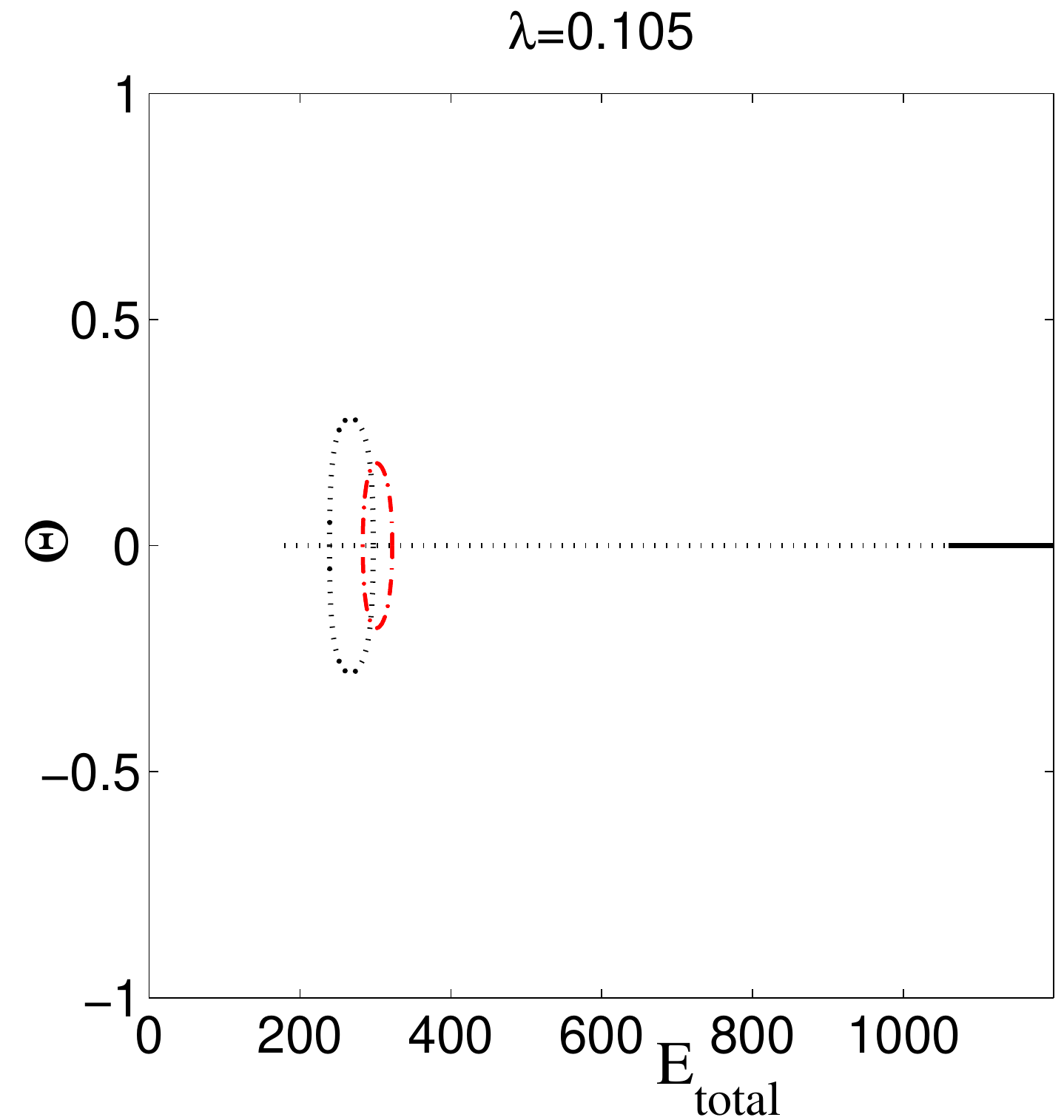}}
\caption{(Color online) The same as in Figs.~\protect\ref{BifLoopS0} and~%
\protect\ref{BifLoopS1}, but for vortex solitons with $s=2$. See also Fig.~%
\protect\ref{Stability_Lambda005_s2-a} below for details of the stability.}
\label{BifLoopS2}
\end{figure}

Similar to its 1D counterpart \cite{Albuch}, the present system features
bifurcation loops accounting for the transition from symmetric solitons to
the asymmetric ones and back. Several generic examples of the bifurcation
loops are shown in Figs.~\ref{BifLoopS0},~\ref{BifLoopS1} and~\ref{BifLoopS2}%
, for $s=0,1$ and $2$, respectively. As in the 1D case, the loops shrink as
the coupling constant, $\lambda $, increases, and they expand as $\lambda $
decreases. In the limit of $\lambda \rightarrow 0$, the loops open up in the
direction of $E\rightarrow \infty $.

For $s=0$, the loop collapses and disappears at $\lambda _{\mathrm{max}%
}^{(s=0)}\approx 0.0964$. Up to $\lambda \approx 0.0852$, both the direct
bifurcation and the reverse one, which closes the loop, are subcritical,
giving rise to two regions of bistability (which may also be called \textit{%
tristability}, as the asymmetric soliton always exists in two copies, which
are specular images to each other). This bifurcation picture is different
from the one obtained in the 1D model, in which the direct bifurcation is
\emph{always} supercritical. In the narrow interval of $0.0852<\lambda
<0.0891$, the direct bifurcation in the present model is supercritical,
while the reverse one remains subcritical. With the further increase of $%
\lambda $ up to the point of the disappearance of the loop, $0.0891<\lambda
<0.0964$, both the direct and reverse bifurcations are supercritical, and
the loop's shape is completely convex, featuring no bistability.

The stability of all the branches of the fundamental ($s=0$) solitons
strictly follows criteria of the elementary bifurcation theory \cite{Bif}.
In particular, the branches generated by super- and subcritical bifurcations
emerge as, respectively, stable and unstable ones, and the character of the
stability changes when a branch passes a turning point. On the other hand,
the \textit{Vakhitov-Kolokolov criterion}, $dE/dk>0$, which in many models
with attractive nonlinearities is a necessary stability condition \cite%
{VK,Agrawal,Barge,Alfano}, does not catch the instability of solitons
related to the symmetry-breaking bifurcations, which is a known fact too
\cite{Albuch,Gubeskys}. The stability properties are different for vortices,
as they may be additionally unstable against azimuthal perturbations \cite%
{Crasovan,Stability,Warchall}. The analysis of the azimuthal instability of
vortex solitons in the present model is reported in the next section.

As seen in Fig. \ref{BifLoopS1}, the bifurcation picture for $s=1$ (without
referring to the stability, for the time being) is very similar to that for $%
s=0$. The solution branches form a loop, with both the direct and reverse
bifurcations being subcritical in the interval of $0<\lambda <0.0998$. At $%
0.0998<\lambda <0.1018$, the direct bifurcation is supercritical, while the
reverse one is still subcritical. The two bifurcations are supercritical,
corresponding to the completely convex loop, at $0.1018<\lambda
<0.110\approx \lambda _{\max }^{(s=1)}$, up to the point where the
bifurcation loop ceases to exist.

The bifurcation loops were also constructed for vortex solitons with spin $%
s=2.$ As seen in Fig. \ref{BifLoopS2}, the direct and reverse bifurcations
are subcritical at $0<\lambda <0.1001$. There is a tiny region ($%
0.1001<\lambda <0.1015$) in which the direct bifurcation is supercritical,
while the reverse one stays subcritical. At $\lambda \approx 0.1015$ the
reverse bifurcation also switches to the supercritical type, making the loop
completely convex. It keeps this shape up to the point of the disappearance
of the loop, at $\lambda _{\max }^{(s=2)}\approx 0.1102$.

\section{The linear-stability analysis}

\label{sec:stability} The stability of stationary solutions (\ref{psiphi})
was explored using the standard approach: we take a perturbed solution with%
\begin{eqnarray}
\psi (r,\theta ,z) &=&\left[ U(r)+\delta U(r,\theta ,z)\right] \exp
(is\theta )\exp (ikz),  \notag \\
\phi (r,\theta ,z) &=&\left[ V(r)+\delta V(r,\theta ,z)\right] \exp
(is\theta )\exp (ikz),  \label{UVpert}
\end{eqnarray}%
where small perturbations are looked for, as usual, in the form of angular
eigenmodes, with an integer azimuthal perturbation index, $n$, and the
respective instability growth rate $\gamma _{n}$ ,%
\begin{eqnarray}
\delta U &=&\left[ U_{+}(r,z)\exp (in\theta )+U_{-}(r,z)\exp (-in\theta )%
\right] \exp (\gamma _{n}z),  \notag \\
\delta V &=&\left[ V_{+}(r,z)\exp (in\theta )+V_{-}(r,z)\exp (-in\theta )%
\right] \exp (\gamma _{n}z).  \label{Perturb}
\end{eqnarray}%
Taking the perturbation in this form leads to a closed system of
linearized equations generated by the substitution of expressions
(\ref{UVpert}) and (\ref{Perturb}) into Eqs.~(\ref{ModelUV}):
\begin{gather}
-kU_{+}+i\gamma _{n}U_{+}+\frac{\partial ^{2}U_{+}}{\partial r^{2}}+\frac{1}{%
r}\frac{\partial U_{+}}{\partial r}-\frac{(s+n)^{2}}{r^{2}}U_{+}  \notag \\
+(2-3U^{2})U^{2}U_{+}+(1-2U^{2})U^{2}U_{-}^{\ast }=-\lambda V_{+};  \notag \\
-kU_{-}+i\gamma _{n}U_{-}+\frac{\partial ^{2}U_{-}}{\partial r^{2}}+\frac{1}{%
r}\frac{\partial U_{-}}{\partial r}-\frac{(s-n)^{2}}{r^{2}}U_{-}  \notag \\
+(2-3U^{2})U^{2}U_{-}+(1-2U^{2})U^{2}U_{+}^{\ast }=-\lambda V_{-};  \notag \\
-kV_{+}+i\gamma _{n}V_{+}+\frac{\partial ^{2}V_{+}}{\partial r^{2}}+\frac{1}{%
r}\frac{\partial V_{+}}{\partial r}-\frac{(s+n)^{2}}{r^{2}}V_{+}  \notag \\
+(2-3V^{2})V^{2}V_{+}+(1-2V^{2})V^{2}V_{-}^{\ast }=-\lambda U_{+};  \notag \\
-kV_{-}+i\gamma _{n}V_{-}+\frac{\partial ^{2}V_{-}}{\partial r^{2}}+\frac{1}{%
r}\frac{\partial V_{-}}{\partial r}-\frac{(s-n)^{2}}{r^{2}}V_{-}  \notag \\
+(2-3V^{2})V^{2}V_{-}+(1-2V^{2})V^{2}V_{+}^{\ast }=-\lambda U_{-}~.
\label{BVP}
\end{gather}%
These equations are to be solved with the boundary conditions, which demand $%
\{U_{\pm },V_{\pm }\}\rightarrow r^{|s\pm n|}$ at $r\rightarrow 0$, and the
exponential decay of the perturbation eigenmodes at $r\rightarrow \infty $.

There are several available numerical methods for solving such a
boundary-value problem and finding the perturbation growth rates, $\gamma
_{n}$ \cite{Enns,Torner,Akhm}. We treated Eqs.~(\ref{BVP}) as an algebraic
eigenvalue problem for $\gamma _{n}$, and solved it directly, using a
finite-difference method. The largest instability-growth rate was identified
as the real part of the most unstable eigenvalue, $\max \{\mathrm{Re}(\gamma
_{n})\}$. This approach has confirmed the azimuthal stability of the
fundamental solitons ($s=0$) and revealed instability regions for vortices
with $s=1,2$.

\begin{figure}[tbp]
\subfigure[]{\includegraphics[width=2.5in]{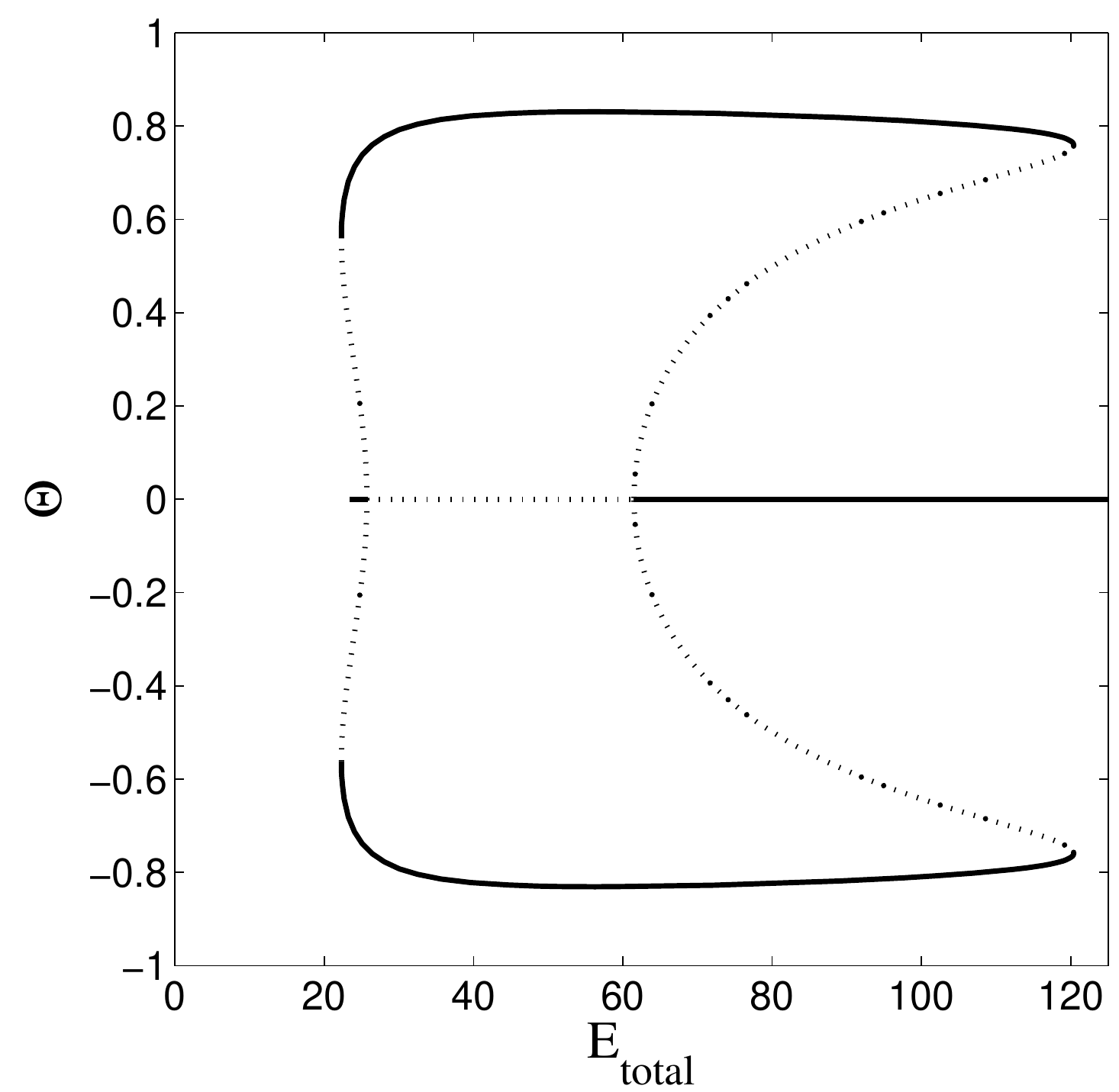}%
\label{Stability_Lambda005_s0-a}} \quad \subfigure[]{%
\includegraphics[width=2.5in]{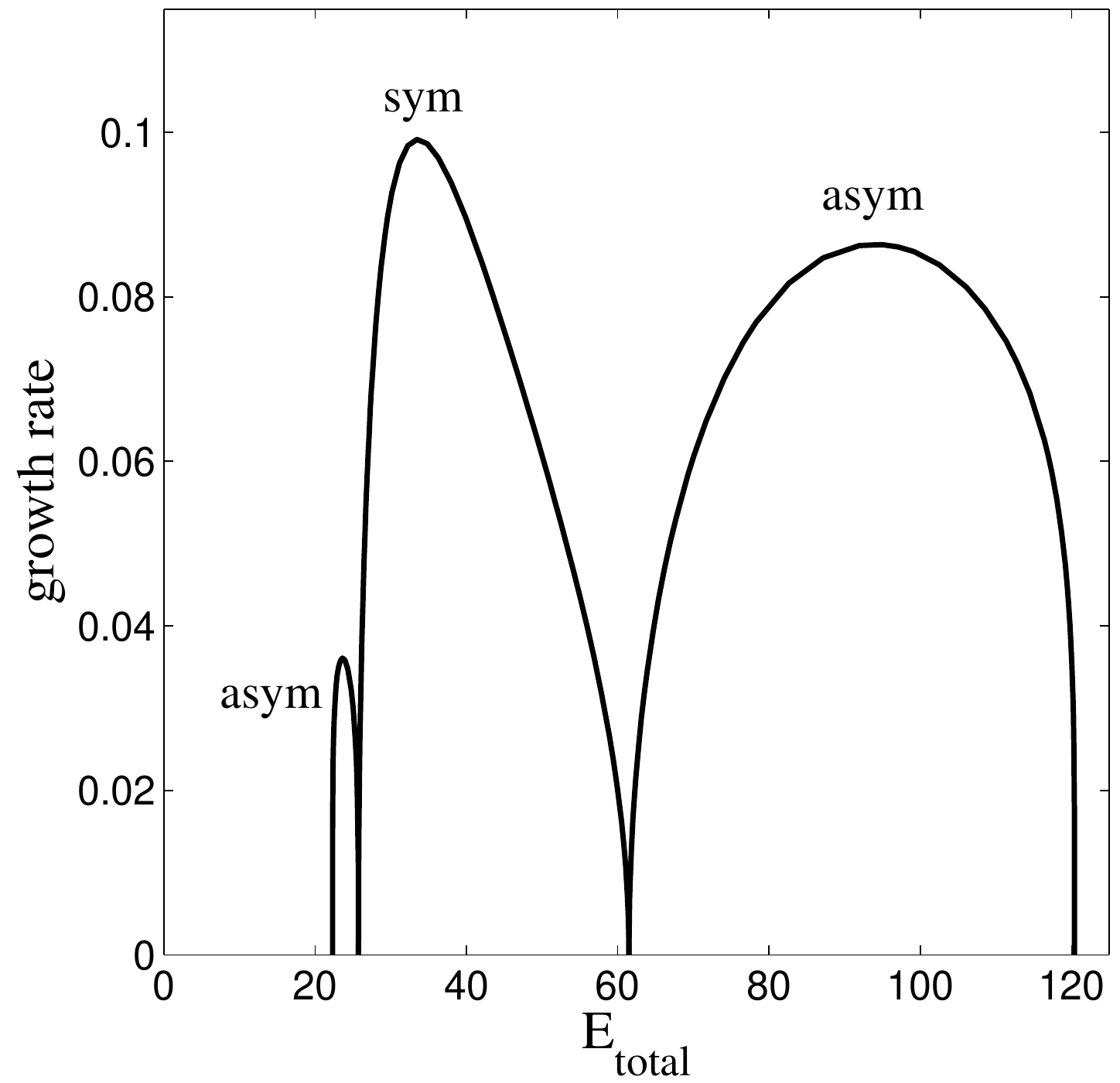}%
\label{Stability_Lambda005_s0-b}}
\caption{(a) The bifurcation diagram, for the fundamental solitons ($s=0$)
at $\protect\lambda =0.05$, cf. Fig.~\protect\ref{BifLoopS0}. (b) The
corresponding maximum growth rate of perturbation eigenmodes for the
symmetric and asymmetric solutions.}
\label{Stability_Lambda005_s0}
\end{figure}

For $s=0$, the most dangerous perturbation azimuthal index is $n=0$ (i.e.,
as said above, the fundamental solitons are not destabilized by azimuthal
perturbations). An example of the ensuing curves which display the maximum
growth rate, in the case corresponding to the bifurcation diagram that
features the double bistability at $\lambda =0.05$ (see Fig.~\ref{BifLoopS0}%
), is shown in Fig.~\ref{Stability_Lambda005_s0}. As expected, unstable are
backward-going portions of the asymmetric solution branches, i.e., in the
region between the bifurcation points and turning points of the bifurcation
curves. As might be expected too, the symmetric solutions are unstable in
the entire region between the points of the direct and reverse bifurcations.

\begin{figure}[tbp]
\subfigure[]{\includegraphics[width=2.3in]{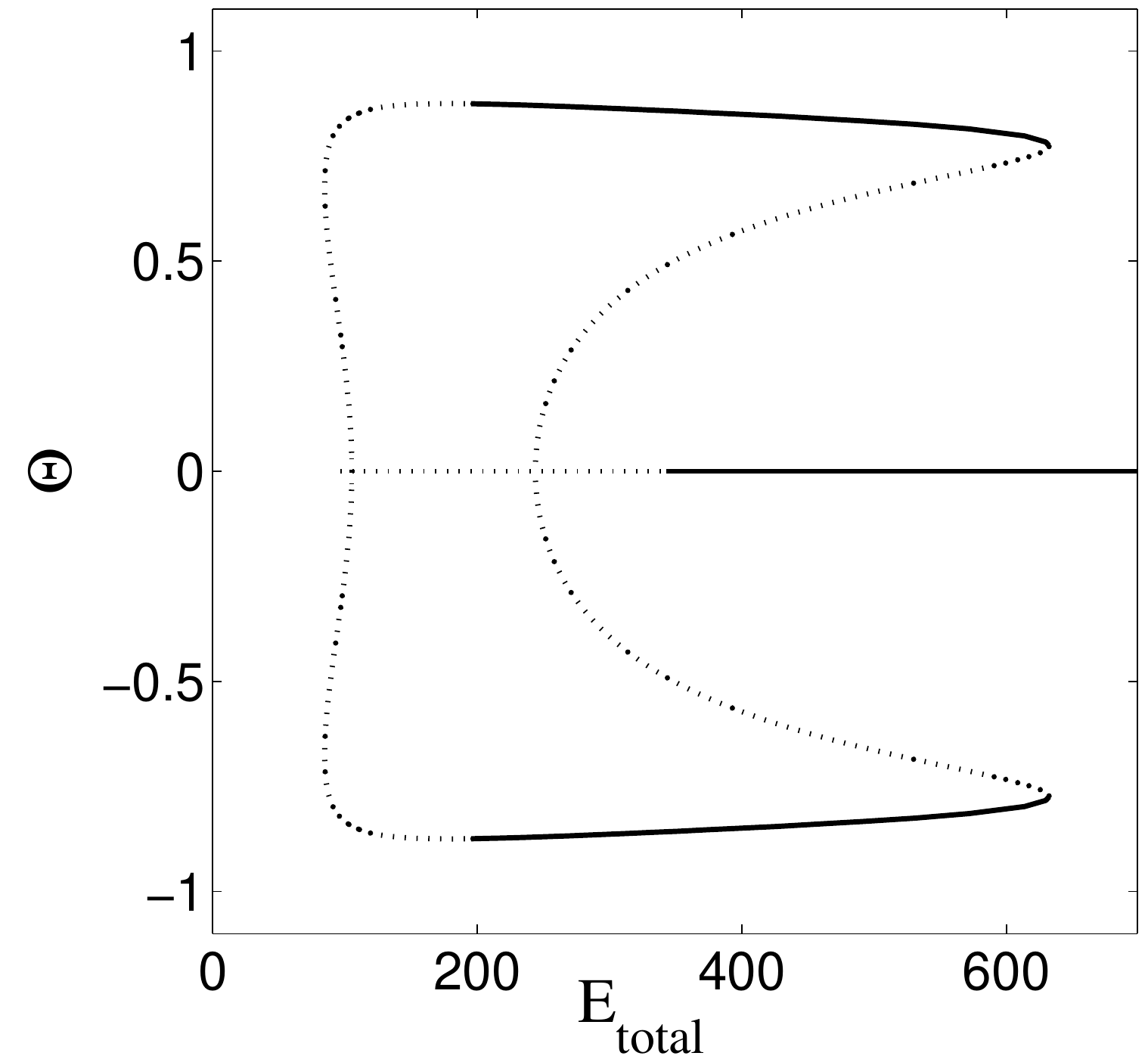} \label{Stability_Lambda005_s1-a}}
\subfigure[]{\includegraphics[width=2.3in]{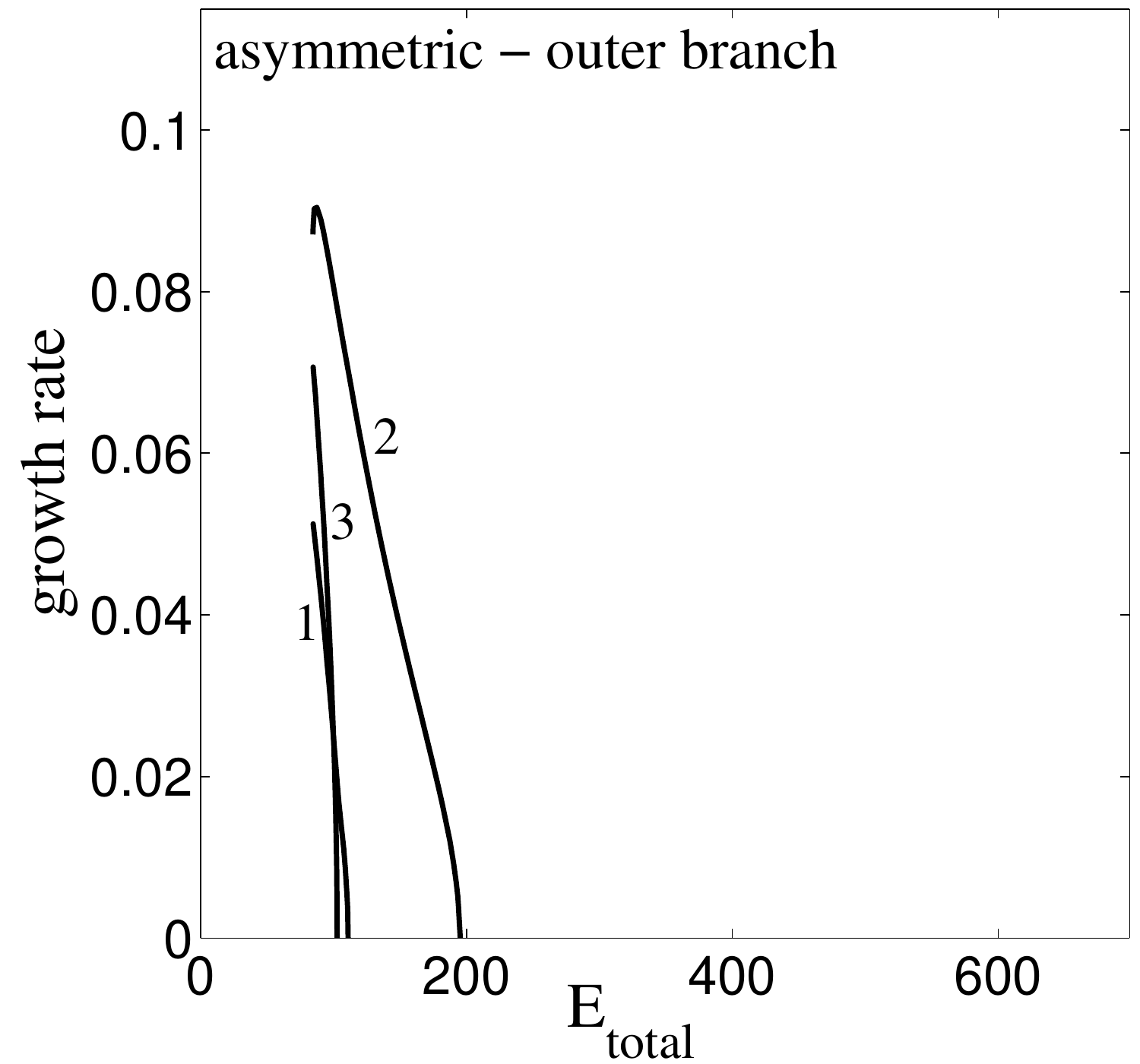} \label{Stability_Lambda005_s1-b}} \\
\subfigure[]{\includegraphics[width=2.3in]{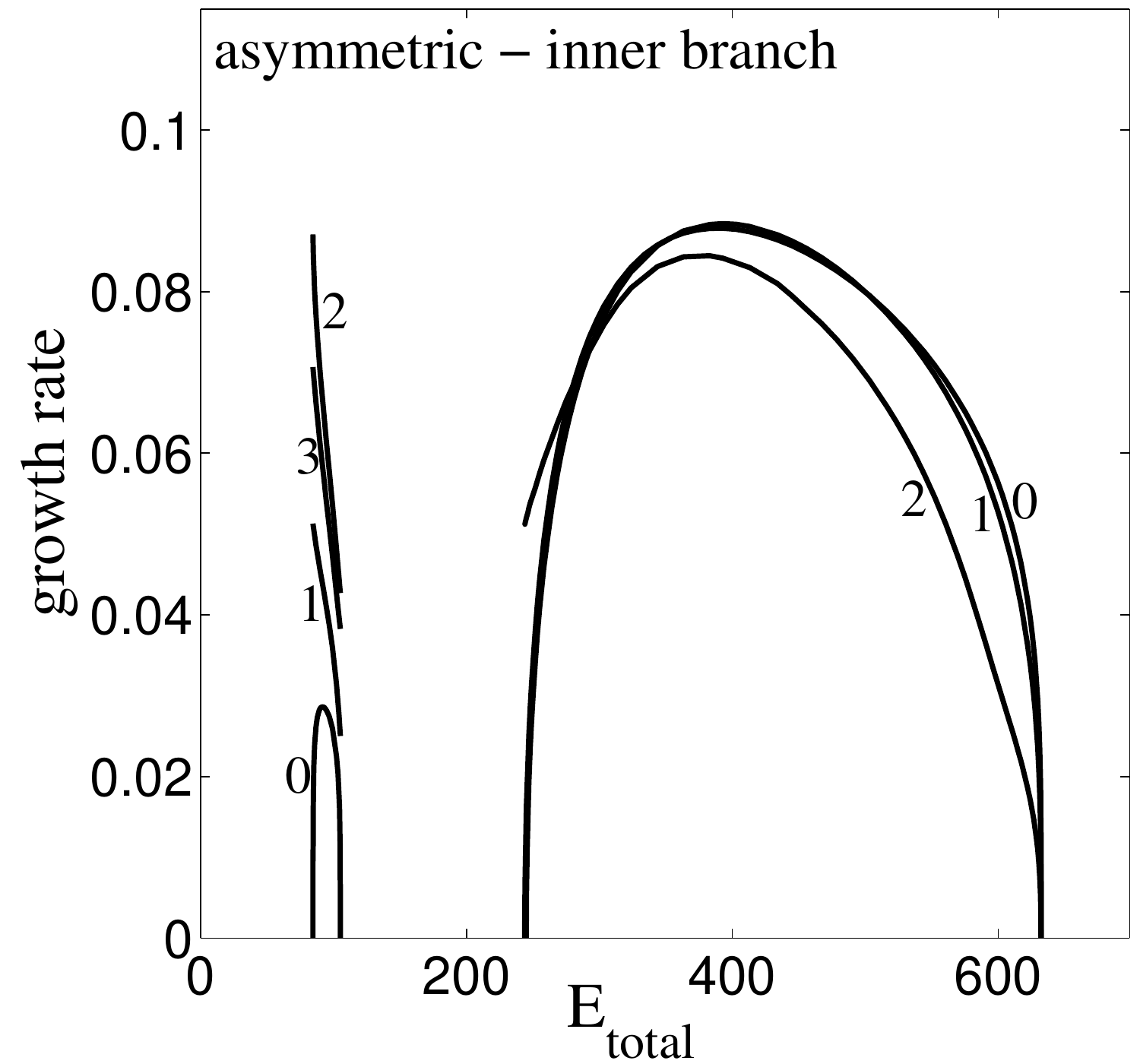} \label{Stability_Lambda005_s1-c}}
\subfigure[]{\includegraphics[width=2.3in]{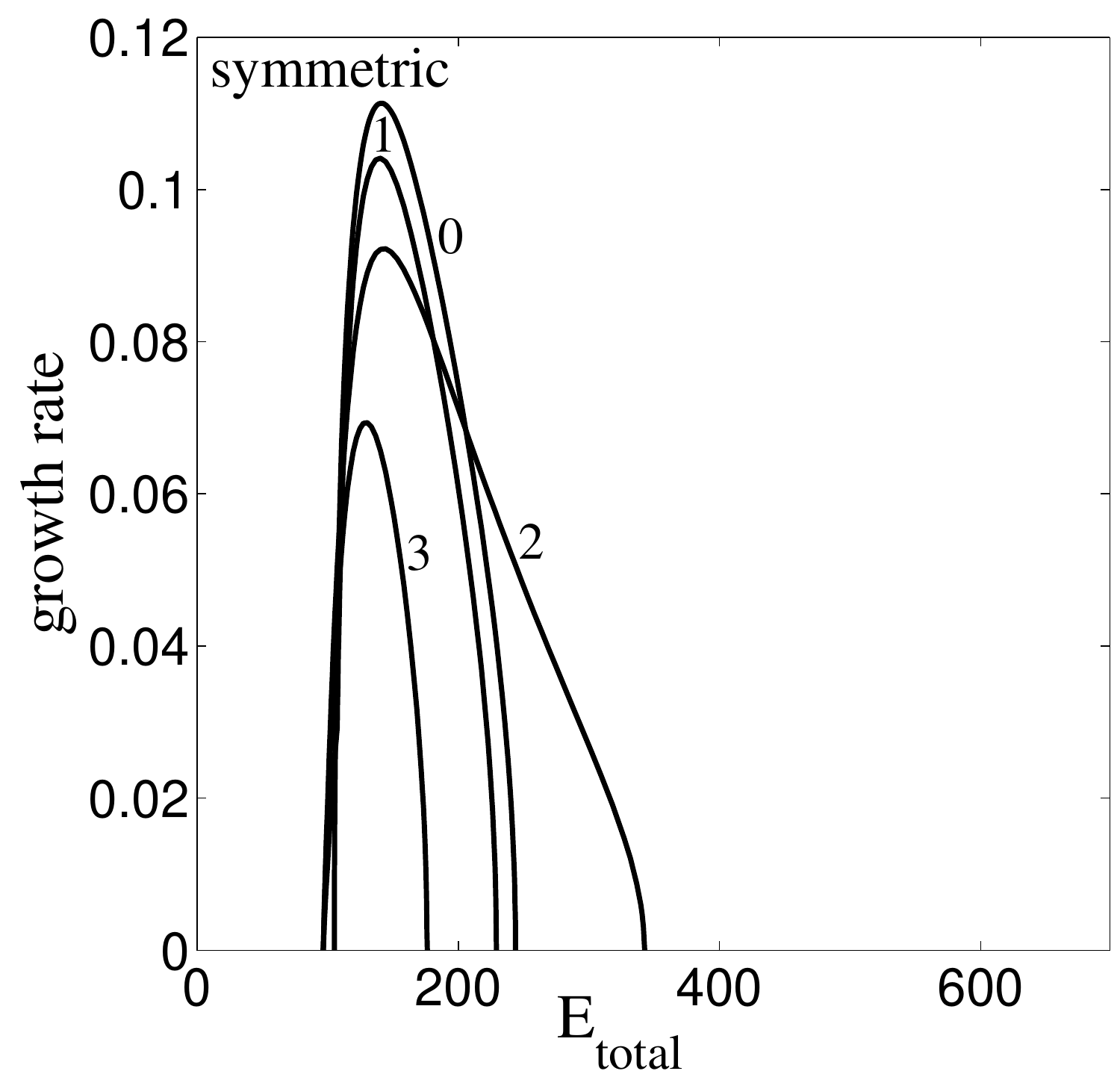} \label{Stability_Lambda005_s1-d}}
\caption{ (a) The bifurcation diagram for the vortices with $s=1$ at $%
\protect\lambda =0.05$, cf. Fig.~\protect\ref{BifLoopS1}. The corresponding
maximum growth rates of perturbation eigenmodes for the inner and outer
asymmetric branches and the symmetric one are shown in panels (b), (c), and
(d), respectively. The labels near the curves indicate the mode's azimuthal
number. Note that the plots which appear aborted in panels (b) and (c) are
actually continuations of each other and of the plots in panel (d). This is
in accordance with the fact that the outer and inner branches of the
asymmetric states are linked at the turning points of the bifurcation
diagram, and the symmetric and inner asymmetric branches are linked at the
bifurcation points.}
\label{Stability_Lambda005_s1}
\end{figure}

For vortices with $s=1$, the growth-rate curves pertaining to $n=0$ feature
the same behavior as for the fundamental solitons. However, additional
unstable eigenmodes, for both the symmetric and asymmetric solutions, were
found with azimuthal indices $n=1,2,3$. On the other hand, no unstable
perturbations were detected for larger $n>3$. A typical example displaying
all the existing unstable eigenvalues for $s=1$ is presented in Fig.~\ref%
{Stability_Lambda005_s1}, for the same coupling constant as in Fig.~\ref%
{Stability_Lambda005_s0}, $\lambda =0.05$. In particular, panel (d) in this
figure shows the results for the symmetric solutions. The eigenmode that
remains the last unstable one with the increase of $E_{\mathrm{total}}$
pertains to $n=2$. This instability ceases when the energy attains value $E_{%
\mathrm{total}}\approx 340$, which corresponds to propagation constant $%
k\approx 0.2$. This value is exactly the expected one, according to relation
$k^{\mathrm{(sta)}}\approx 0.15+\lambda $, where $k\approx 0.15$ is the
known stability threshold in the single-component model ($\lambda =0$), for $%
s=1$ \cite{Stability}. Further, the results for unstable eigenmodes
disturbing the two inner asymmetric branches (the portions of the branches
that commence at the bifurcation point and end at the turning points) are
shown in panel \ref{Stability_Lambda005_s1}(c). At the respective edge
points, these curves are linked to their counterparts (continuations) in
panels~\ref{Stability_Lambda005_s1}(d) and~\ref{Stability_Lambda005_s1}(b),
the latter panel pertaining to the outer asymmetric branches. The conclusion
is that the inner portions are completely unstable (as well as in the case
of $s=0$, cf. Fig. \ref{Stability_Lambda005_s0}), while parts of the outer
branches are \emph{stable}. In the case shown in Fig. \ref%
{Stability_Lambda005_s1} (recall it pertains to $\lambda =0.05$), the
stability segment of the outer branch of the asymmetric solutions with $s=1$
is $195<E_{\mathrm{total}}<630$, which corresponds to $0.165<k<0.186$.

With the increase of $\lambda $, the stable section of the asymmetric branch
with $s=1$ shrinks, and it disappears at $\lambda \approx 0.067$ (which
still corresponds to the double-concave shape of the loop with both the
direct and reverse bifurcations of the subcritical type). At larger values
of $\lambda $, the instability accounted for by the perturbation mode with $%
n=2$ covers the entire asymmetric branch. With the further increase of $%
\lambda $, the instability regions corresponding to the eigenmodes with the
other values of $n$ also expand and gradually cover the entire asymmetric
branch. In all the cases that we have examined, the growth rate
corresponding to $n=2$ is always the largest one for the upper asymmetric
branch.

\begin{figure}[tbp]
\subfigure[]{\includegraphics[width=2.3in]{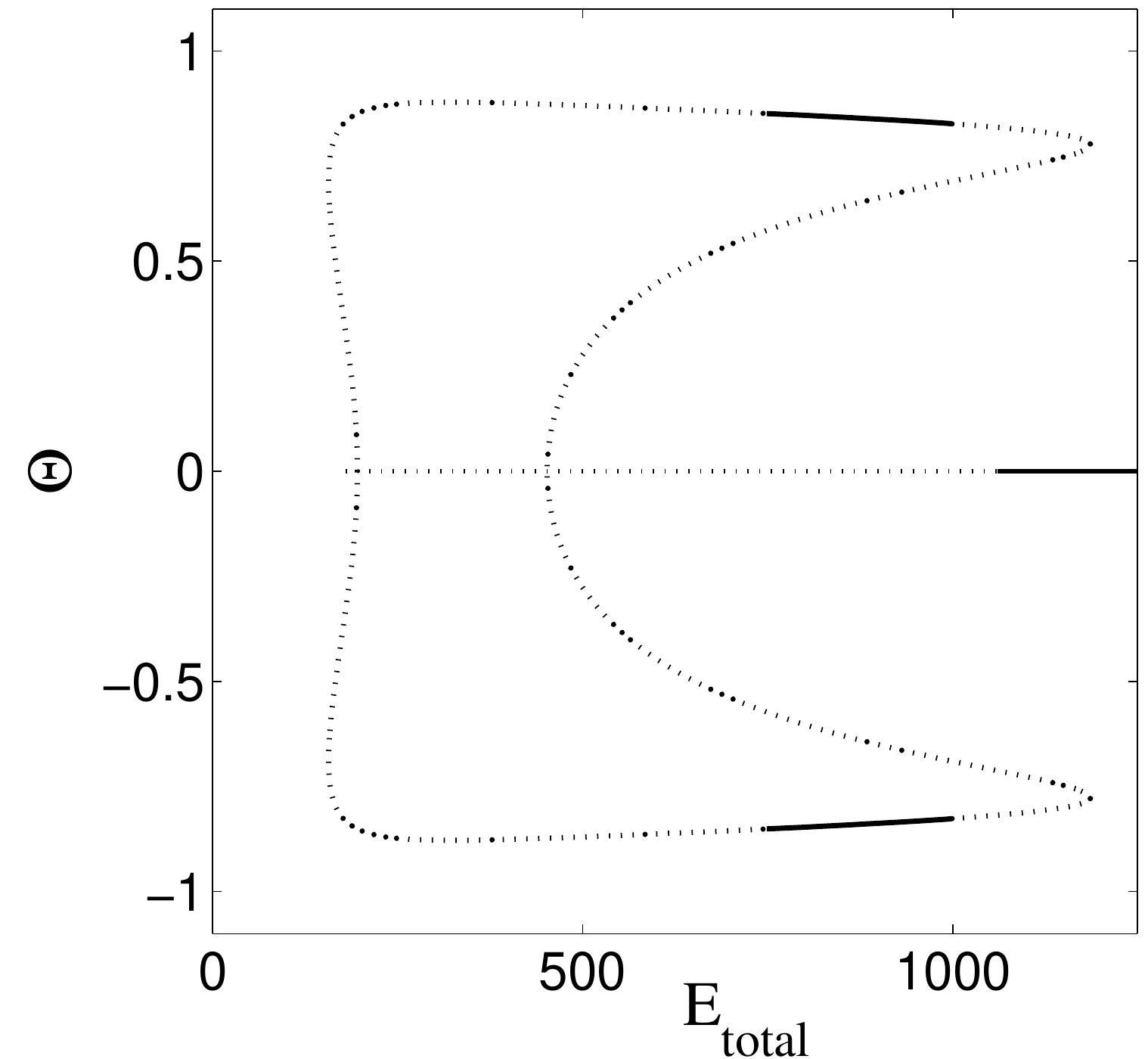}%
\label{Stability_Lambda005_s2-a}} \subfigure[]{%
\includegraphics[width=2.3in]{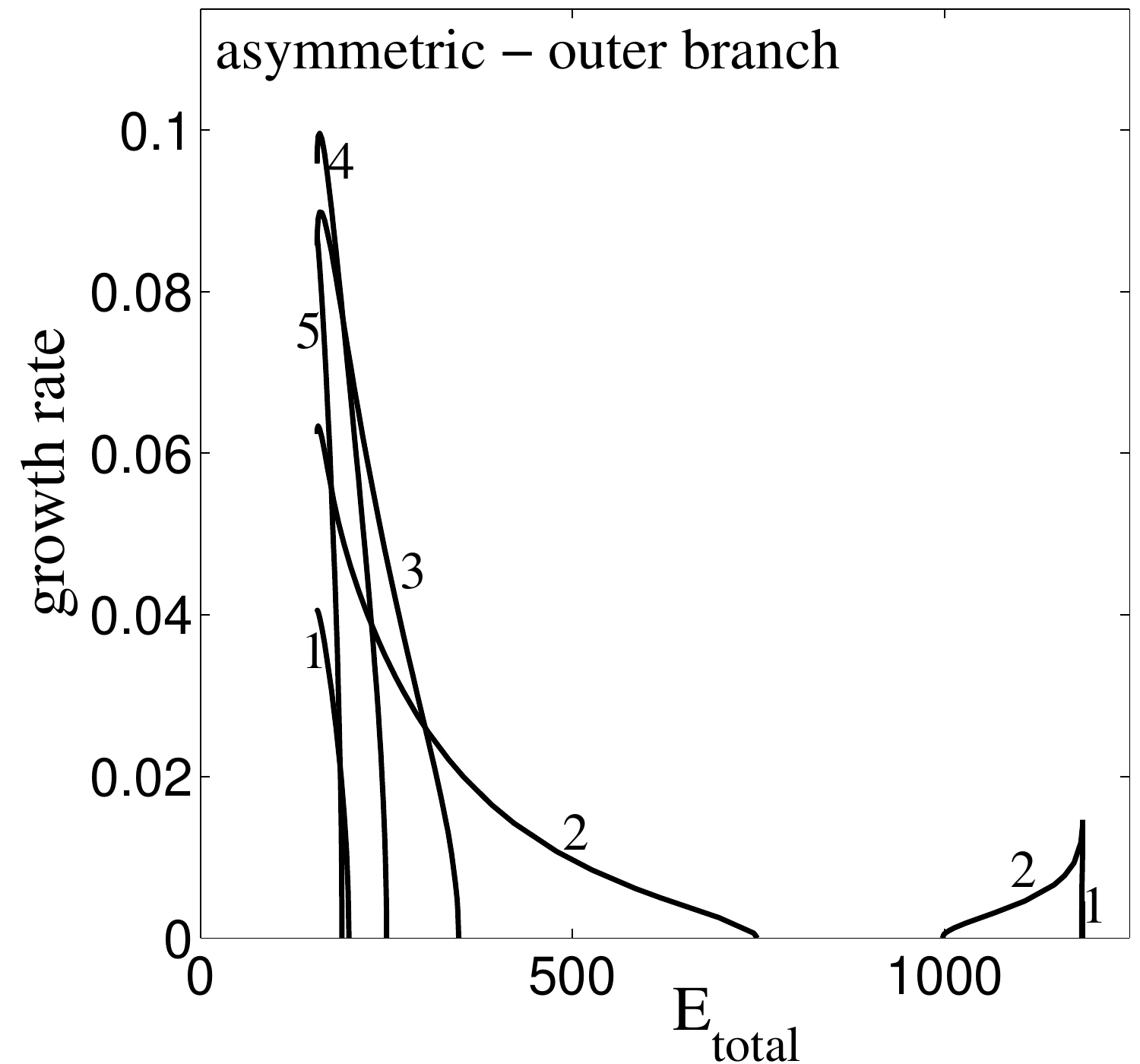}%
\label{Stability_Lambda005_s2-b}} \\
\subfigure[]{\includegraphics[width=2.3in]{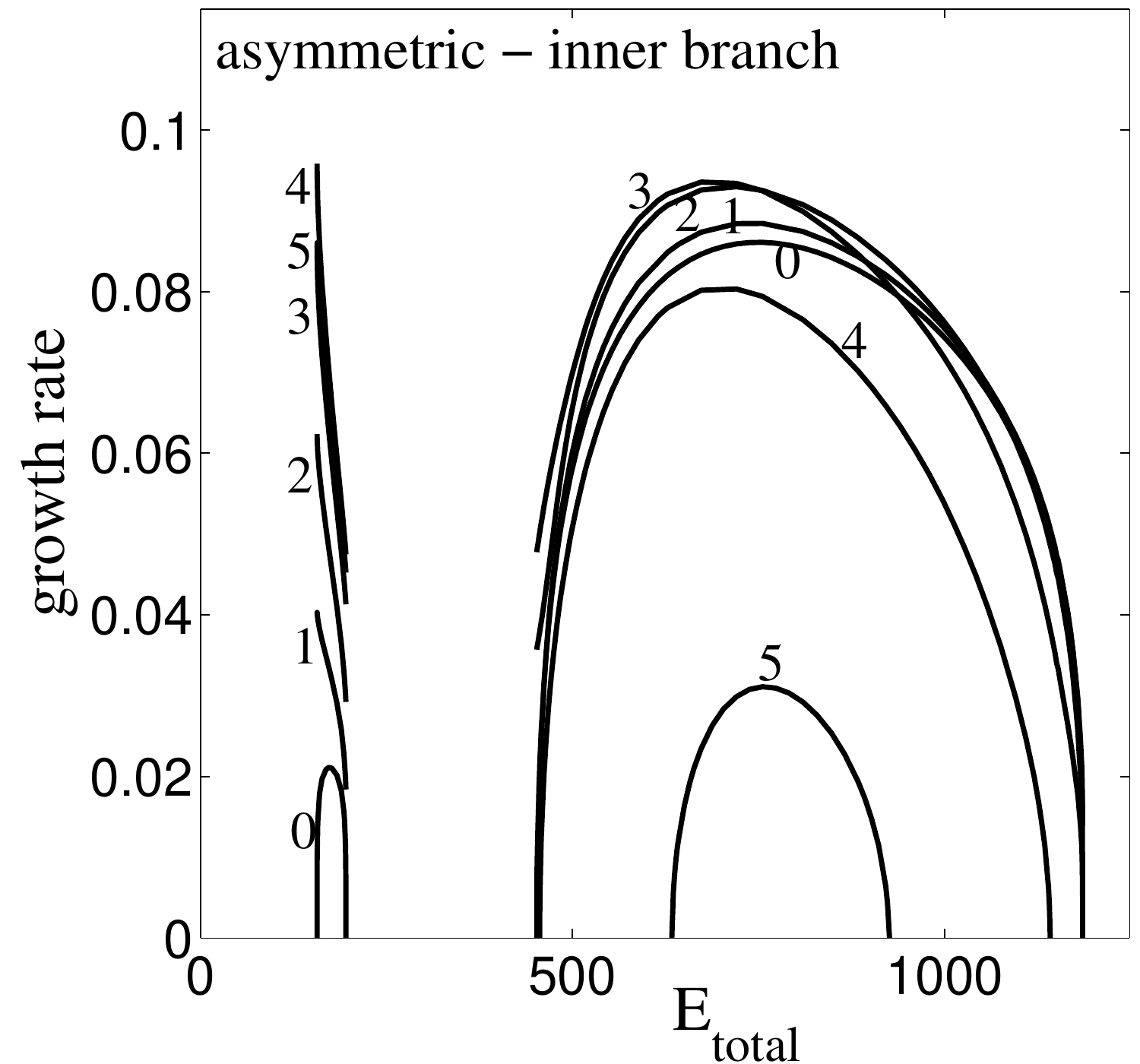}%
\label{Stability_Lambda005_s2-c}} \subfigure[]{%
\includegraphics[width=2.3in]{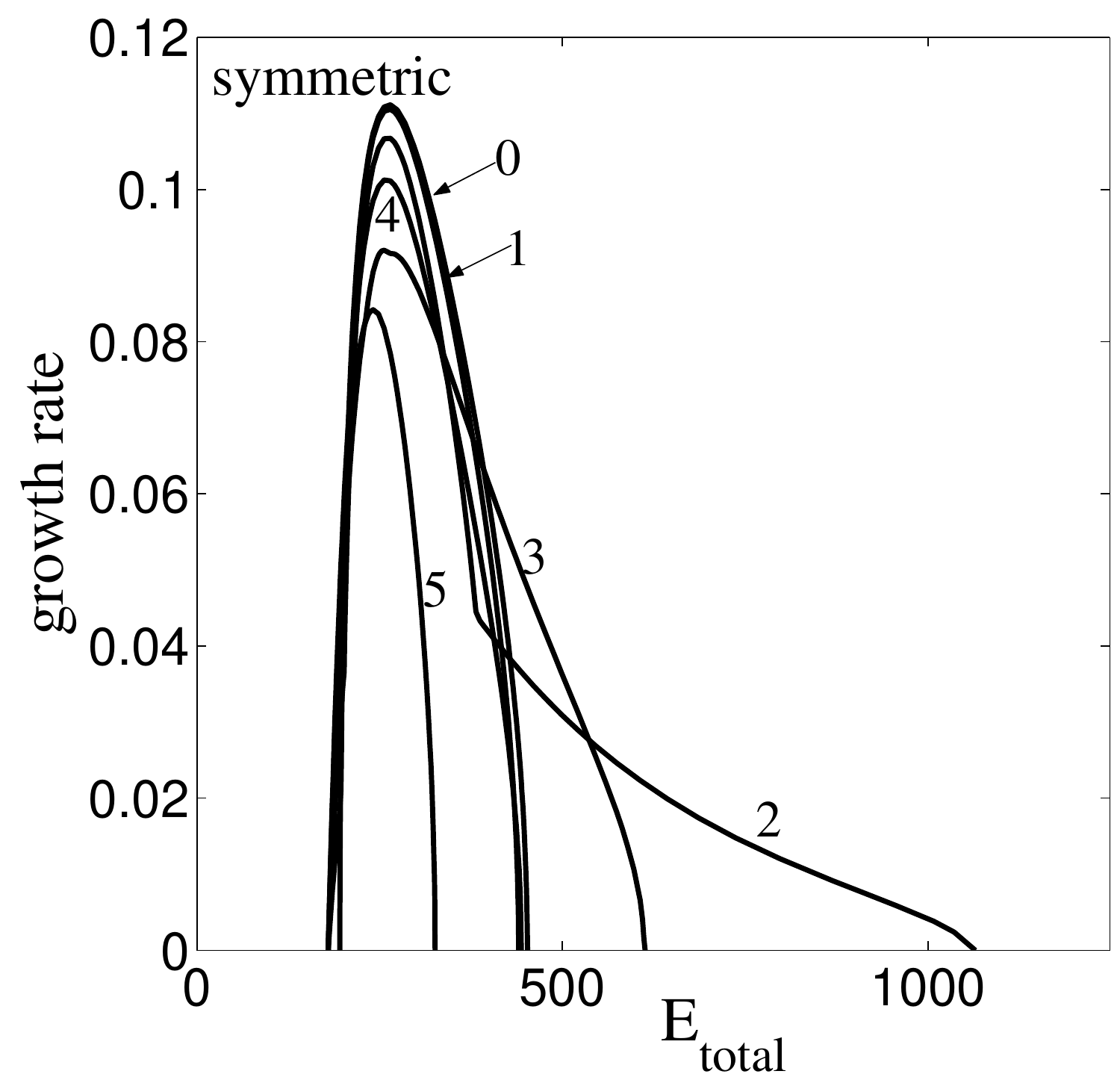}%
\label{Stability_Lambda005_s2-d}}
\caption{The same as in Fig.~\protect\ref{Stability_Lambda005_s1}, but for
the double vortices ($s=2$) at $\protect\lambda =0.05$.}
\label{Stability_Lambda005_s2}
\end{figure}

In the case of the double vortex, with $s=2$, results of the stability
analysis are presented in Fig.~\ref{Stability_Lambda005_s2}, again for $%
\lambda =0.05$. As in the case of $s=1$, the eigenmode that determines the
stability boundaries has $n=2$. However, in this case the growth rate for
perturbations with $n=2$ is not necessarily the highest for the outer
asymmetric branch, and the azimuthal index of the dominant perturbation
eigenmode switches from $n=4$ to $n=3$ and then to $n=2$. Similar to the
behavior of the vortex with $s=1$, the stability region of the asymmetric
stable solutions diminishes with the increase of $\lambda $. The entire
diagram is totally unstable for $\lambda >0.0505$, which is slightly larger
than $\lambda =0.05$ for which Fig.~\ref{Stability_Lambda005_s2} is
displayed. As well as in the case of $s=1$, the ultimate destabilization
occurs when the shape of the bifurcation loop is still double-concave.

Finally, the symmetric vortices with $s=2$ are unstable at $\lambda
<k<0.162+\lambda $, where $k\approx 0.162$ is the stability threshold for $%
s=2$ in the single-component model \cite{Stability}. This threshold
corresponds to the total energy $E\approx 1060$~of the symmetric vortex.

\section{The variational analysis}

\label{sec:VA} The stationary solutions can also be studied analytically by
means of the variational approximation (VA), cf. Ref. \cite{Gubeskys},
taking into regard that stationary equations (\ref{ModelUV}) can be derived
from the Lagrangian,
\begin{eqnarray}
\frac{L}{\pi } &=&\int_{0}^{\infty }r\left\{ -k(U^{2}+V^{2})-\left[ \left(
\frac{dU}{dr}\right) ^{2}+\left( \frac{dV}{dr}\right) ^{2}\right] -\frac{%
s^{2}}{r^{2}}(U^{2}+V^{2})\right.  \notag \\
&&\left. +\frac{1}{2}(U^{4}+V^{4})-\frac{1}{3}(U^{6}+V^{6})-2\lambda
UV\right\} \,dr.  \label{Lagrangian}
\end{eqnarray}%
To approximate solutions to Eqs.~(\ref{ModelUV}) (generally, asymmetric
ones), the following \textit{ansatz }was adopted, with common width $W$ of
both components, but different amplitudes, $A$ and $B$:
\begin{equation}
\{U(r),V(r)\}_{\mathrm{ansatz}}=\{A,B\}r^{s}\exp \left( -\frac{r^{2}}{2W^{2}}%
\right) ,  \label{Ansatz}
\end{equation}%
where $s=0,1,2$ is the same spin as above. The energies of the two
components of this ansatz, defined according to~(\ref{Nuv}), are
\begin{equation}
\{E_{U,V}\}_{\mathrm{ansatz}}=\pi s!\{A^{2},B^{2}\}W^{2(s+1)}
\label{Nuv_Ansatz}
\end{equation}%
The substitution of the ansatz into Lagrangian (\ref{Lagrangian}) and the
integration yield the effective Lagrangian:
\begin{gather}
\frac{2}{\pi }L_{\mathrm{eff}%
}=-s!k(A^{2}+B^{2})W^{2(1+s)}-(s+1)s!(A^{2}+B^{2})W^{2s}  \notag \\
+\frac{(2s)!}{2^{2(1+s)}}(A^{4}+B^{4})W^{2(1+2s)}-\frac{(3s)!}{3^{3s+2}}%
(A^{6}+B^{6})W^{2(1+3s)}+2s!\lambda ABW^{2(1+s)}.  \label{EffLagrangian}
\end{gather}%
It is convenient to redefine the variational parameters as
\begin{equation*}
\alpha \equiv \left( W^{2}/\sqrt{2}\right) (A+B),~\beta \equiv \left( W^{2}/%
\sqrt{2}\right) (A-B),
\end{equation*}%
in terms of which effective Lagrangian (\ref{EffLagrangian}) takes the form
of
\begin{gather}
\frac{2}{\pi }L_{\mathrm{eff}}=-(s+1)s!(\alpha ^{2}+\beta ^{2})+\left[
-s!k(\alpha ^{2}+\beta ^{2})+\frac{(2s)!}{2^{2s+3}}(\alpha ^{4}+\beta
^{4}+6\alpha ^{2}\beta ^{2})\right.  \notag \\
\left. -\frac{(3s)!}{4\cdot 3^{3s+2}}(\alpha ^{6}+\beta ^{6}+15\alpha
^{4}\beta ^{2}+15\alpha ^{2}\beta ^{4})+s!\lambda (\alpha ^{2}-\beta ^{2})%
\right] W^{2}.  \label{NewEffLagrangian}
\end{gather}%
Values of the variational parameters corresponding to stationary solutions, $%
\alpha $, $\beta $ and $W$, are determined by the Euler-Lagrange equations, $%
\partial L_{\mathrm{eff}}/\partial (W^{2})=\partial L_{\mathrm{eff}%
}/\partial (\alpha ^{2})=\partial L_{\mathrm{eff}}/\partial (\beta ^{2})=0$,
i.e.,
\begin{gather}
-k(\alpha ^{2}+\beta ^{2})+\frac{(2s)!}{2^{2s+3}s!}(\alpha ^{4}+\beta
^{4}+6\alpha ^{2}\beta ^{2})  \notag \\
-\frac{(3s)!}{4\cdot 3^{3s+2}s!}(\alpha ^{6}+\beta ^{6}+15\alpha ^{4}\beta
^{2}+15\alpha ^{2}\beta ^{4})+\lambda (\alpha ^{2}-\beta ^{2})=0,
\label{EL_1} \\
-kW^{2}-(s+1)+\frac{(2s)!}{2^{2s+2}s!}(\alpha ^{2}+3\beta ^{2})W^{2}  \notag
\\
-\frac{(3s)!}{4\cdot 3^{3s+1}s!}(\alpha ^{4}+10\alpha ^{2}\beta ^{2}+5\beta
^{4})W^{2}+\lambda W^{2}=0,  \label{EL_2} \\
-kW^{2}-(s+1)+\frac{(2s)!}{2^{2s+2}s!}(\beta ^{2}+3\alpha ^{2})W^{2}  \notag
\\
-\frac{(3s)!}{4\cdot 3^{3s+1}s!}(\beta ^{4}+10\alpha ^{2}\beta ^{2}+5\alpha
^{4})W^{2}-\lambda W^{2}=0.  \label{EL_3}
\end{gather}%
The variational solutions are obtained by numerically solving the system of
equations,~(\ref{EL_1})-(\ref{EL_3}) for $\alpha $, $\beta $ and $W$, for
given $s$, $\lambda $ and $k$. In this way, several VA-predicted bifurcation
loops were constructed, for different values of $\lambda $ and for $s=0,1$
and~$2$, as shown above in Figs.~\ref{BifLoopS0},~\ref{BifLoopS1} and~\ref%
{BifLoopS2} by dashed-dotted curves, alongside the numerically found loops.
The figures show that the VA quite accurately predicts the transformation of
the bifurcation loop from the concave shape to the convex one with the
increase of $\lambda $. An adequate indication of the accuracy of the VA is
given by comparing critical values of $\lambda $ at which the
symmetry-breaking bifurcations disappear, along with the bifurcation loops.
For that purpose, we set $\beta =0$ in Eqs.~(\ref{EL_1})-(\ref{EL_3}) and
subtract the second equation from the third, which yields
\begin{equation}
\alpha ^{2}=\frac{3^{3s+1}s!}{(3s)!}\left( \frac{(2s)!}{2^{2s+2}s!}\pm \sqrt{%
\left( \frac{(2s)!}{2^{2s+2}s!}\right) ^{2}-2\frac{(3s)!}{3^{3s+1}s!}\lambda
}\right) .  \label{Alpha_for_Beta0}
\end{equation}%
Within the framework of of the VA, the asymmetric solutions exist under the
condition that expression~(\ref{Alpha_for_Beta0}) yields real values:
\begin{equation}
\lambda <\lambda _{s}\equiv \frac{3^{3s+1}((2s)!)^{2}}{2^{4s+5}(3s)!s!}.
\label{lambda}
\end{equation}%
For $s=0,1$ and~$2$, Eq. (\ref{lambda}) predicts critical values $\lambda
_{0}=0.09375$, $\lambda _{1}=0.10547$, and $\lambda _{2}=0.10679$. The
comparison with their numerically found counterparts shows that the
difference is $2.8\%$, $4.3\%$ and $3.9\%$, respectively.

\begin{figure}[tbp]
\subfigure[]{\includegraphics[width=2.25in]{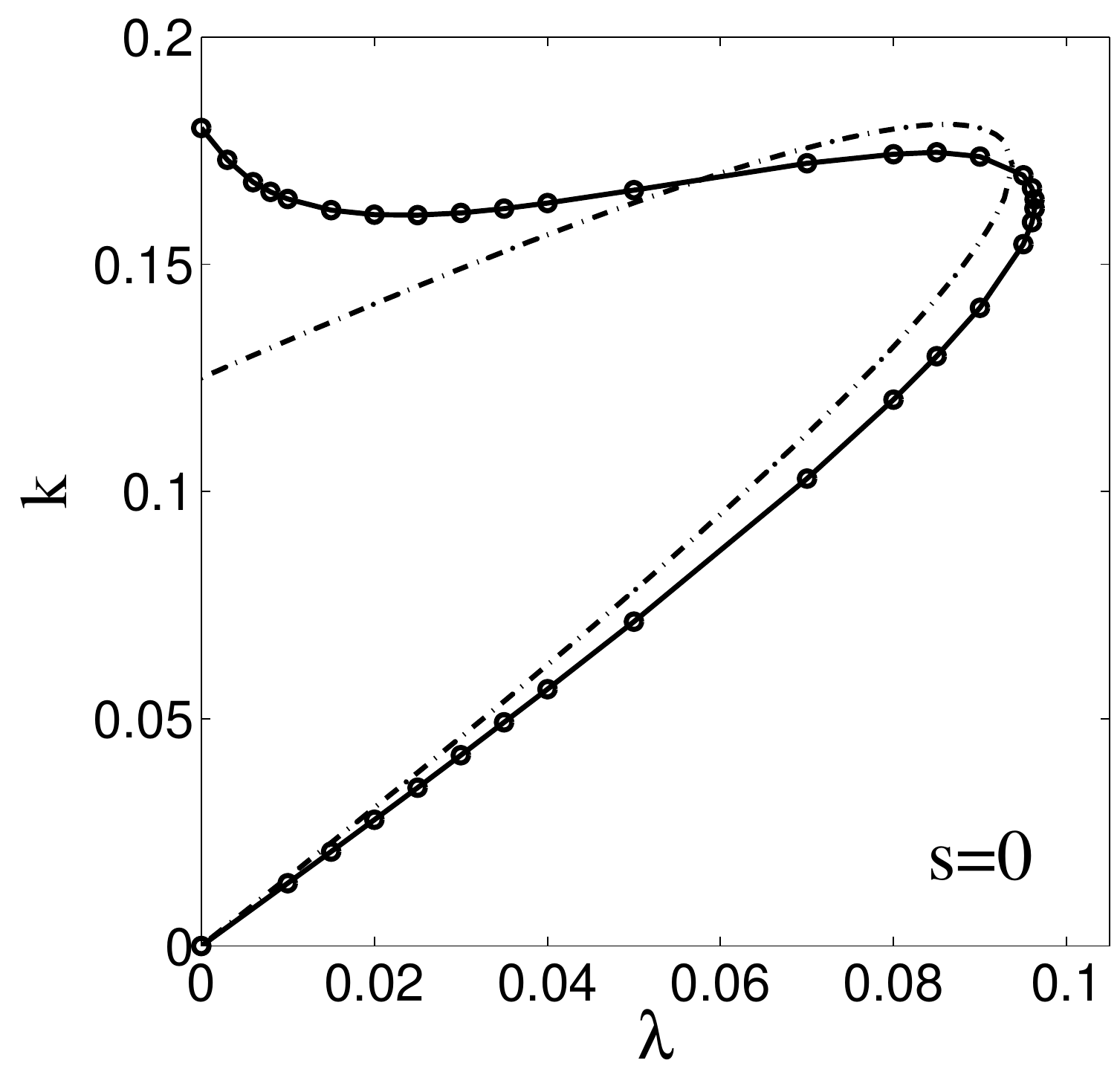}%
\label{kbif_vs_lambda_s0}} \subfigure[]{%
\includegraphics[width=2.25in]{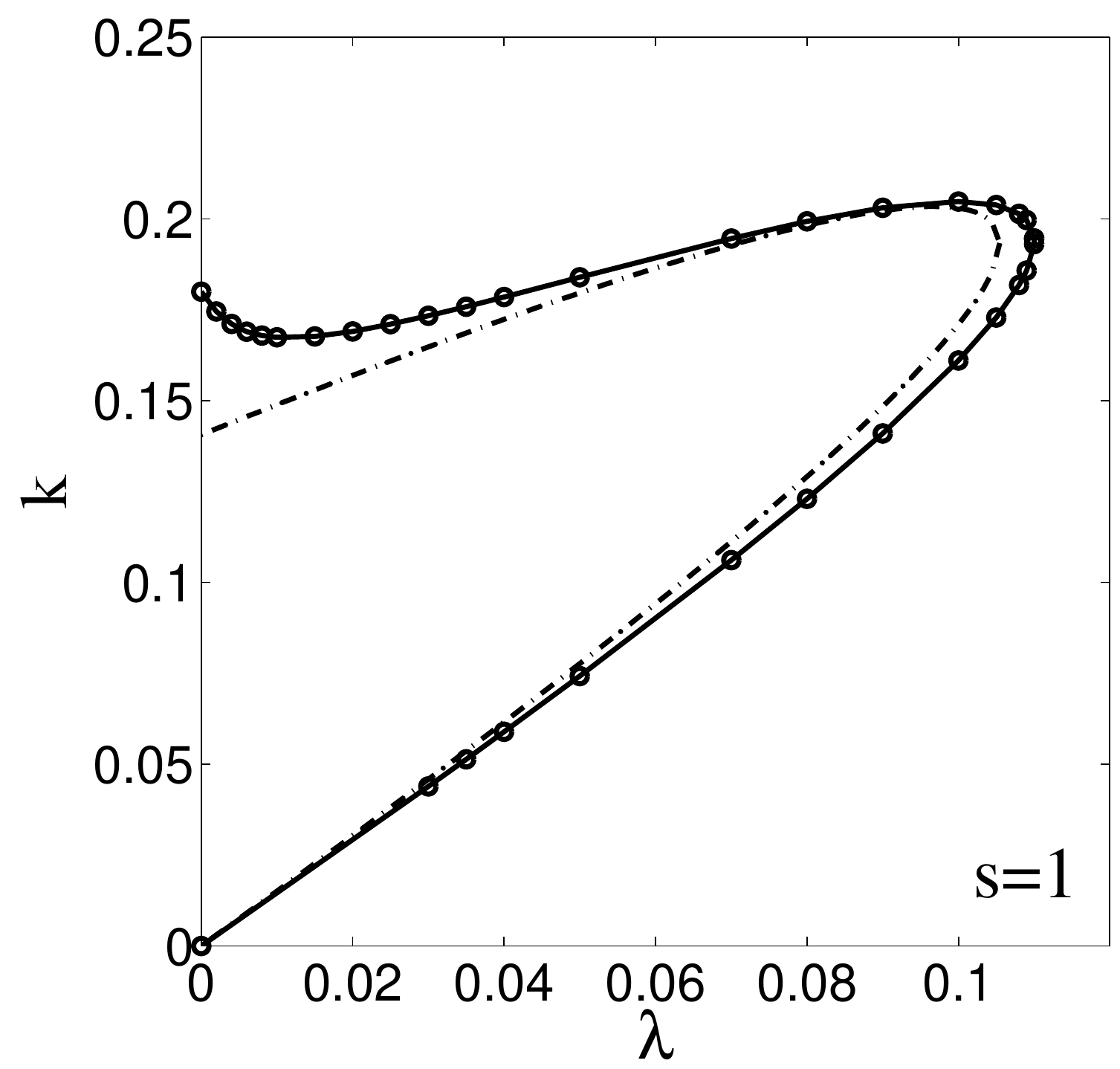}%
\label{kbif_vs_lambda_s1}} \subfigure[]{%
\includegraphics[width=2.25in]{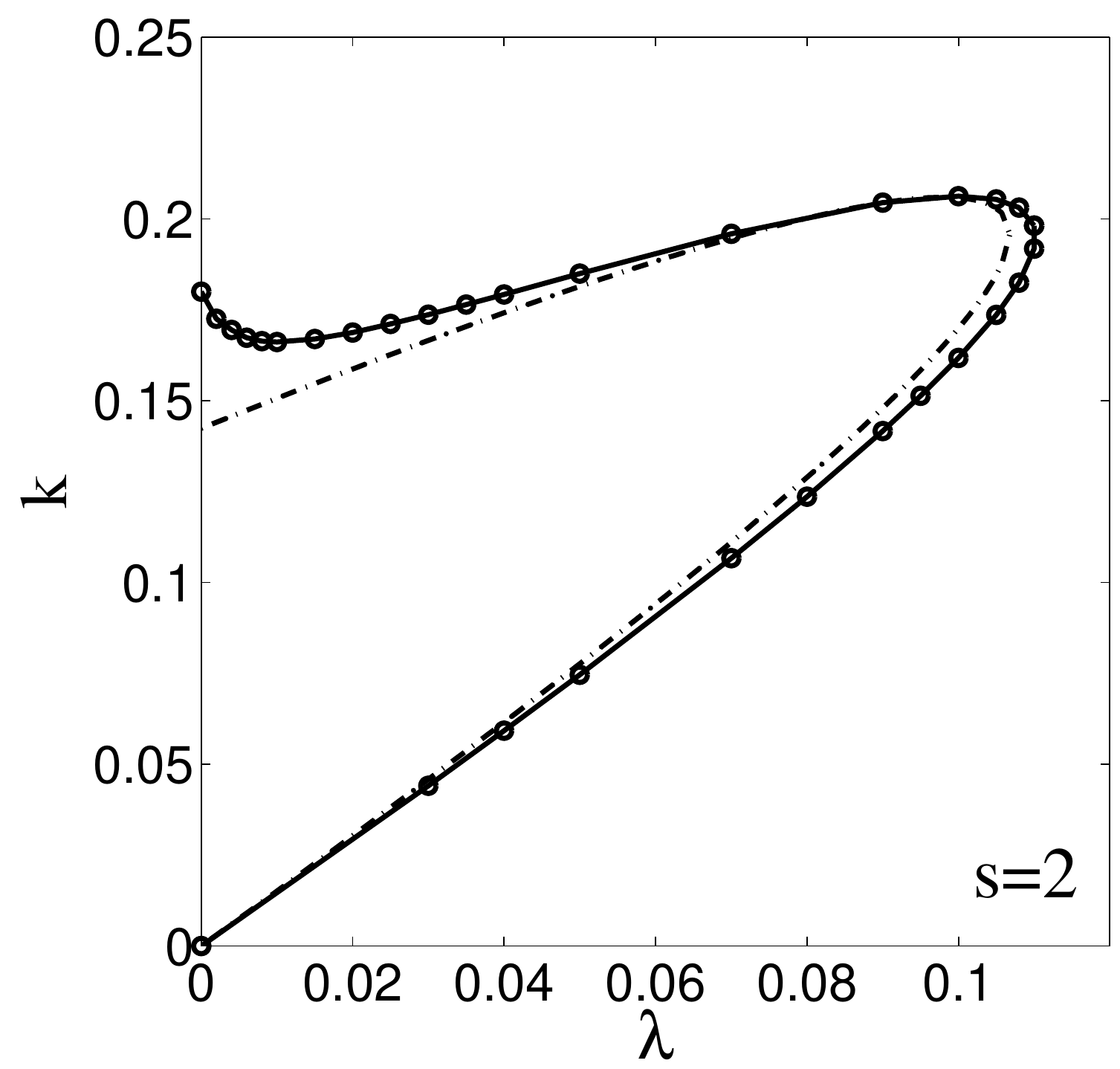}%
\label{kbif_vs_lambda_s2}} \newline
\subfigure[]{\includegraphics[width=2.25in]{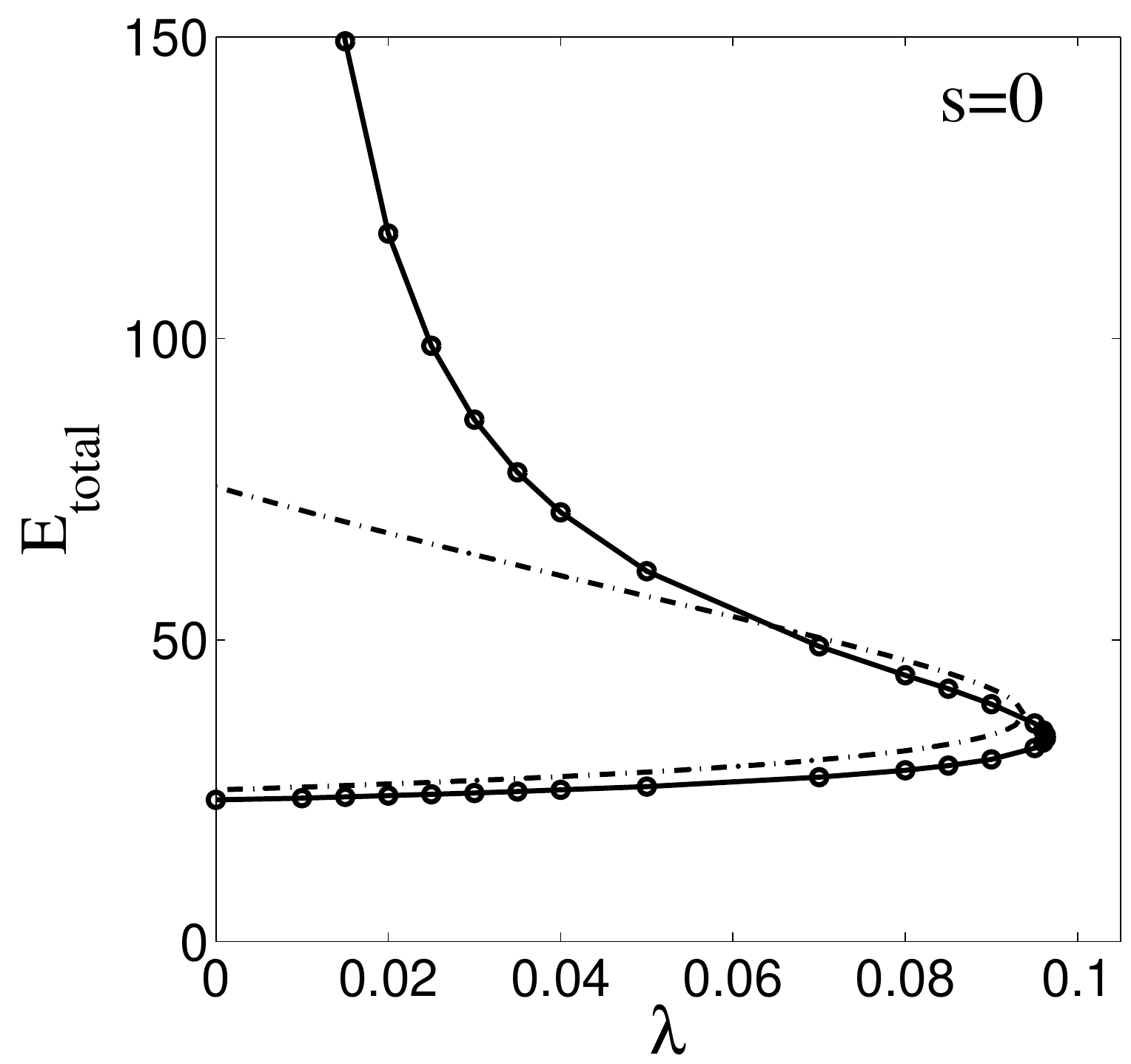}%
\label{Nbif_vs_lambda_s0}} \subfigure[]{%
\includegraphics[width=2.25in]{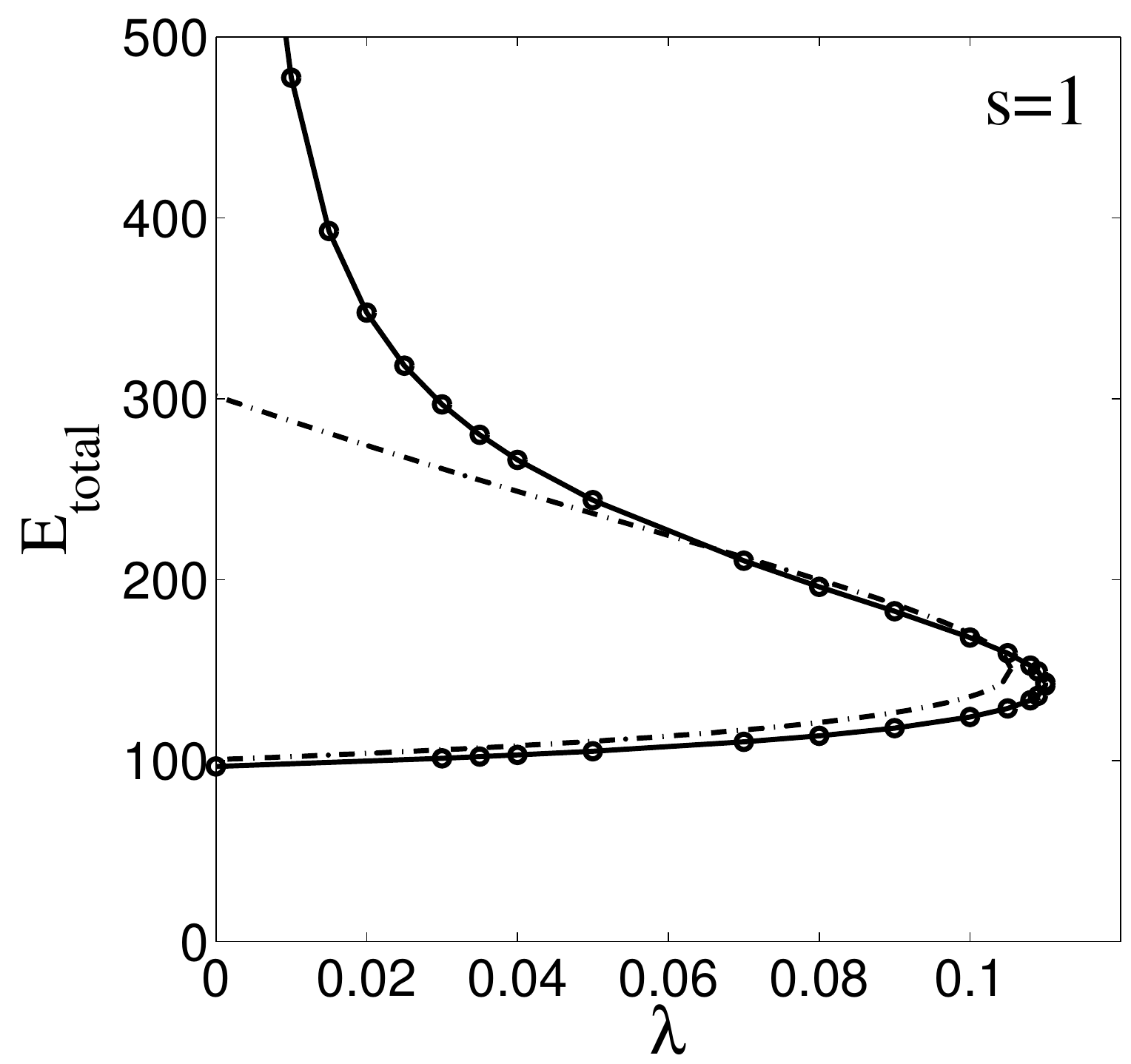}%
\label{Nbif_vs_lambda_s1}} \subfigure[]{%
\includegraphics[width=2.25in]{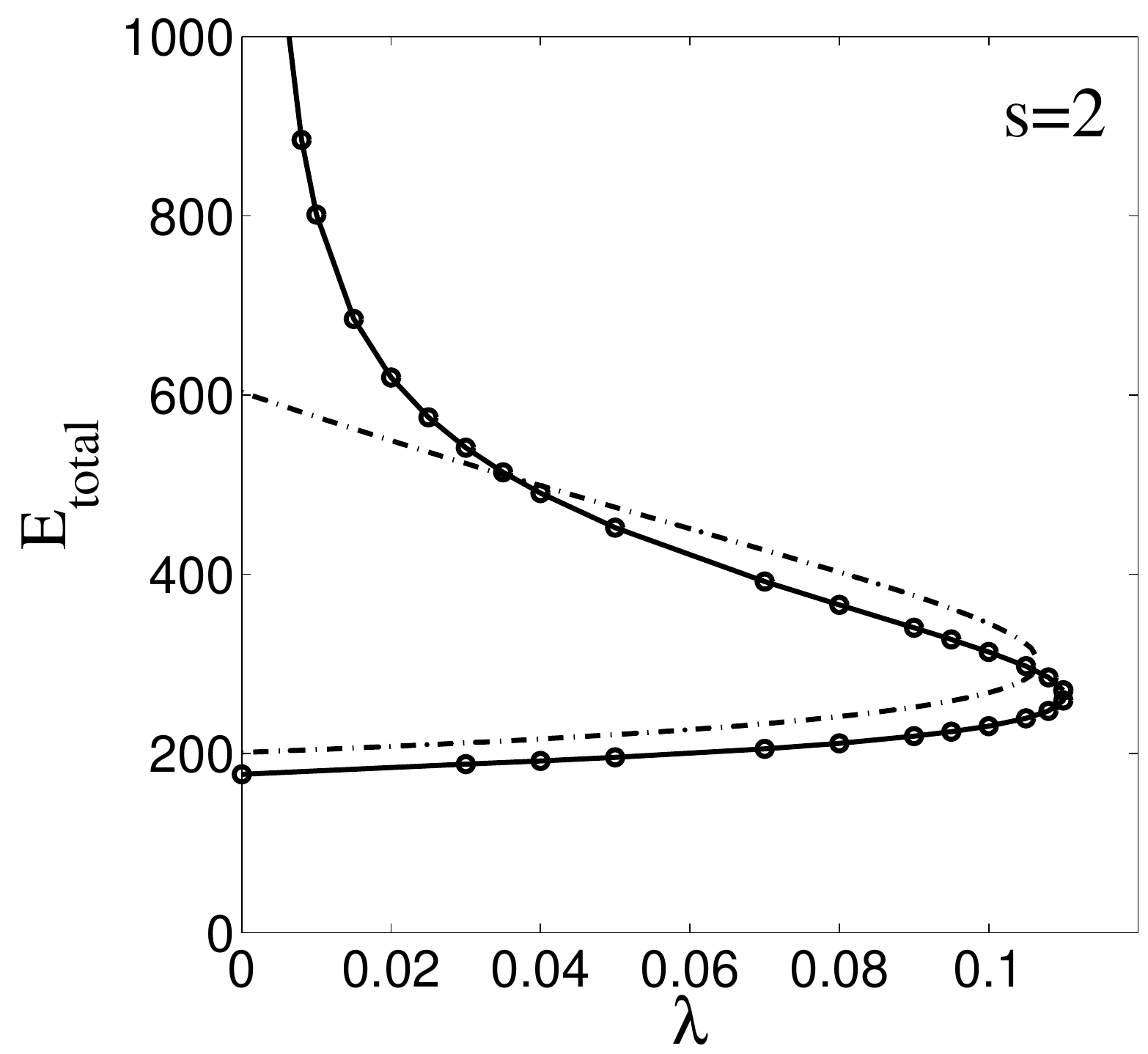}%
\label{Nbif_vs_lambda_s2}}
\caption{The comparison of the numerically generated (circles connected by
continues lines) and variationally predicted (dashed-dotted lines) points of
the direct and reverse bifurcations. (a)-(c) The propagation constant, $k$,
at which the bifurcations occur, as a function of $\protect\lambda $, for $%
s=0,1$ and~$2$. (d)-(f) The same for the total energy, $E_{\mathrm{total}}$.}
\label{Nk_vrs_Lambda}
\end{figure}

To calculate coordinates of the VA-predicted bifurcation points, we
substitute~expression (\ref{Alpha_for_Beta0}) into Eq.~(\ref{EL_3}) with $%
\beta =0$. Both the variational and the numerically generated plots for
values of $E$ and $k$ at the bifurcation points are shown, versus the
coupling constant, $\lambda $, in Fig.~\ref{Nk_vrs_Lambda}.
It is seen that the variational and numerical results are always in good
agreement for the direct bifurcation. On the other hand, the approximation
for the reverse bifurcation becomes inaccurate for very small values of $%
\lambda $. This difference is explained by the fact that, near the reverse
bifurcation, the actual profiles of the soliton components become
increasingly rectangular-like, i.e., different from the shape assumed by
ansatz (\ref{Ansatz}).

\section{Development of the instability of vortex rings}

\label{sec:splitting} To explore results of the instability development,
direct simulations of Eqs.~(\ref{ModelPsiPhi}) were performed by means of
the standard pseudospectral split-step method, for initial conditions
corresponding to the stationary solutions presented in section~\ref%
{sec:loops}. Perturbations were not explicitly added to unstable solitons,
the instability being initiated by truncation errors of the numerical code.
First, we demonstrate the splitting of vortex solitons which are unstable
against azimuthal perturbations. For the single-component model, a similar
numerical analysis was reported in Ref.~\cite{Stability}, where it was
concluded that, generally, the azimuthal index, $n$, of the most unstable
eigenmode determines the number of fragments produced by the splitting.

\begin{figure}[tbp]
\centering
\parbox{3.4in}{
\begin{minipage}{3.3in}
\includegraphics[width=3.4in]{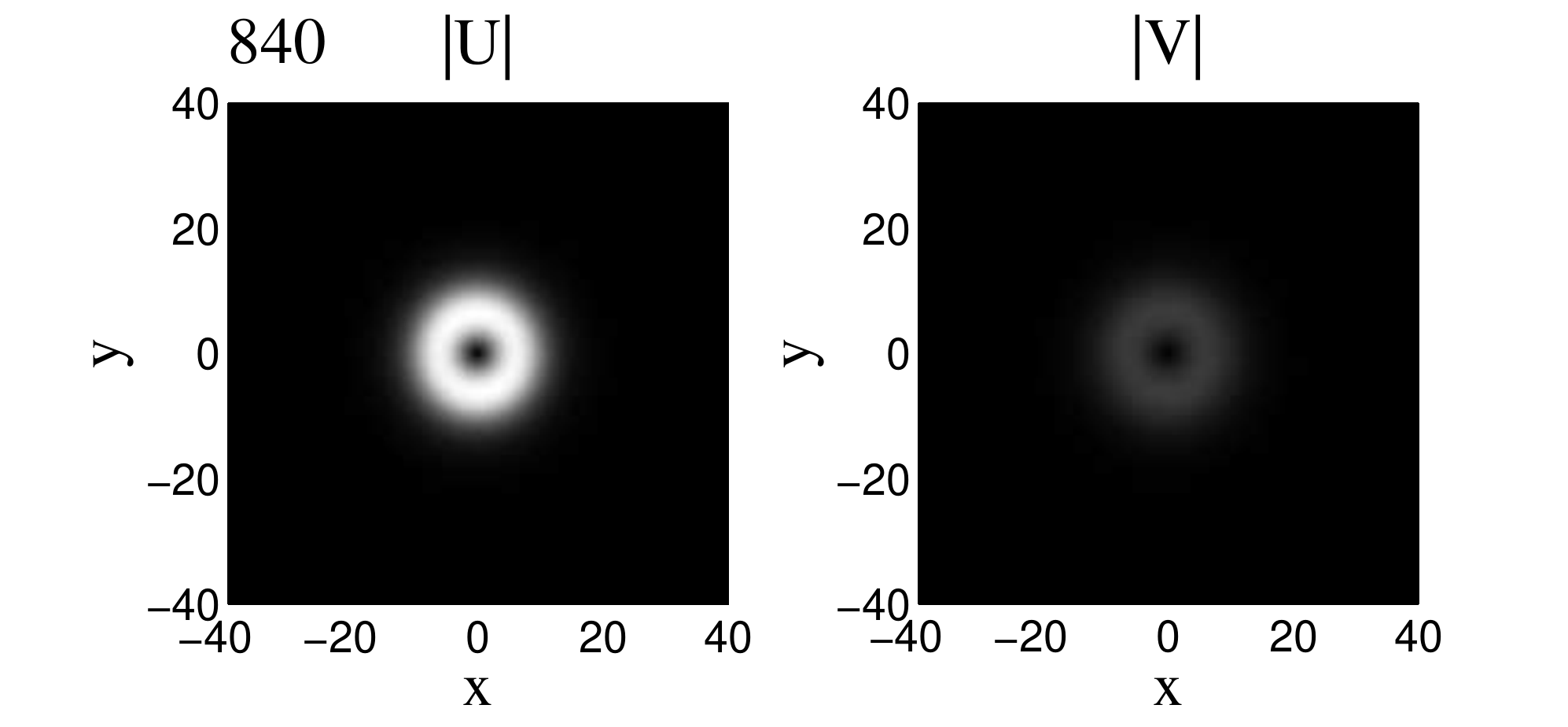}\label{aa} \\
\begin{minipage}{3.3in}
\includegraphics[width=3.4in]{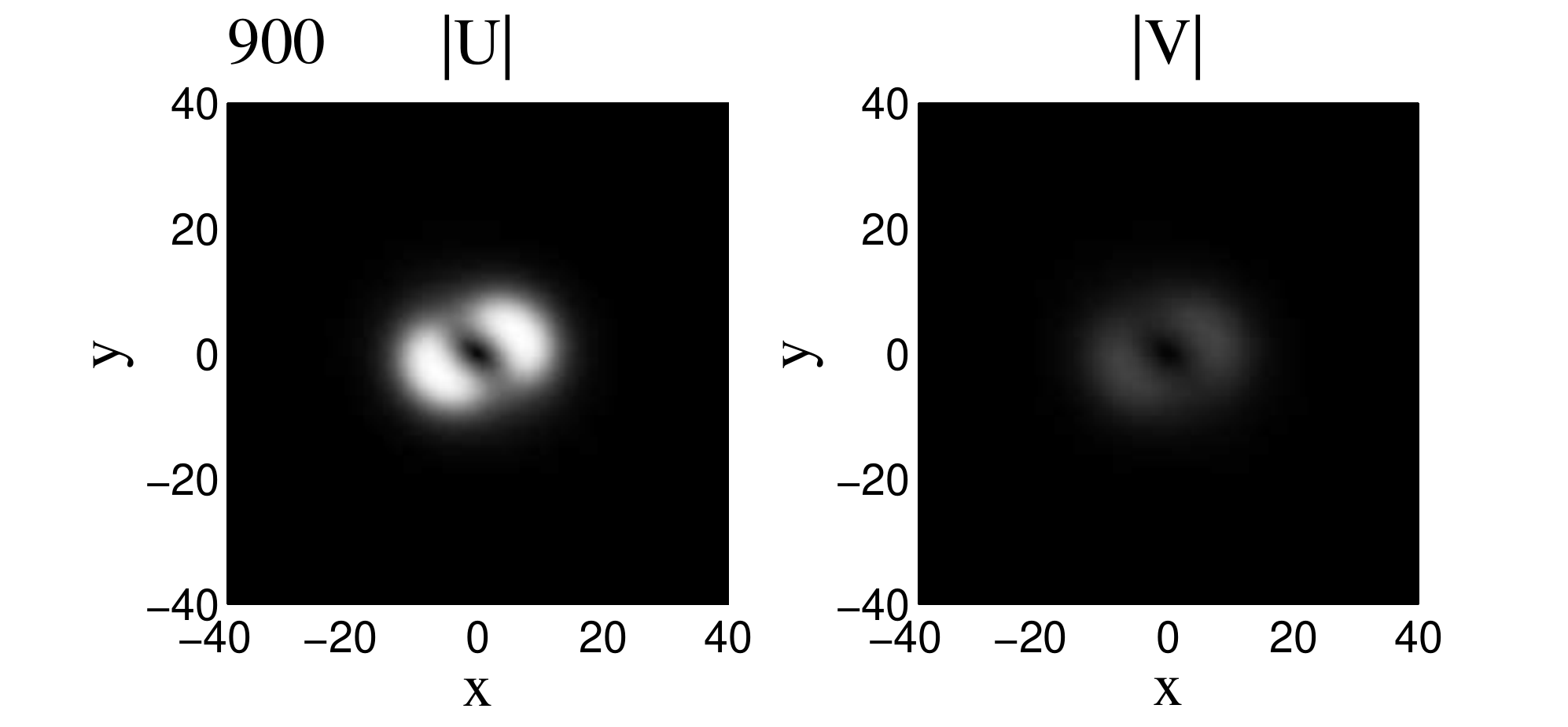}\label{aa}\\
\begin{minipage}{3.3in}
\includegraphics[width=3.4in]{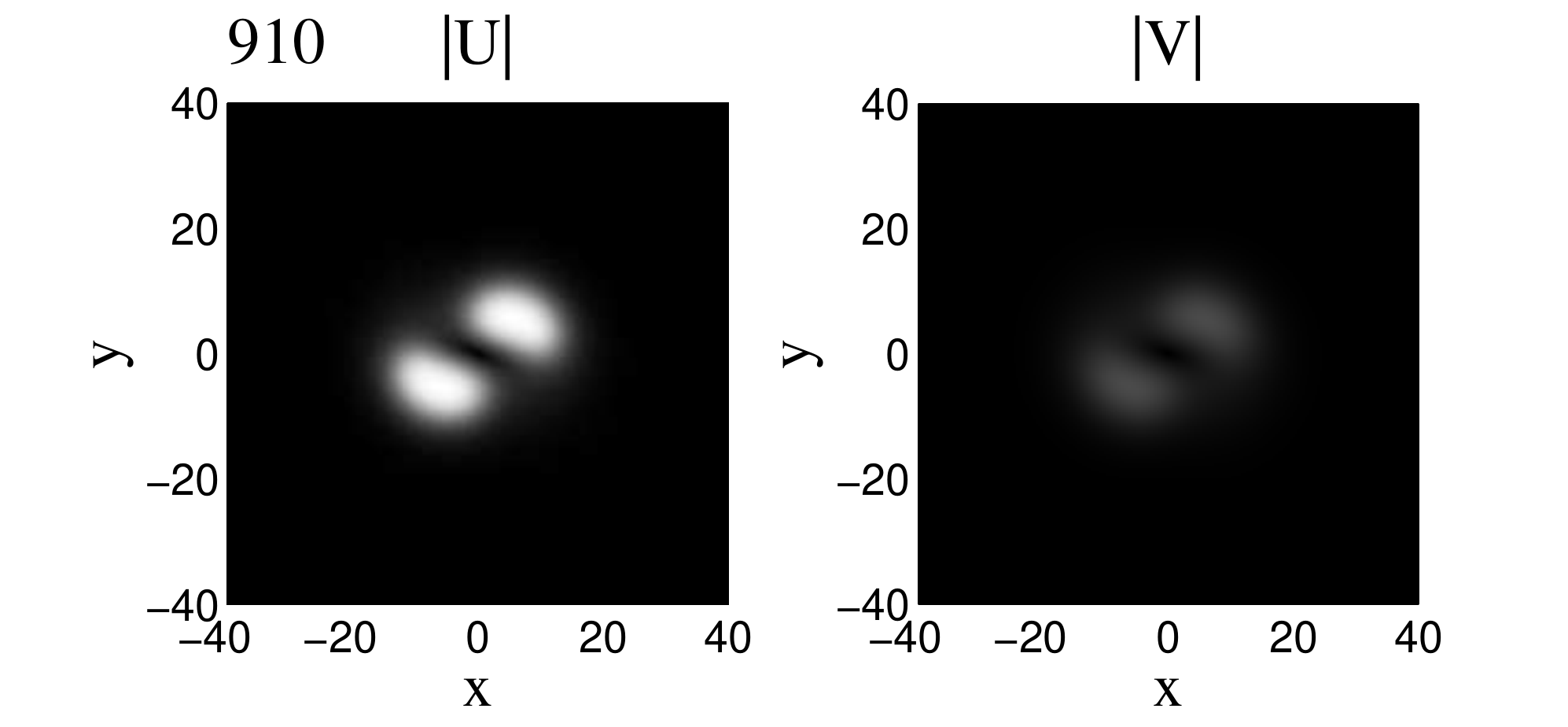}\label{aa}\\
\begin{minipage}{3.3in}
\includegraphics[width=3.4in]{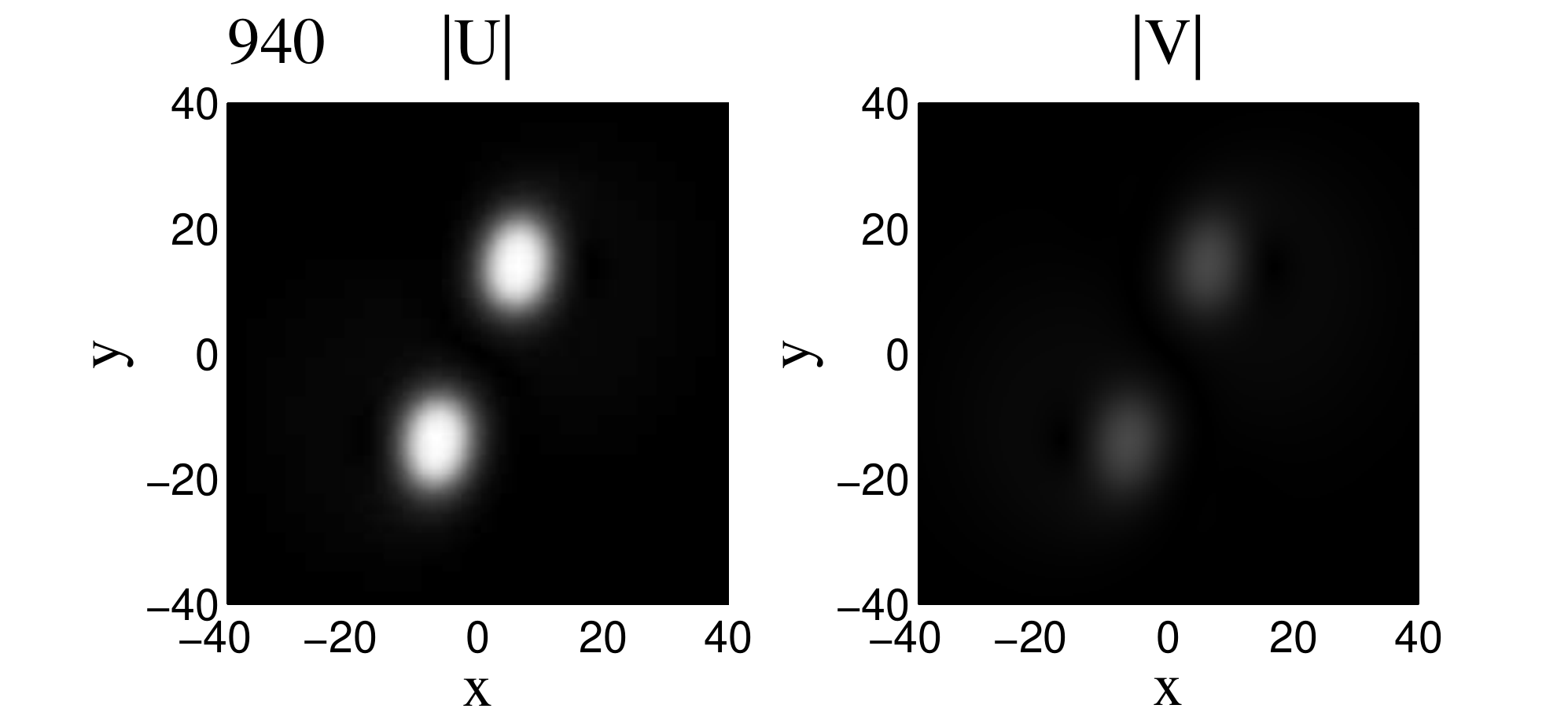}\label{aa}
\end{minipage}
\end{minipage}
\end{minipage}
\end{minipage}
\caption{Gray-scale plots illustrating the splitting of an unstable
vortex solution, at $s=1$ and $k=0.155$ ($E_{total}\approx150$).
Values of the propagation distance are labeled above the left
frames.} \label{Split_s1_k0155}} \qquad
\begin{minipage}{3.3in}
\includegraphics[width=3.4in]{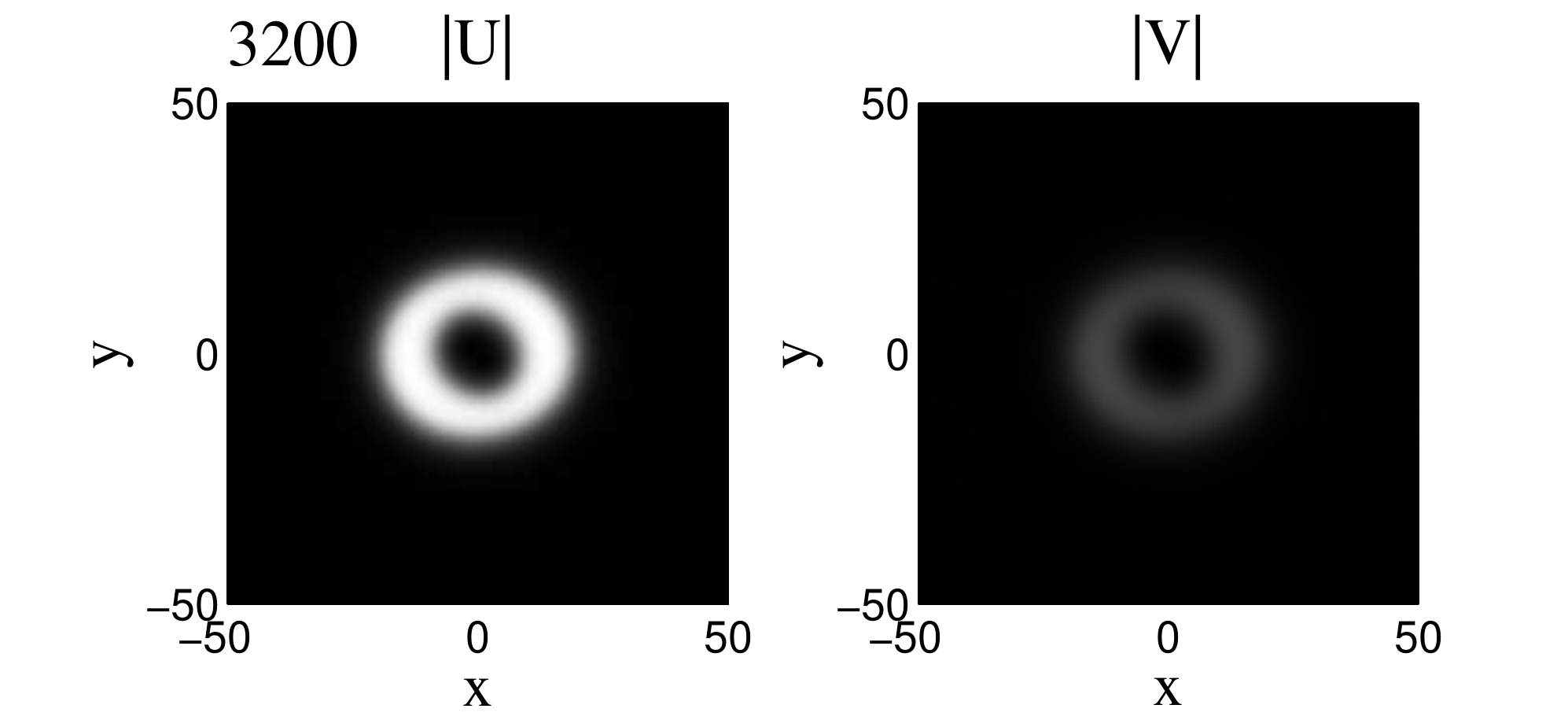}\label{aa} \\
\begin{minipage}{3.3in}
\includegraphics[width=3.4in]{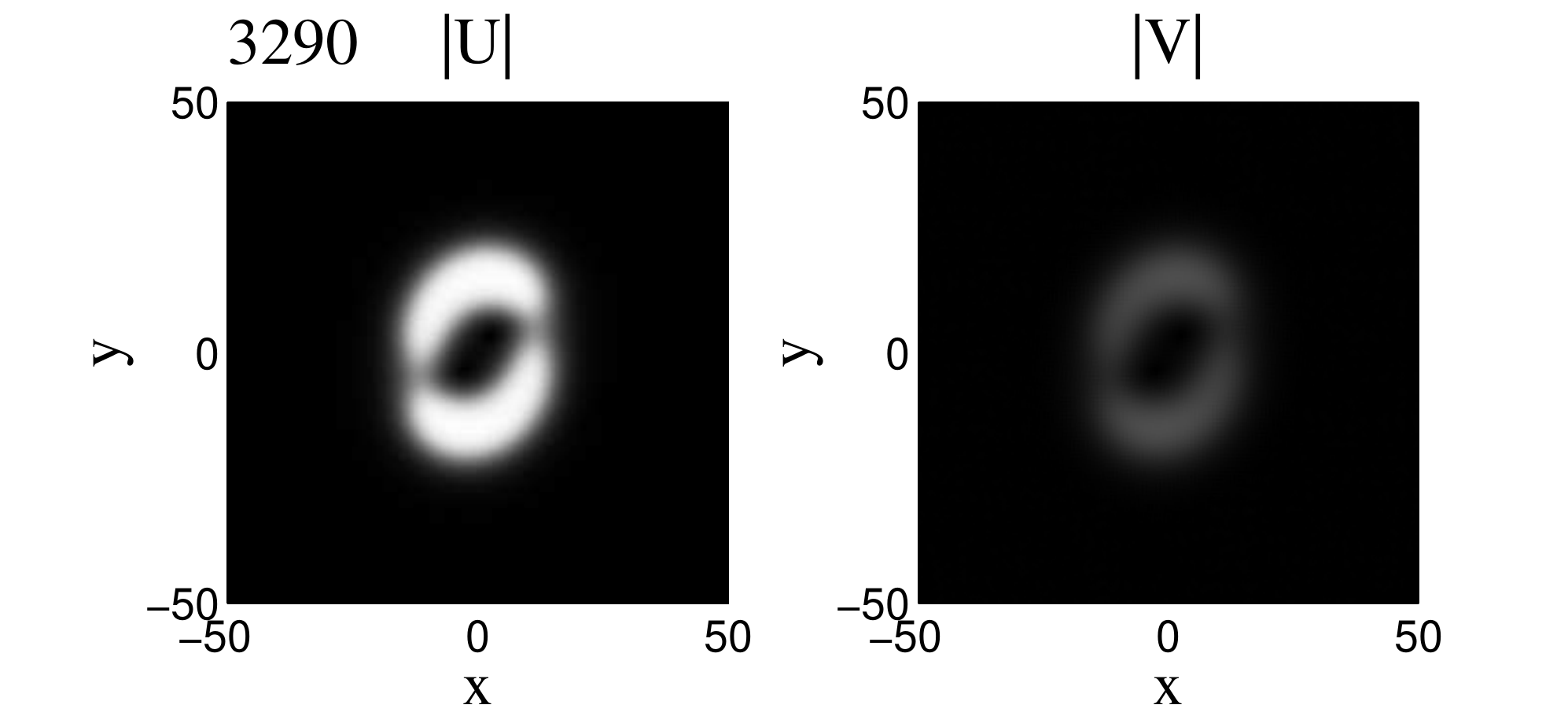}\label{aa}\\
\begin{minipage}{3.3in}
\includegraphics[width=3.4in]{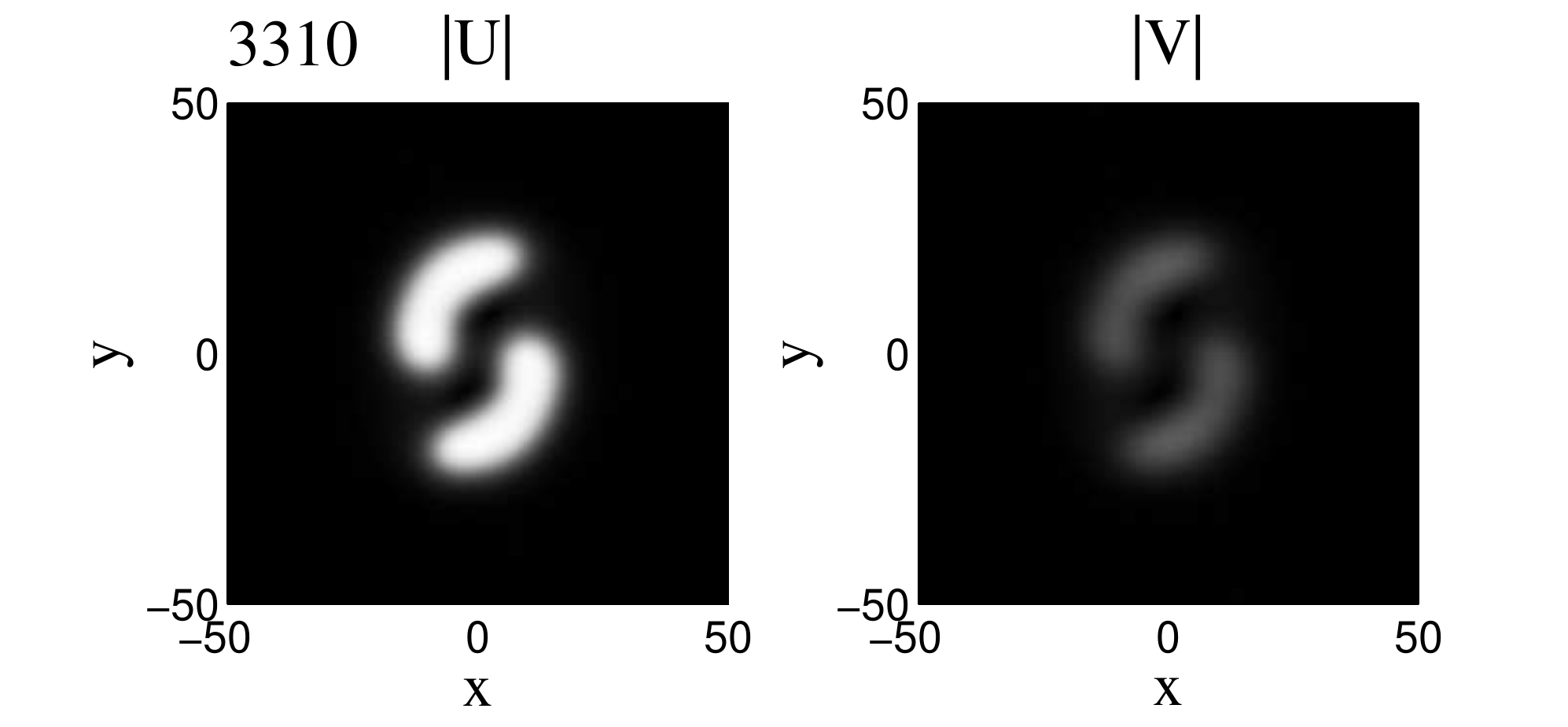}\label{aa}\\
\begin{minipage}{3.3in}
\includegraphics[width=3.4in]{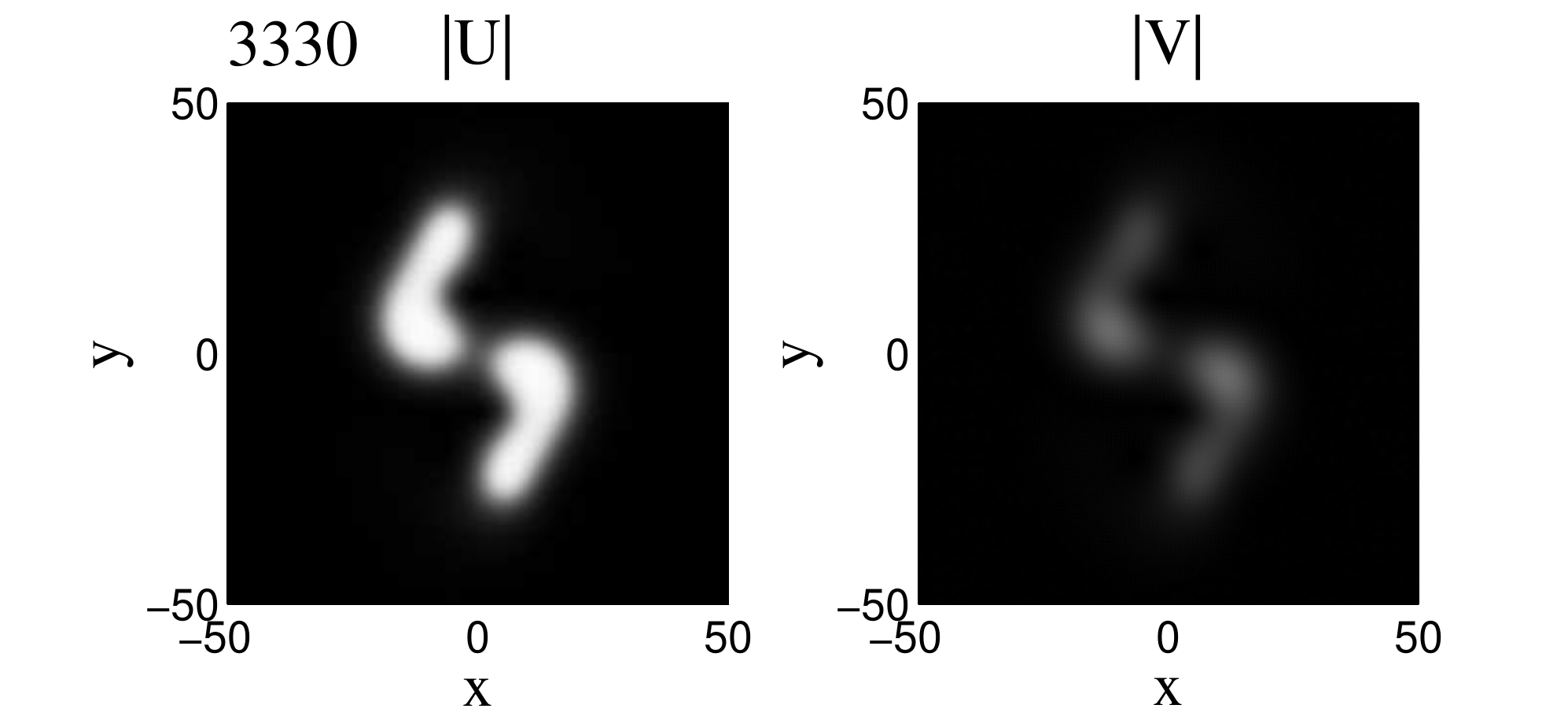}\label{aa}
\caption{The same as in Fig.~\ref{Split_s1_k0155}, but for $s=2$ and
$k=0.174$ ($E_{total}\approx500$).}
\label{Split_s2_k0174}
\end{minipage}
\end{minipage} \end{minipage} \end{minipage}
\end{figure}

Figure~\ref{Split_s1_k0155} displays the numerically simulated evolution of
the asymmetric vortex ring with $s=1$, $\lambda =0.05$, $k=0.155$ and $E_{%
\mathrm{total}}\approx 150$, for which the single unstable perturbation
eigenmode has $n=2$, as per Fig.~\ref{Stability_Lambda005_s1-b}. In this
case, the breakup of the vortex becomes conspicuous at $z_{\mathrm{split}%
}\approx 900$, giving rise to two fragments, in accordance with the
linear-stability analysis.

\begin{figure}[tbp]
\centering
\parbox{3.4in}{
\begin{minipage}{3.3in}
\includegraphics[width=3.4in]{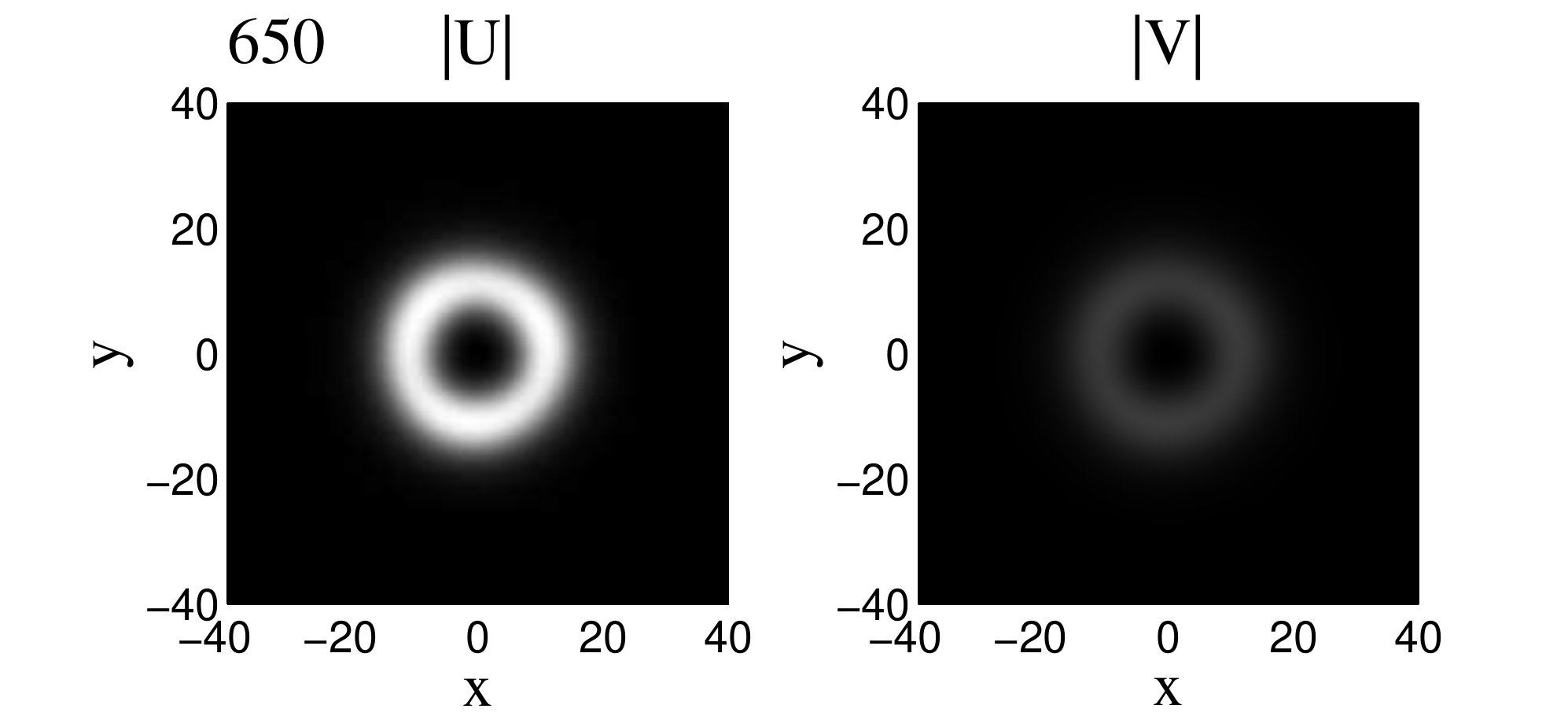}\\
\begin{minipage}{3.3in}
\includegraphics[width=3.4in]{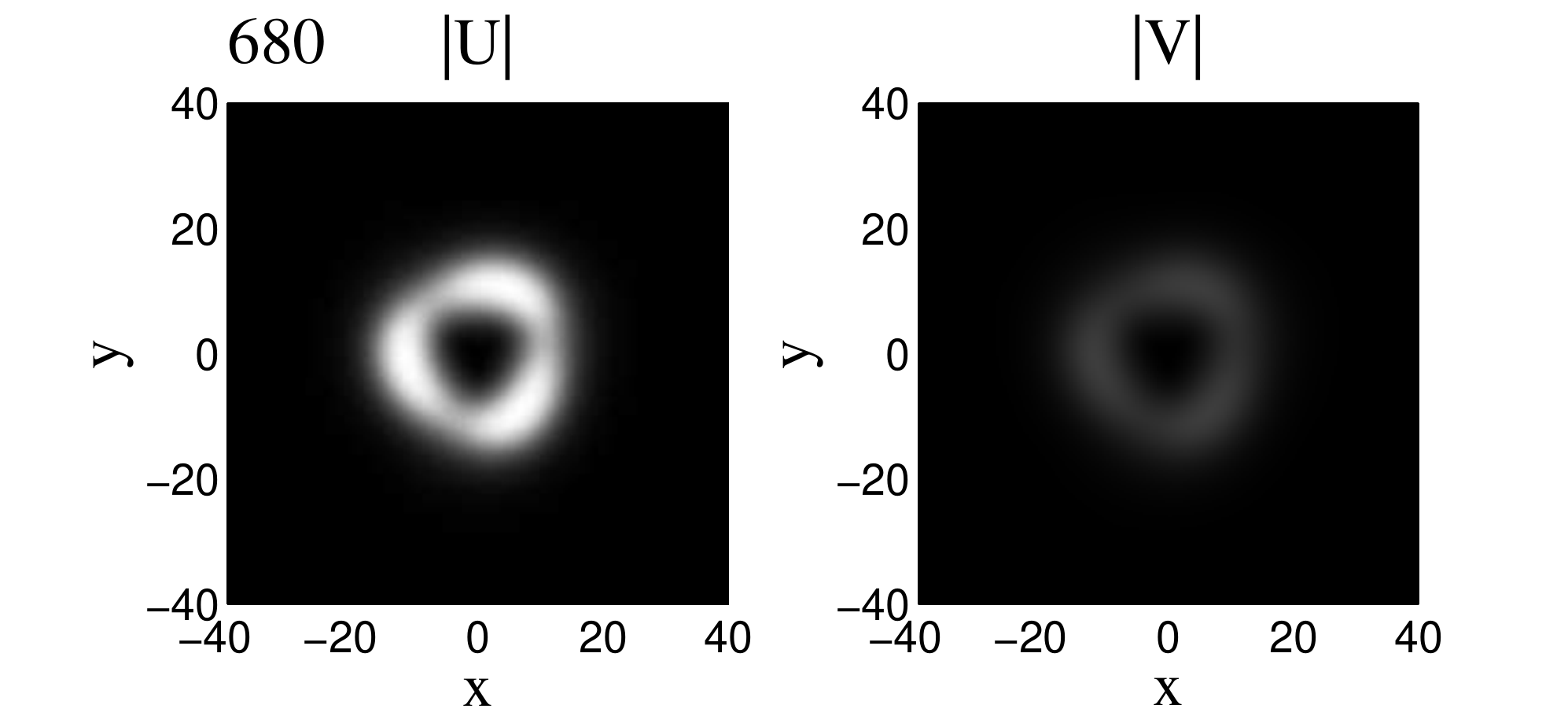}\\
\begin{minipage}{3.3in}
\includegraphics[width=3.4in]{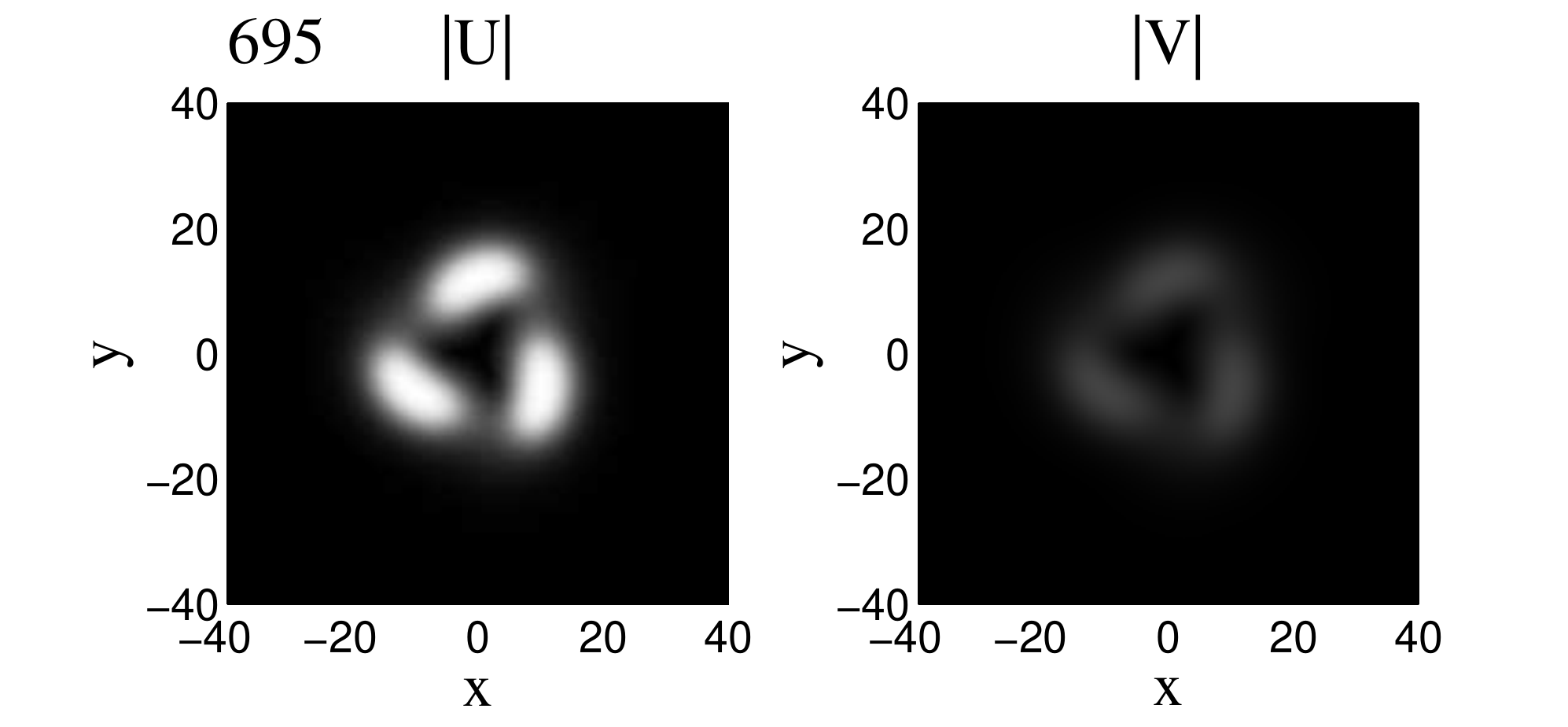}\\
\begin{minipage}{3.3in}
\includegraphics[width=3.4in]{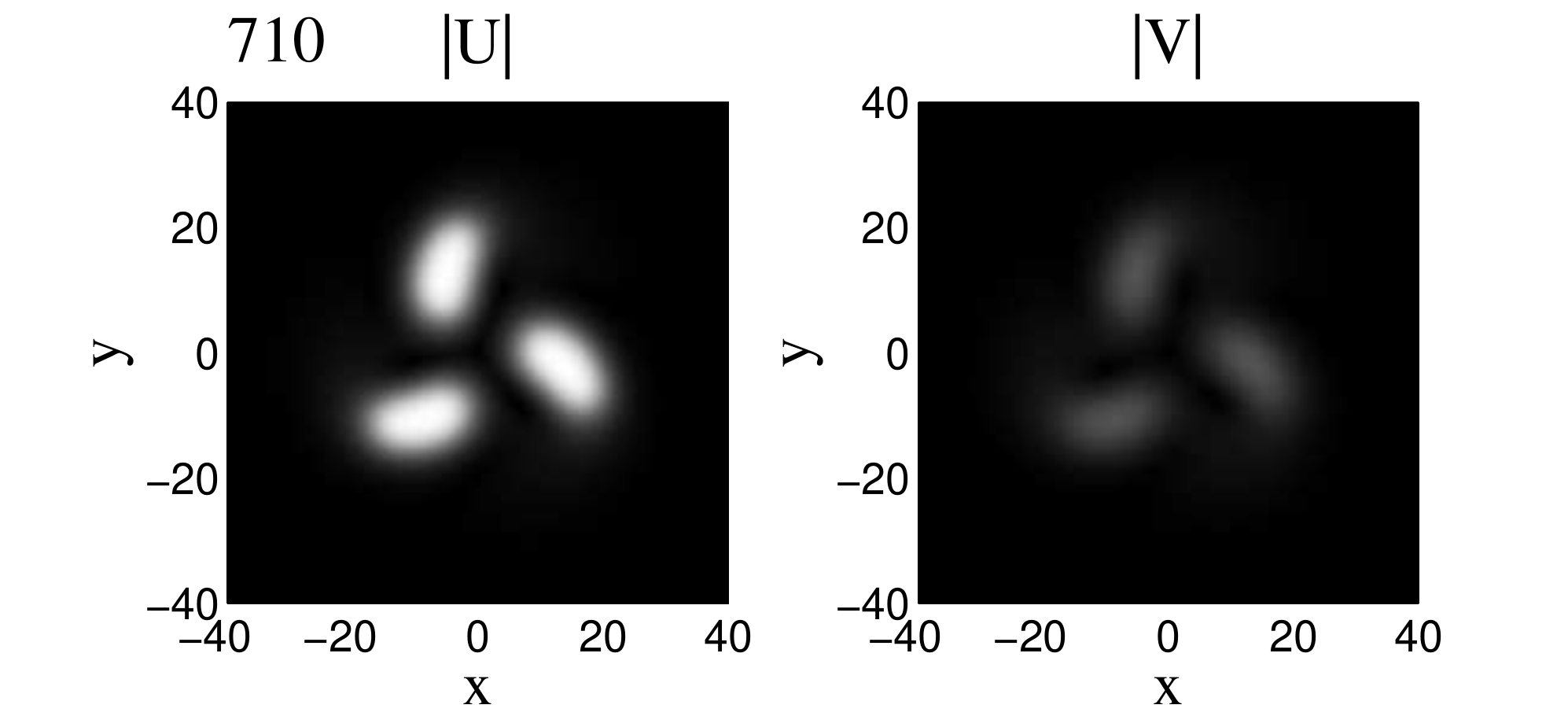}
\end{minipage}
\end{minipage}
\end{minipage}
\end{minipage}
\caption{The same as in Figs.~\ref{Split_s1_k0155} and~\ref{Split_s2_k0174},
but for $s=2$ and $k=0.1505$ ($E_{total}\approx250$).}
\label{Split_s2_k01505}} \qquad
\begin{minipage}{3.3in} %
\includegraphics[width=3.4in]{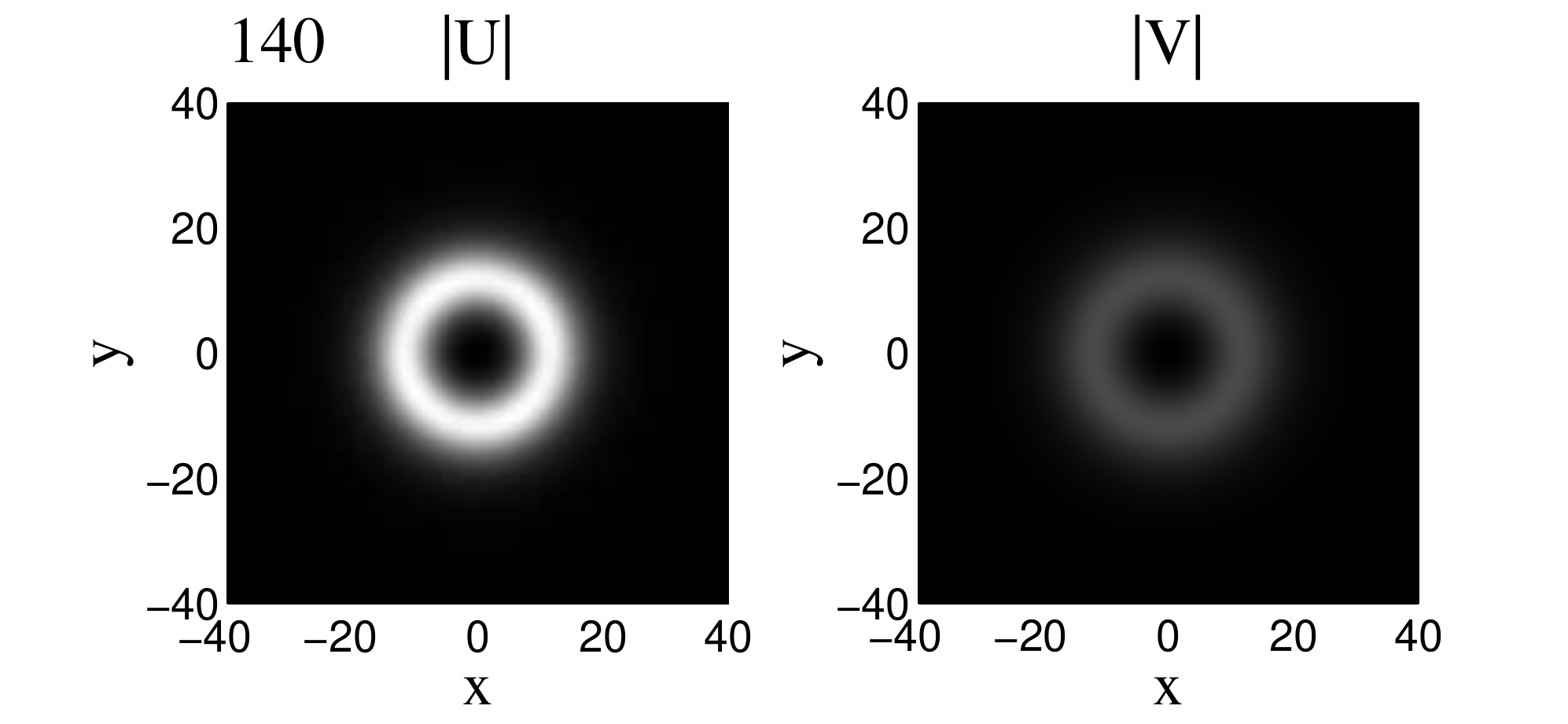}\newline
\begin{minipage}{3.3in} %
\includegraphics[width=3.4in]{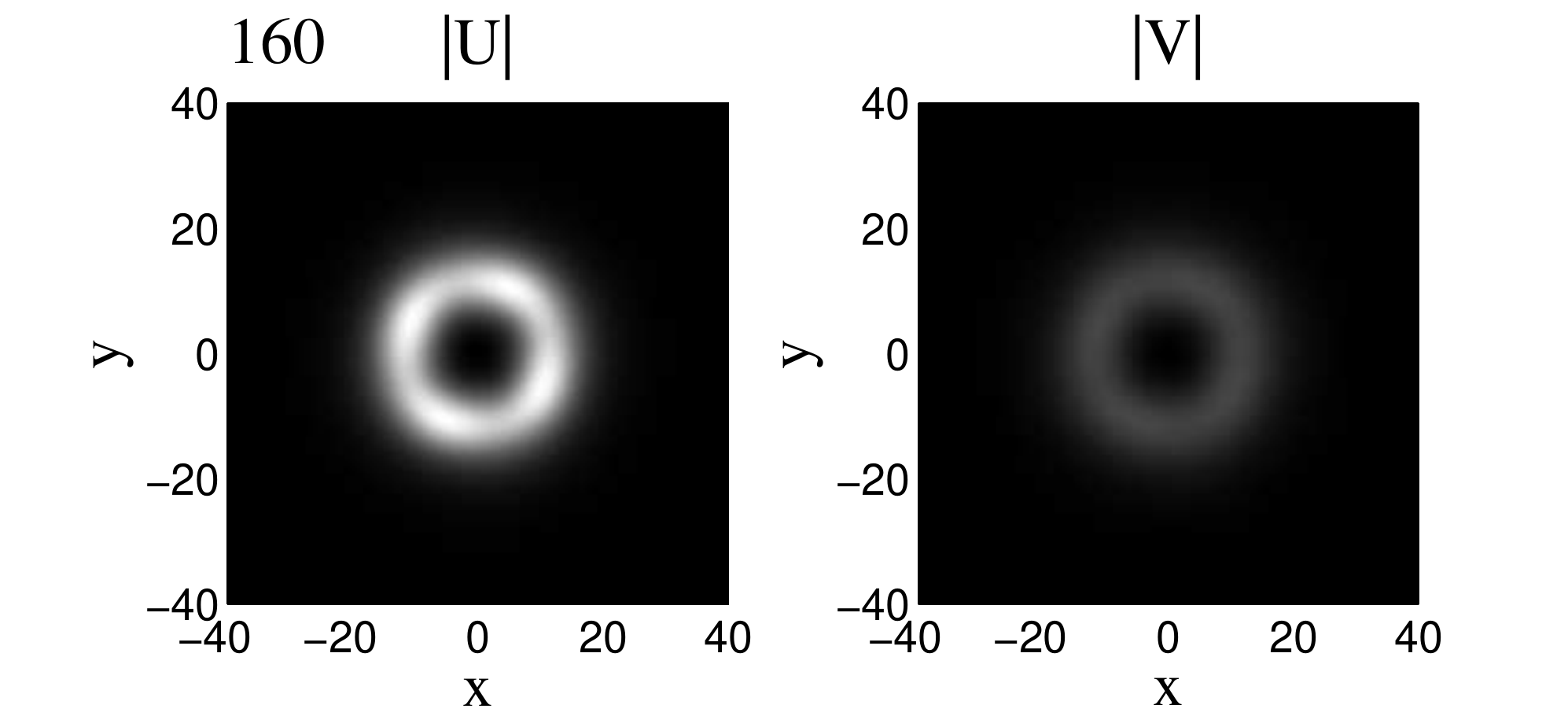}\newline
\begin{minipage}{3.3in} %
\includegraphics[width=3.4in]{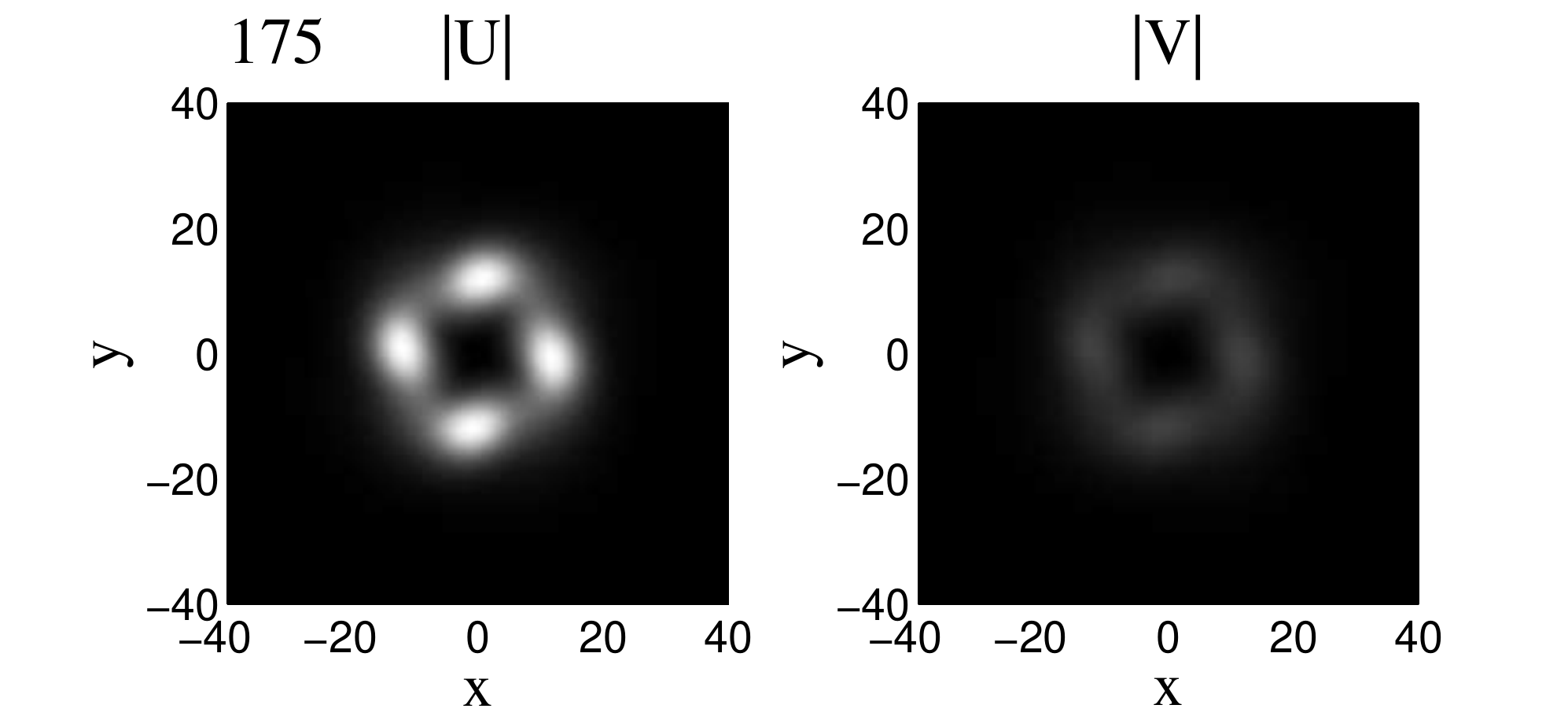}\newline
\begin{minipage}{3.3in}
\includegraphics[width=3.4in]{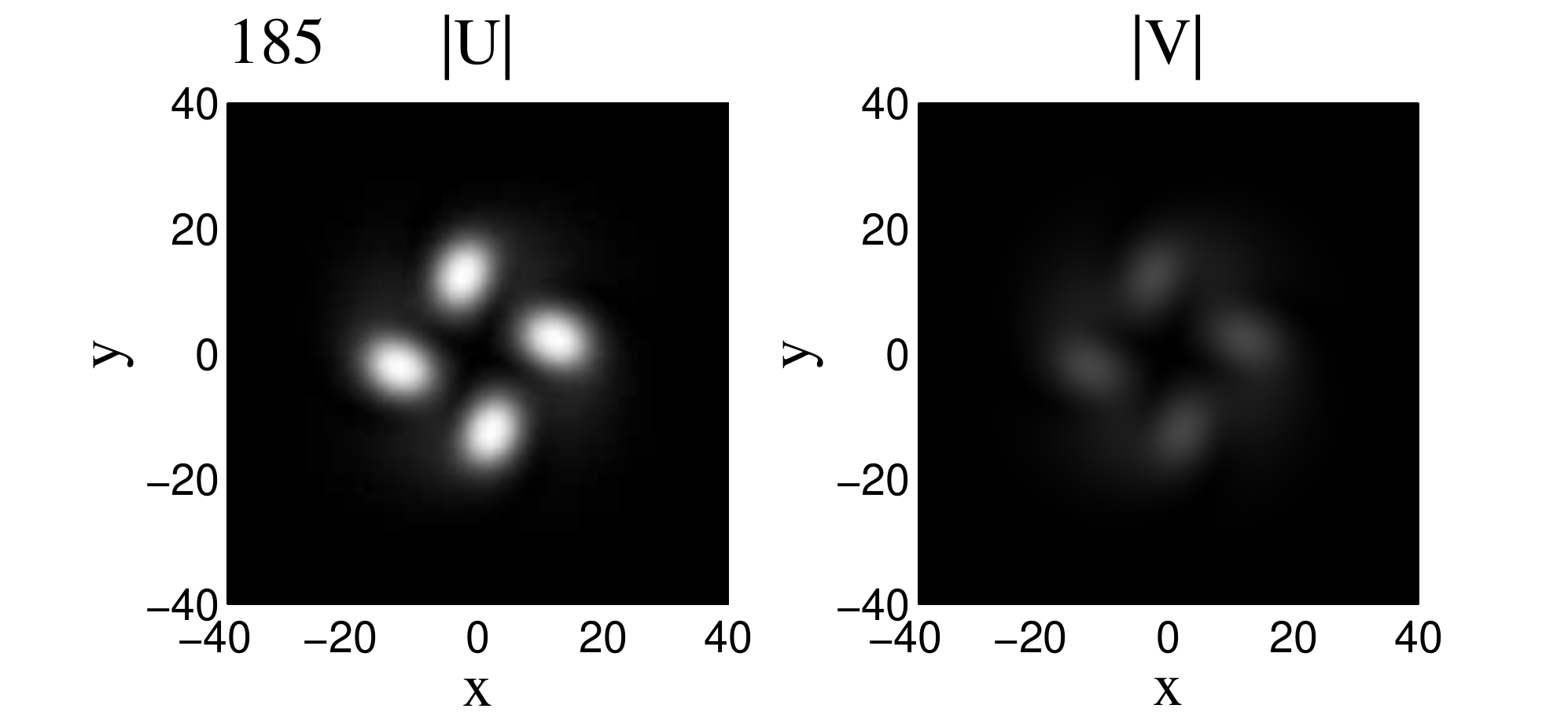}
\caption{The same as in Figs.~\ref{Split_s1_k0155}--\ref{Split_s2_k01505}, but
for $s=2$ and $k=0.112$ ($E_{total}\approx165$).}
\label{Split_s2_k0112}
\end{minipage}
\end{minipage} \end{minipage} \end{minipage}
\end{figure}

Similar results for the double asymmetric vortices ($s=2$) and $\lambda
=0.05 $ are presented in Figs.~\ref{Split_s2_k0174}--\ref{Split_s2_k0112}.
The initial states were chosen so as to have, in each case, the largest
growth rate at a different value of $n$. To this end, we took $%
k=0.112,~0.1505,~0.174$, which correspond to the vortices with $E_{\mathrm{%
total}}\approx 165,~250,~500$, the corresponding largest instability growth
rates being $\gamma _{n=4}\approx 0.098,\gamma _{n=3}\approx 0.047,\gamma
_{n=2}\approx 0.009$, respectively. As expected, the numbers of fragments
generated by the breakup are consistent with these values of $n$. We stress
that, in all the cases presented here, there is a well-pronounced dominant
eigenmode. As mentioned in Ref.~\cite{Stability}, when the parameters are
taken close to borders between regions dominated by unstable eigenmodes with
different azimuthal indices $n$, it is difficult to predict which one will
determine the outcome of the splitting.

Values of the propagation distance needed for the splitting to commence are
also in agreement with the predictions based on the growth rates of the
linear instability. In particular, a large growth rate was found for $%
k=0.112 $ ($E_{\mathrm{total}}\approx 165$) and, accordingly, in that case
the splitting starts very early, at $z_{\mathrm{split}}\approx 170$. For $%
k=0.1505$ ($E_{\mathrm{total}}\approx 250$), the breakup starts later, at $%
z_{\mathrm{split}}\approx 680$, and when the growth rate is small -- for
instance, at $k=0.174$ ($E_{\mathrm{total}}\approx 500$) -- the splitting
sets in after a very long evolution, at $z_{\mathrm{split}}\approx 3250$. In
all the cases that we have examined, the fragments maintain the asymmetry of
the original unstable vortex rings.

Note that unstable asymmetric solutions could transform into stable
symmetric ones (and vise versa) if the only unstable perturbation eigenmode
were the one with $n=0$. In fact, this happens solely for $s=0$ (see Fig. %
\ref{Stability_Lambda005_s0}, for example). In all the cases that we have
examined, the azimuthal instability of the vortices with $s=1$ and $2$
destabilizes and destroys the solutions, before they could be reshaped into
stable symmetric or asymmetric structures with the same~$s$.

\section{Interactions between solitons}

\label{sec:Interactions} Direct simulations were also used to study
interactions between two initially quiescent solitons separated by a
relatively small distance. In the 2D single-component model, a similar
investigation was reported in Ref.~\cite{Michinel}, for vortices with $s=1$.
Collisions between asymmetric solitons in the 1D dual-core model were
studied earlier in Ref.~\cite{Chu}. Here we focus on pairs of initially
quiescent solitons (rather than moving ones), as the interaction effects are
strongest in such a case.

Our analysis was performed for several combinations of stable asymmetric
solutions, for all the three values of the spin considered here, $s=0,1,2$.
First, we examined the interaction between identical solitons separated by
distance $\Delta x$, namely: $\left\{ U,V\right\} _{\mathrm{initial}%
}(x,y)=\left\{ U,V\right\} _{\mathrm{stationary}}(x-\Delta x,y)+\left\{
U,V\right\} _{\mathrm{stationary}}(x+\Delta x,y)$. Next, we considered pairs
of cross-identical asymmetric solitons (one being a specular counterpart of
the other): $\left\{ U,V\right\} _{\mathrm{initial}}(x,y)=\left\{
U,V\right\} _{\mathrm{stationary}}(x-\Delta x,y)+\left\{ V,U\right\} _{%
\mathrm{stationary}}(x+\Delta x,y)$. We have also performed the simulations
for the soliton pairs with the phase shift of $\Delta \theta =\pi $.

\begin{figure}[tbp]
\subfigure[]{
\begin{minipage}{3.3in}
\includegraphics[width=3.2in]{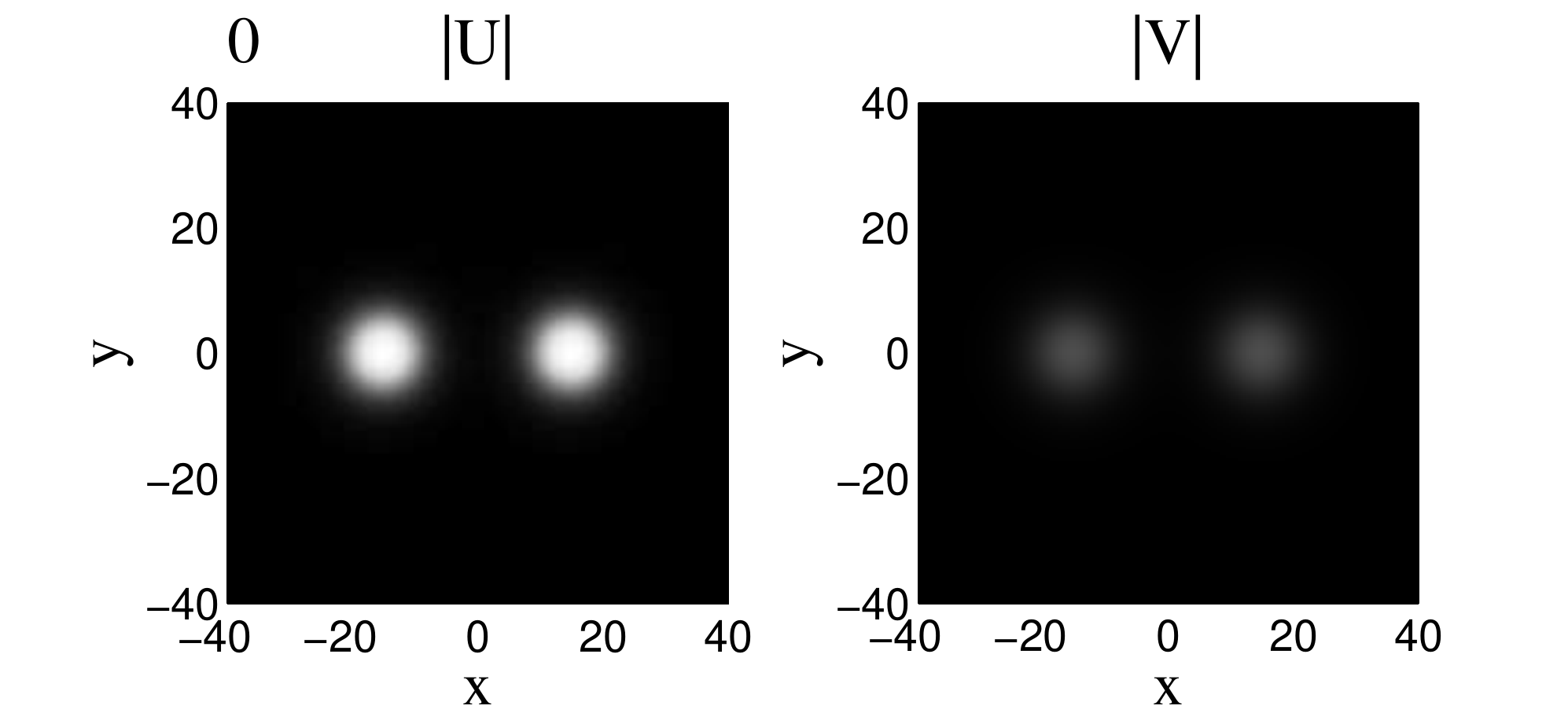}\\
\begin{minipage}{3.3in}
\includegraphics[width=3.2in]{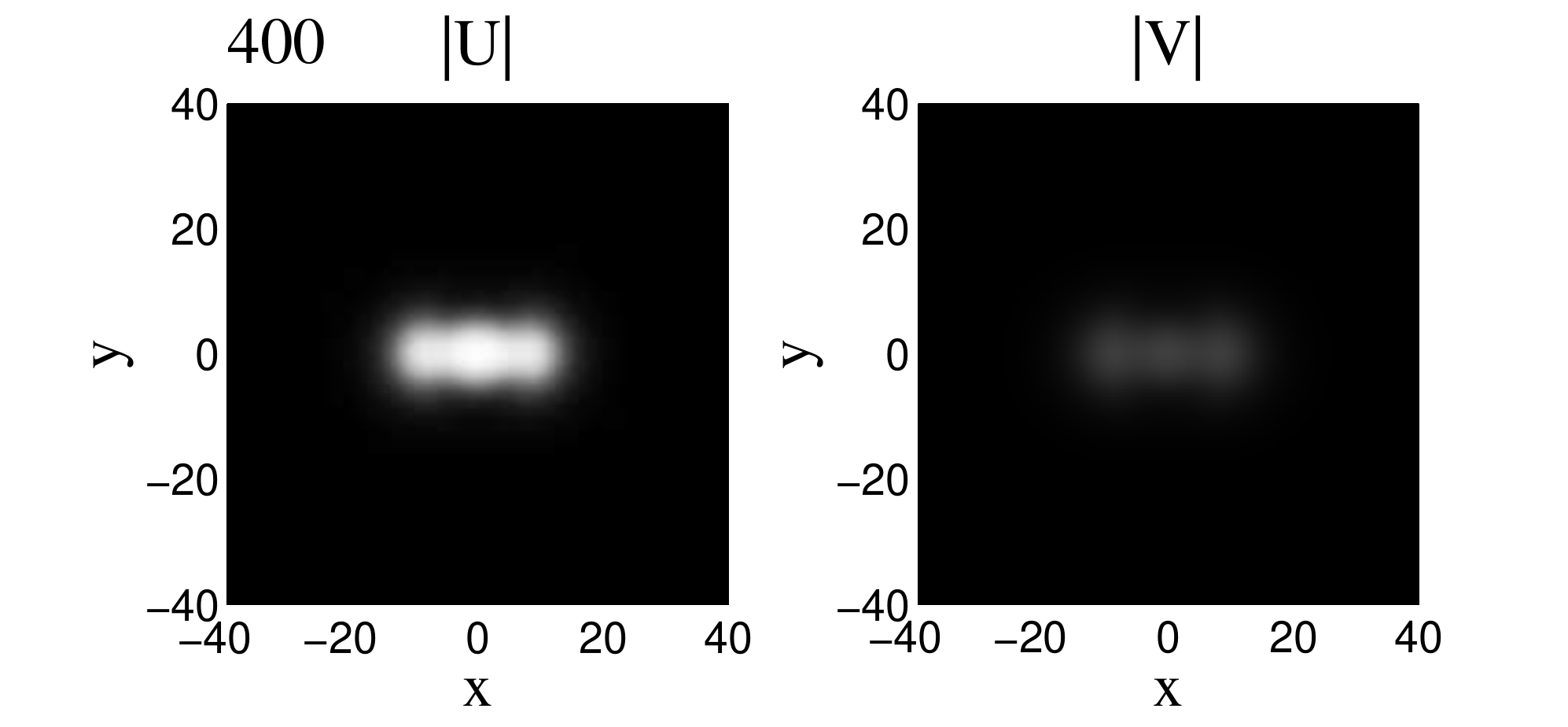}\\
\begin{minipage}{3.3in}
\includegraphics[width=3.2in]{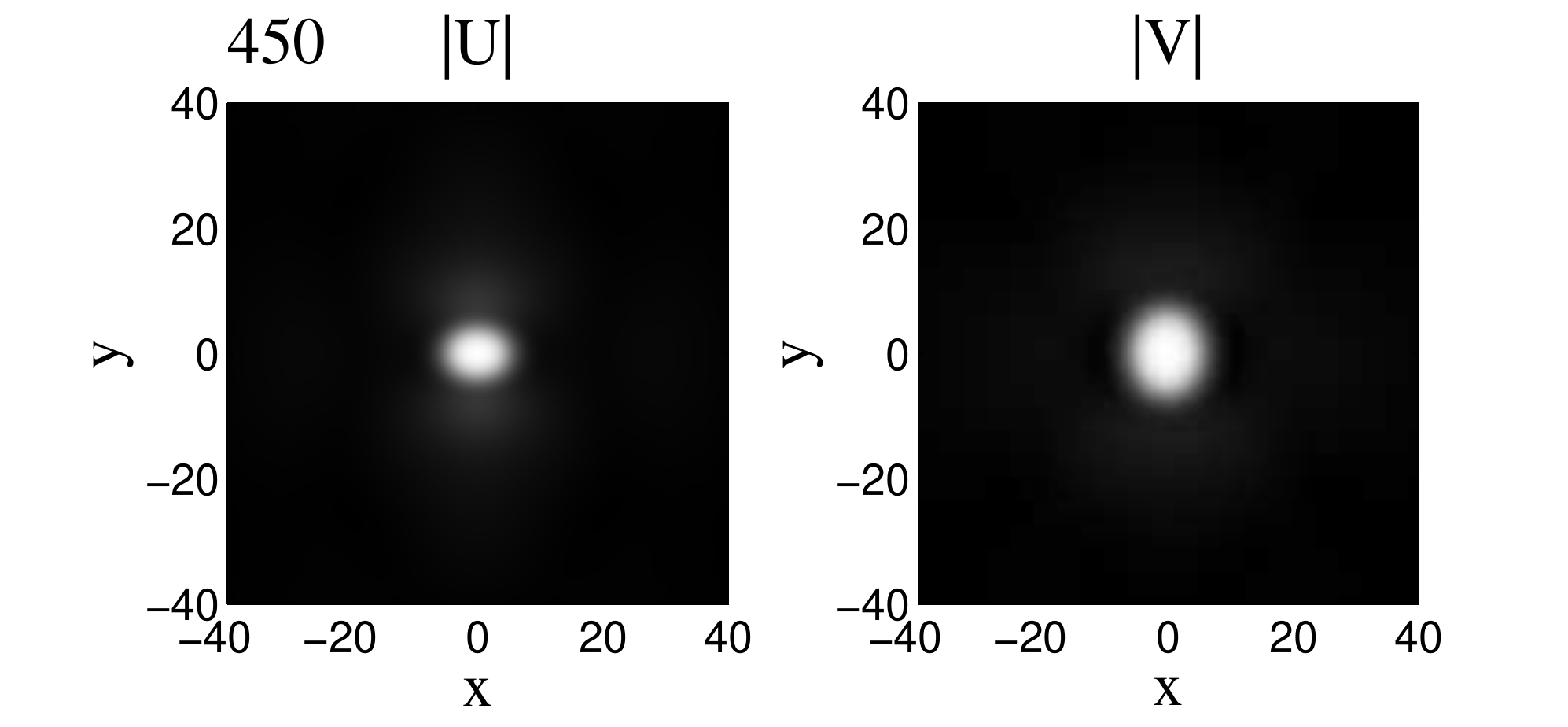}\label{collisions_s0-a}
\end{minipage}
\end{minipage}
\end{minipage}} \quad
\subfigure[]{
\begin{minipage}{3.3in}
\includegraphics[width=3.2in]{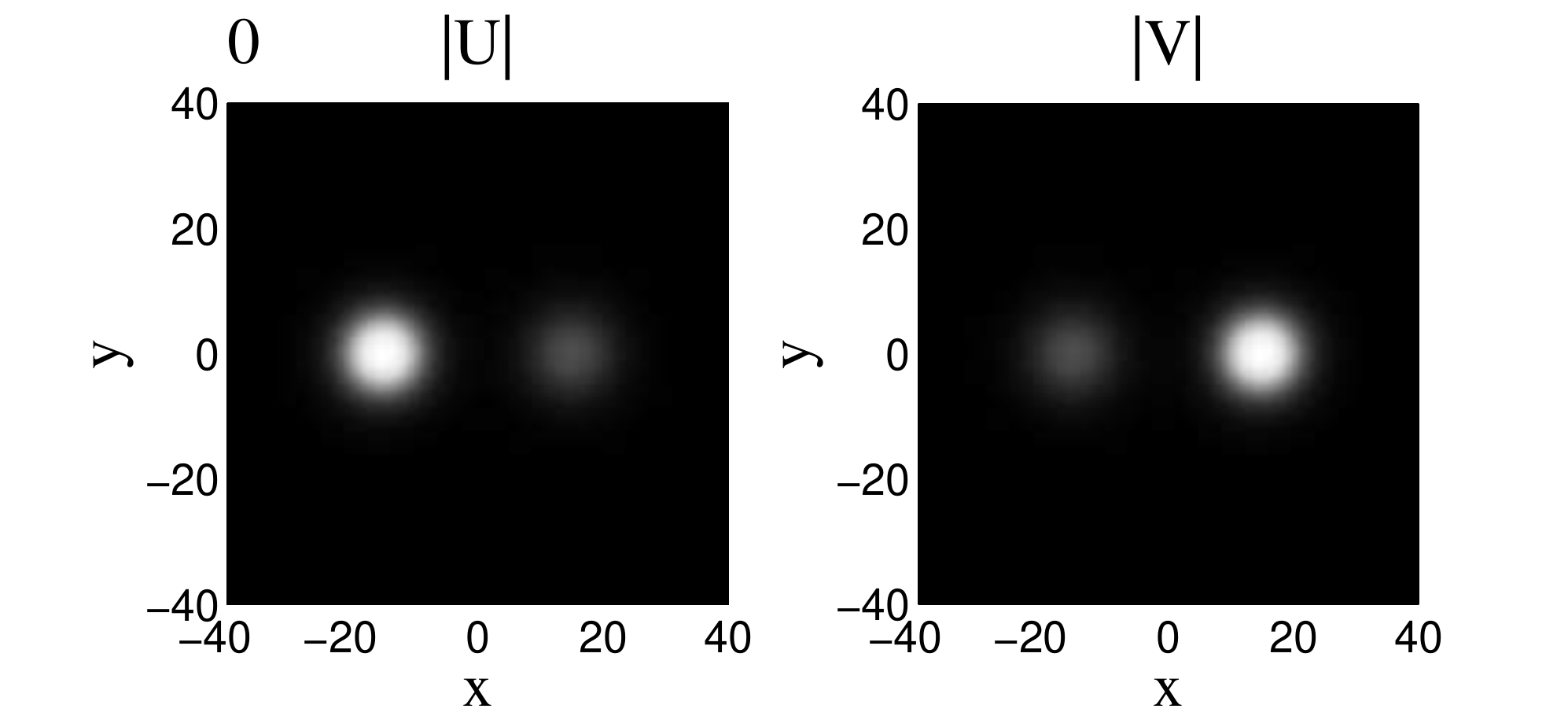}\\
\begin{minipage}{3.3in}
\includegraphics[width=3.2in]{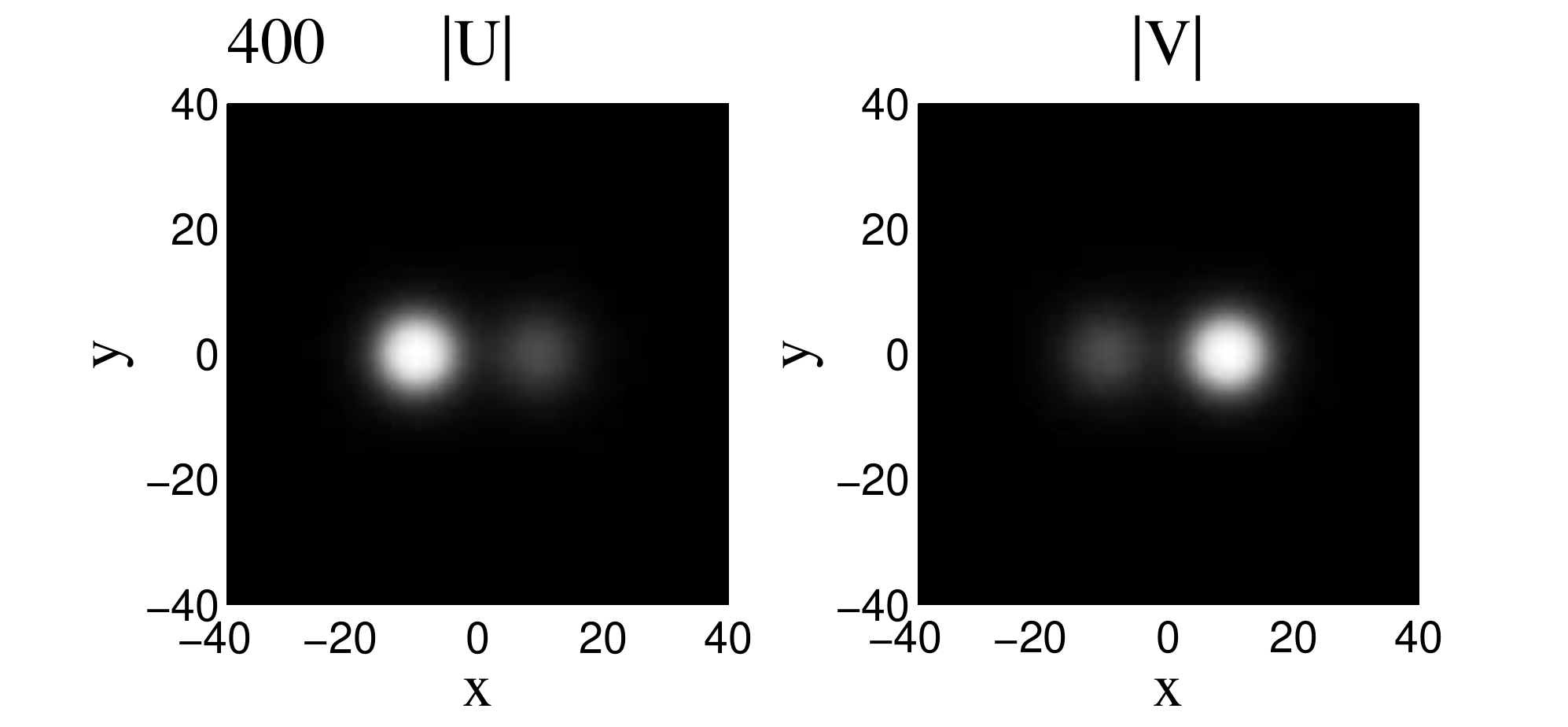}\\
\begin{minipage}{3.3in}
\includegraphics[width=3.2in]{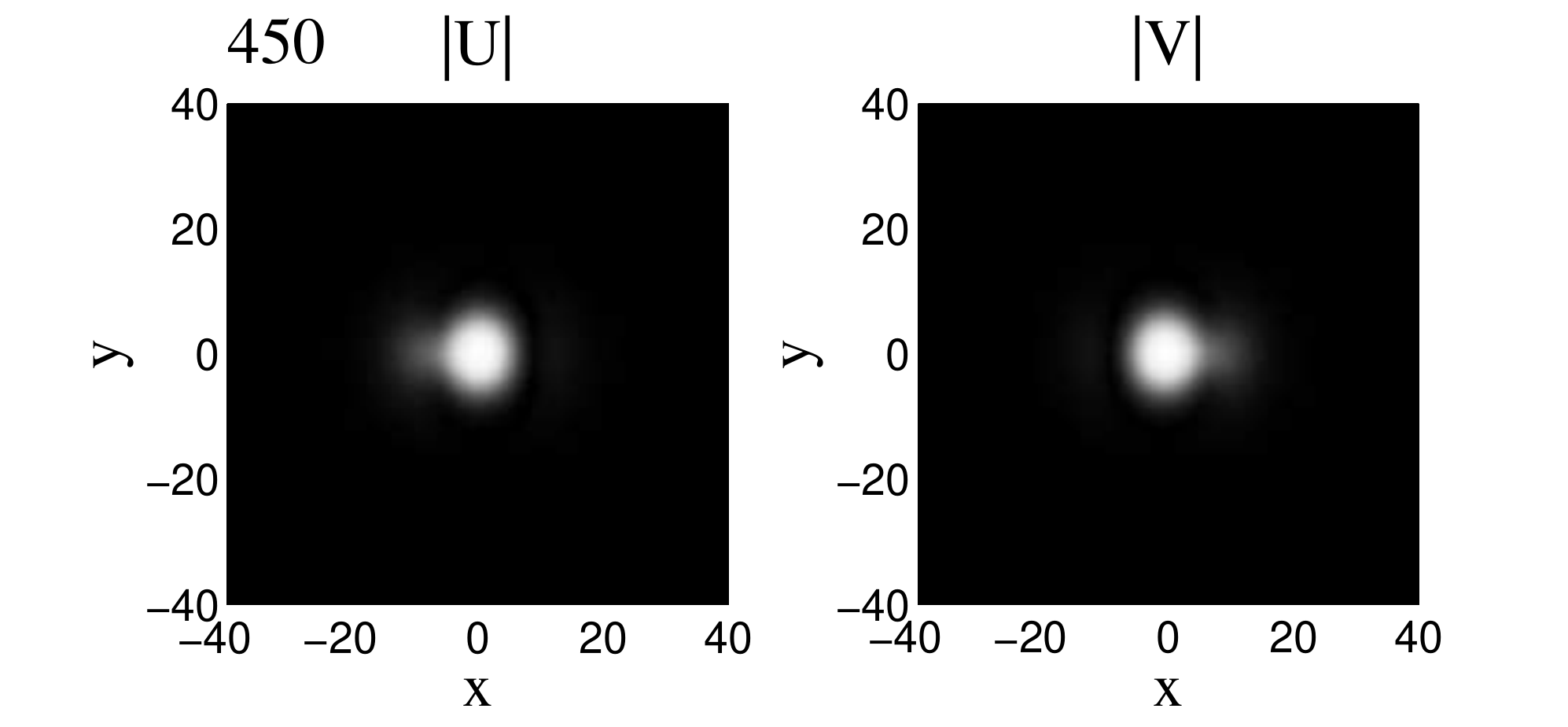}\label{collisions_s0-b}
\end{minipage}
\end{minipage}
\end{minipage}}
\caption{The interaction between fundamental ($s=0$) in-phase ($\Delta
\protect\theta =0$) asymmetric solitons, with $\protect\lambda =0.05$, $%
k=0.16$, and initial separation $\Delta x=30$. In panel (a) the solitons are
identical, while in (b) they are cross-symmetric. The evolution distances, $%
z $, are marked above each frame.}
\label{collisions_s0}
\end{figure}

\begin{figure}[tbp]
\subfigure[]{
\begin{minipage}{3.3in}
\includegraphics[width=3.2in]{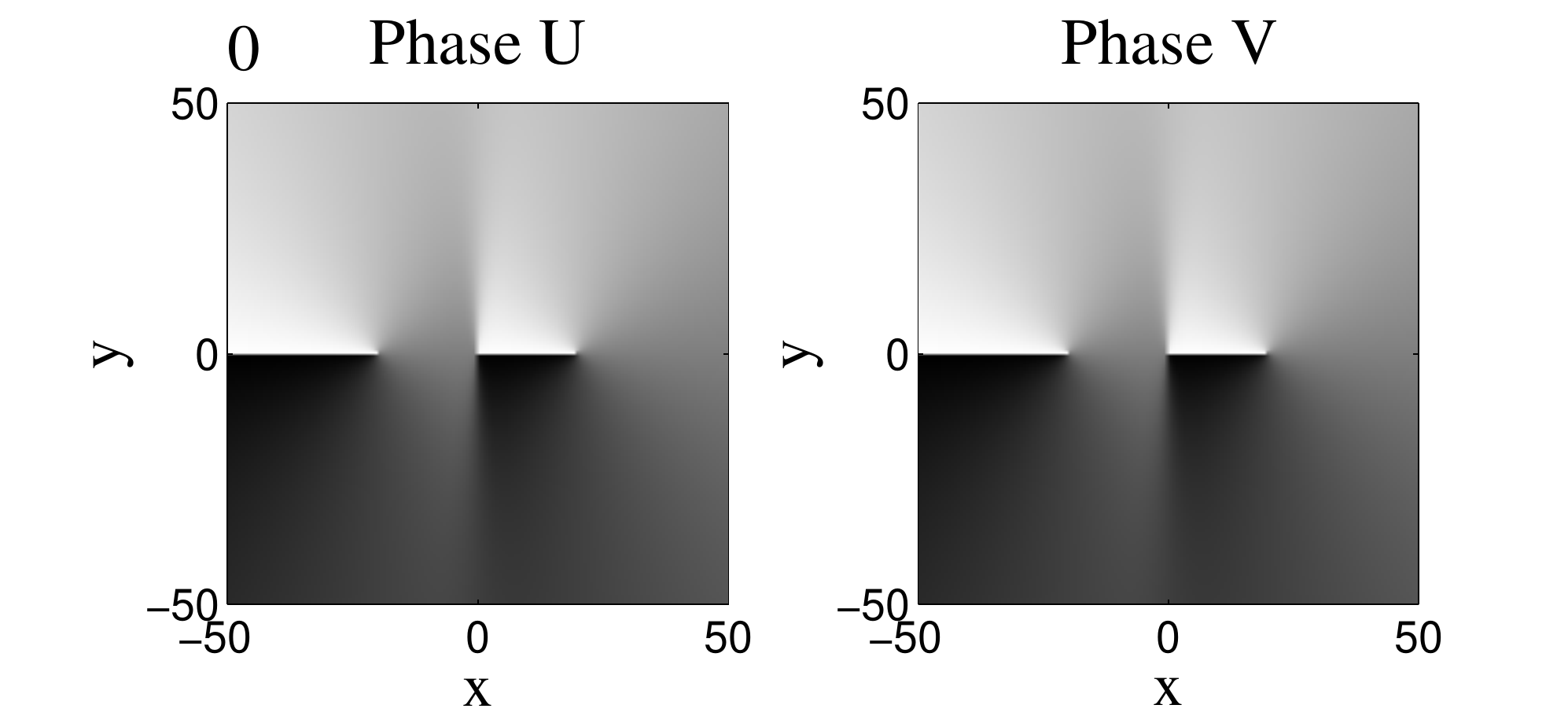} \\
\begin{minipage}{3.3in}
\includegraphics[width=3.2in]{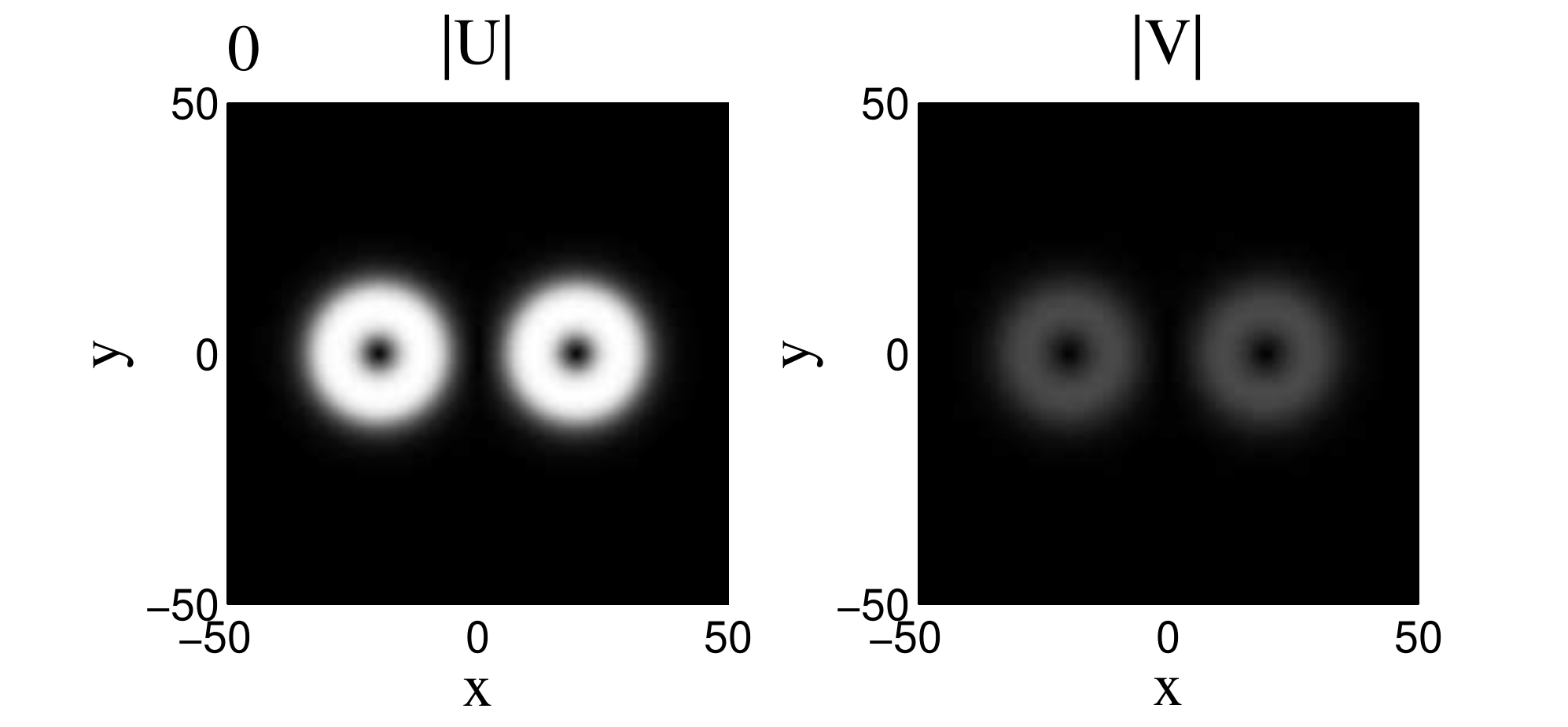}\\
\begin{minipage}{3.3in}
\includegraphics[width=3.2in]{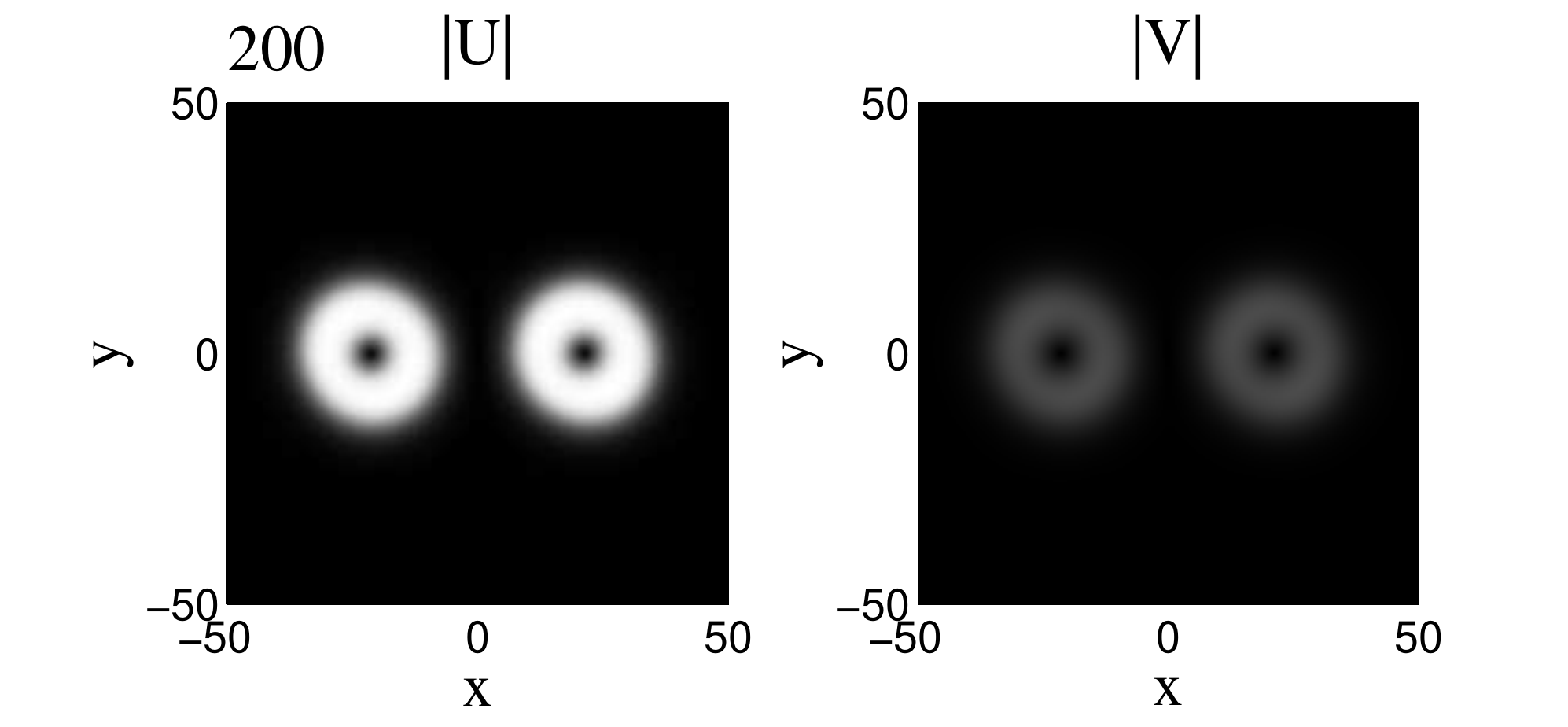}\\
\begin{minipage}{3.3in}
\includegraphics[width=3.2in]{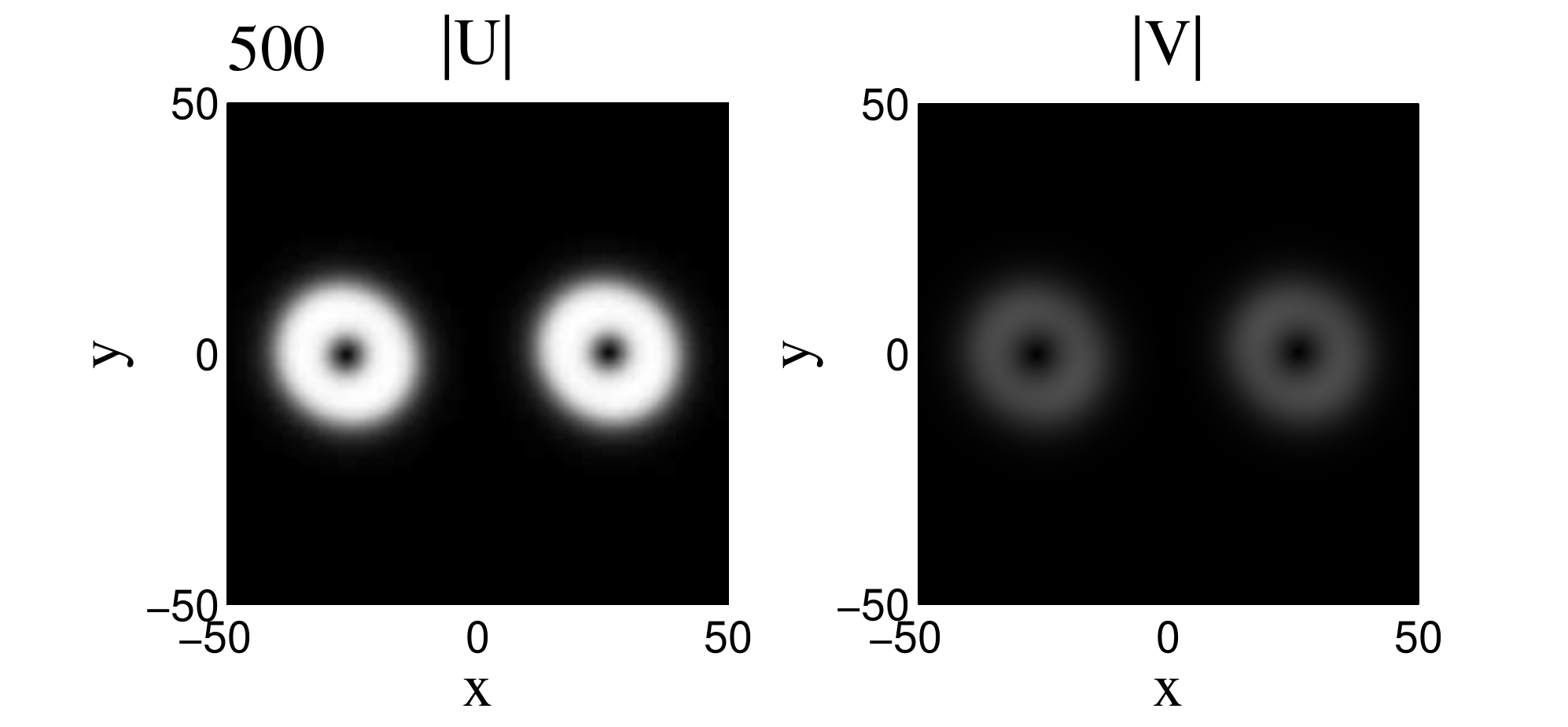}\label{collisions_s1-a}
\end{minipage}
\end{minipage}
\end{minipage}
\end{minipage}} \quad
\subfigure[]{
\begin{minipage}{3.3in}
\includegraphics[width=3.2in]{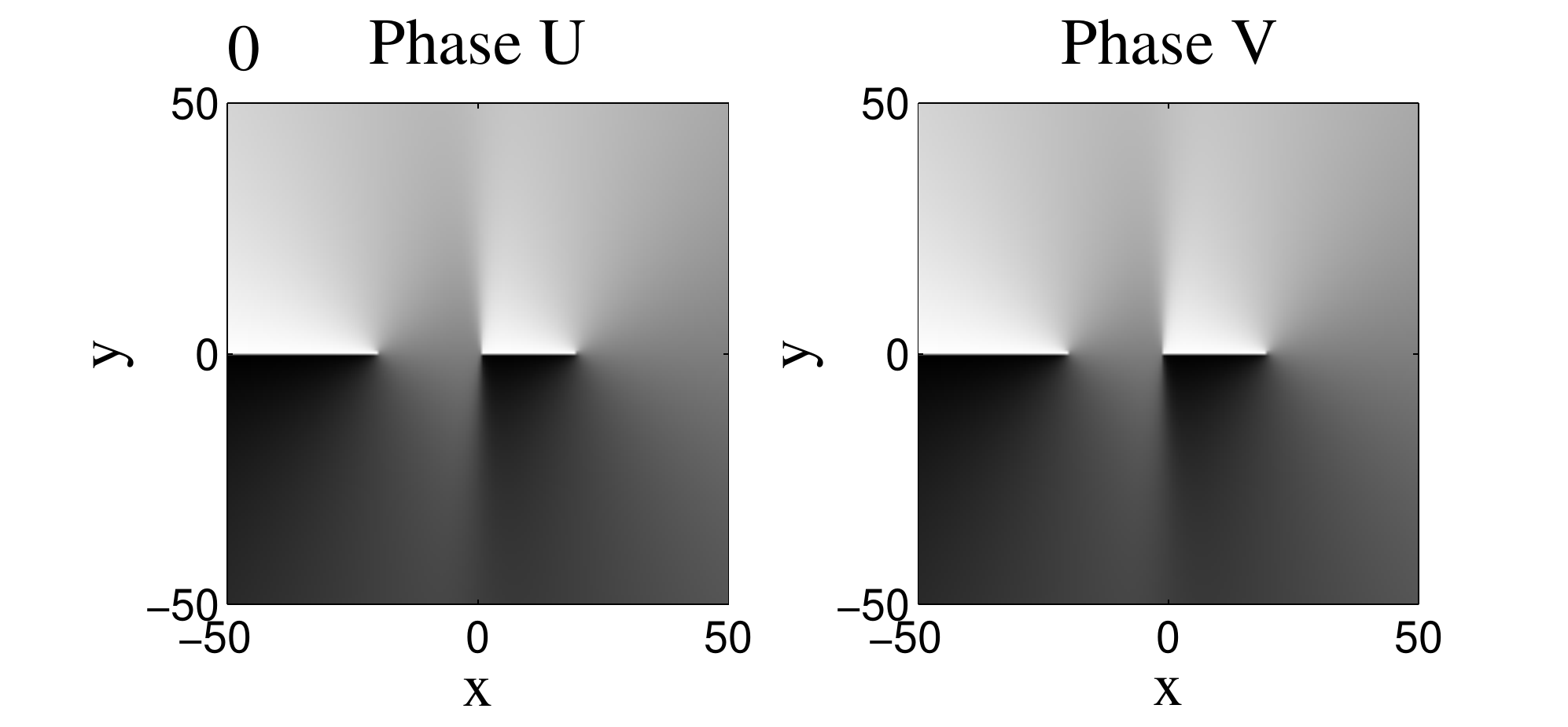} \\
\begin{minipage}{3.3in}
\includegraphics[width=3.2in]{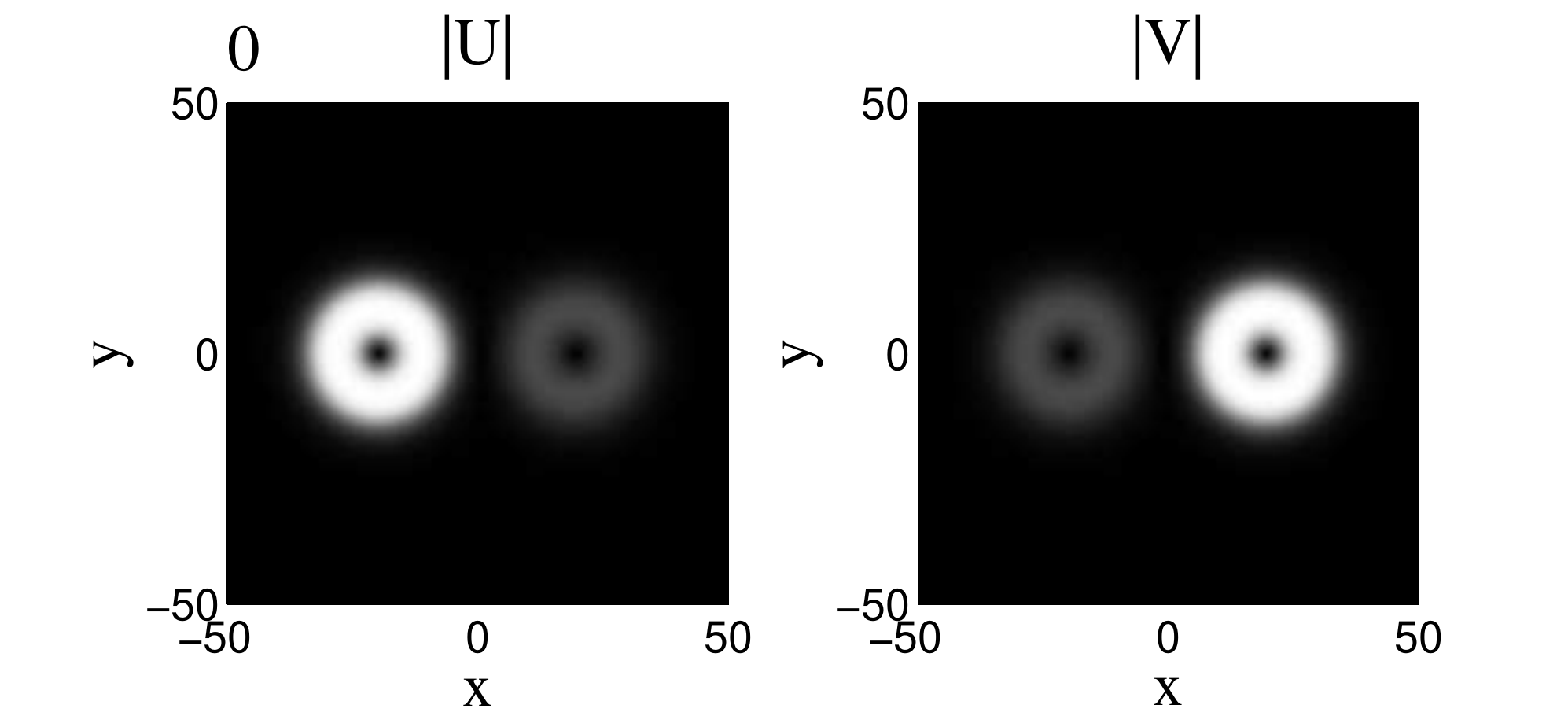}\\
\begin{minipage}{3.3in}
\includegraphics[width=3.2in]{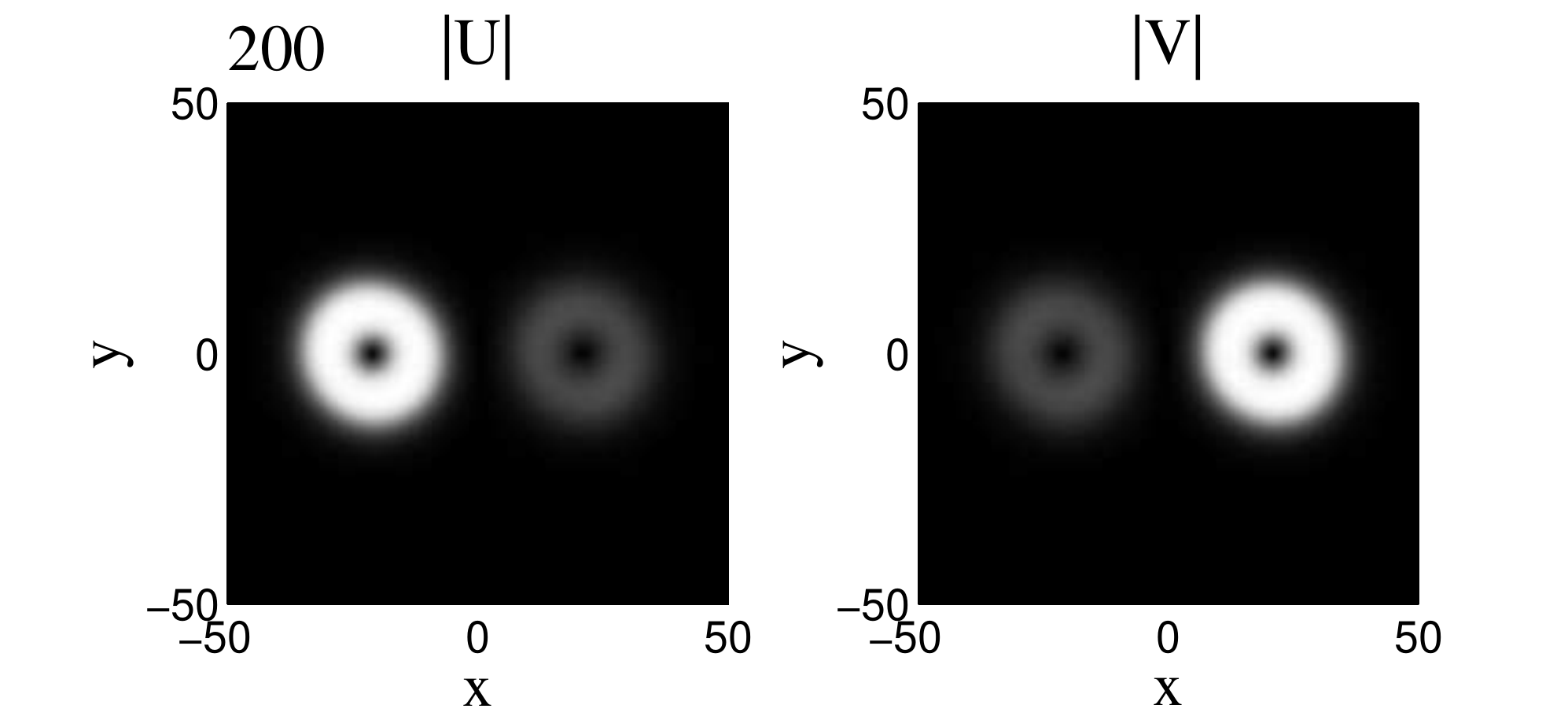}\\
\begin{minipage}{3.3in}
\includegraphics[width=3.2in]{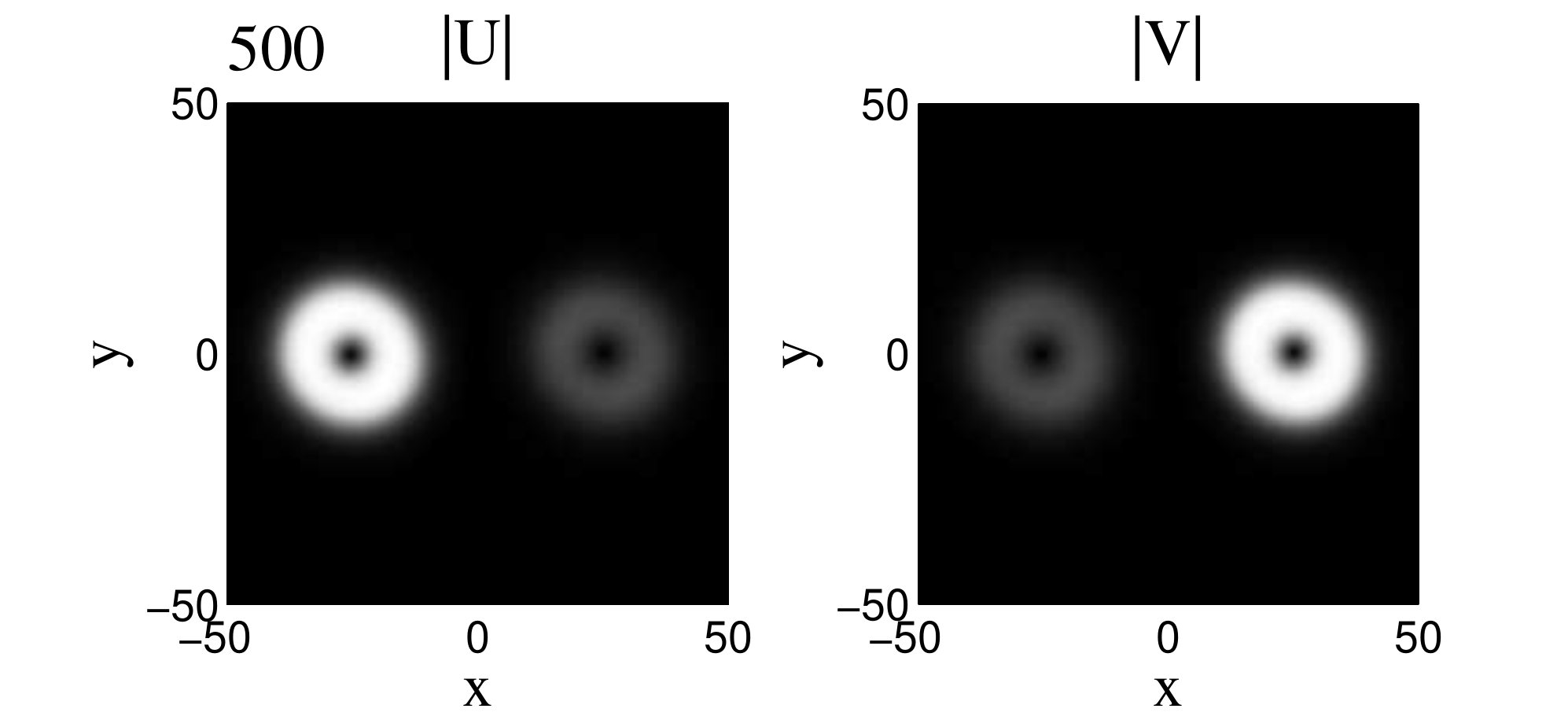}\label{collisions_s1-b}
\end{minipage}
\end{minipage}
\end{minipage}
\end{minipage}}
\caption{The interaction between asymmetric in-phase vortices with $s=1$, $%
\protect\lambda =0.05$, $k=0.18$, and $\Delta x=40$. Notice the \emph{%
repulsion} between the in-phase vortex solitons in this case.}
\label{collisions_s1}
\end{figure}

\begin{figure}[tbp]
\subfigure[]{
\begin{minipage}{3.3in}
\includegraphics[width=3.2in]{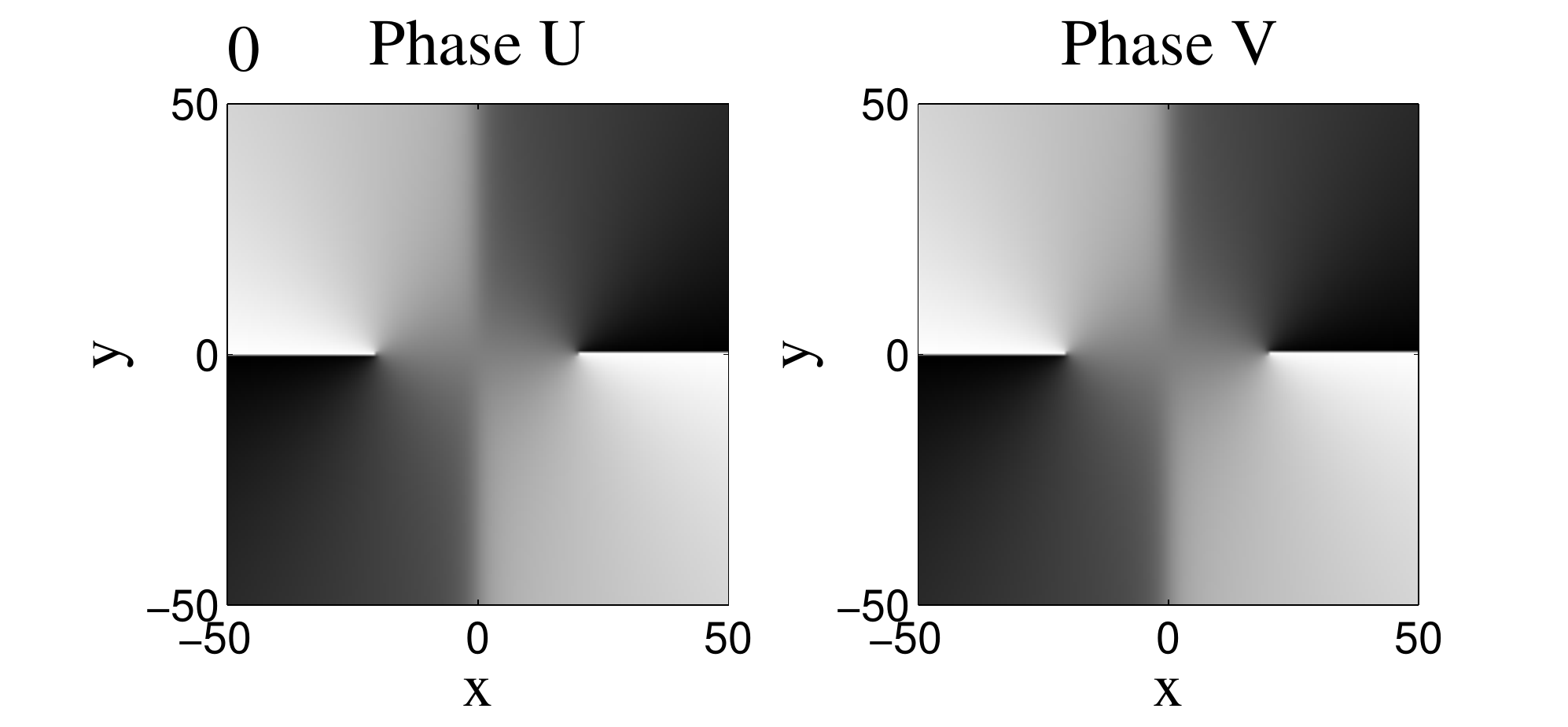} \\
\begin{minipage}{3.3in}
\includegraphics[width=3.2in]{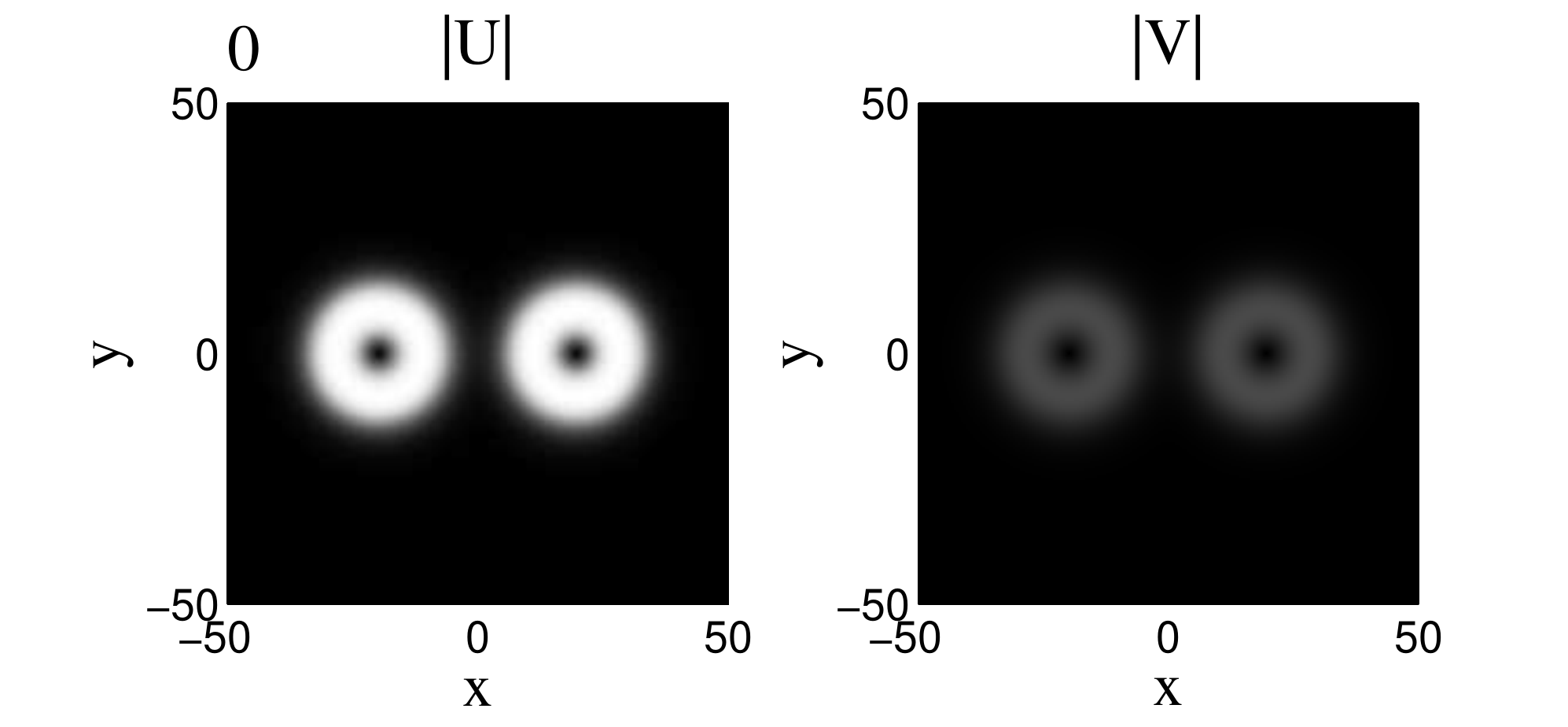}\\
\begin{minipage}{3.3in}
\includegraphics[width=3.2in]{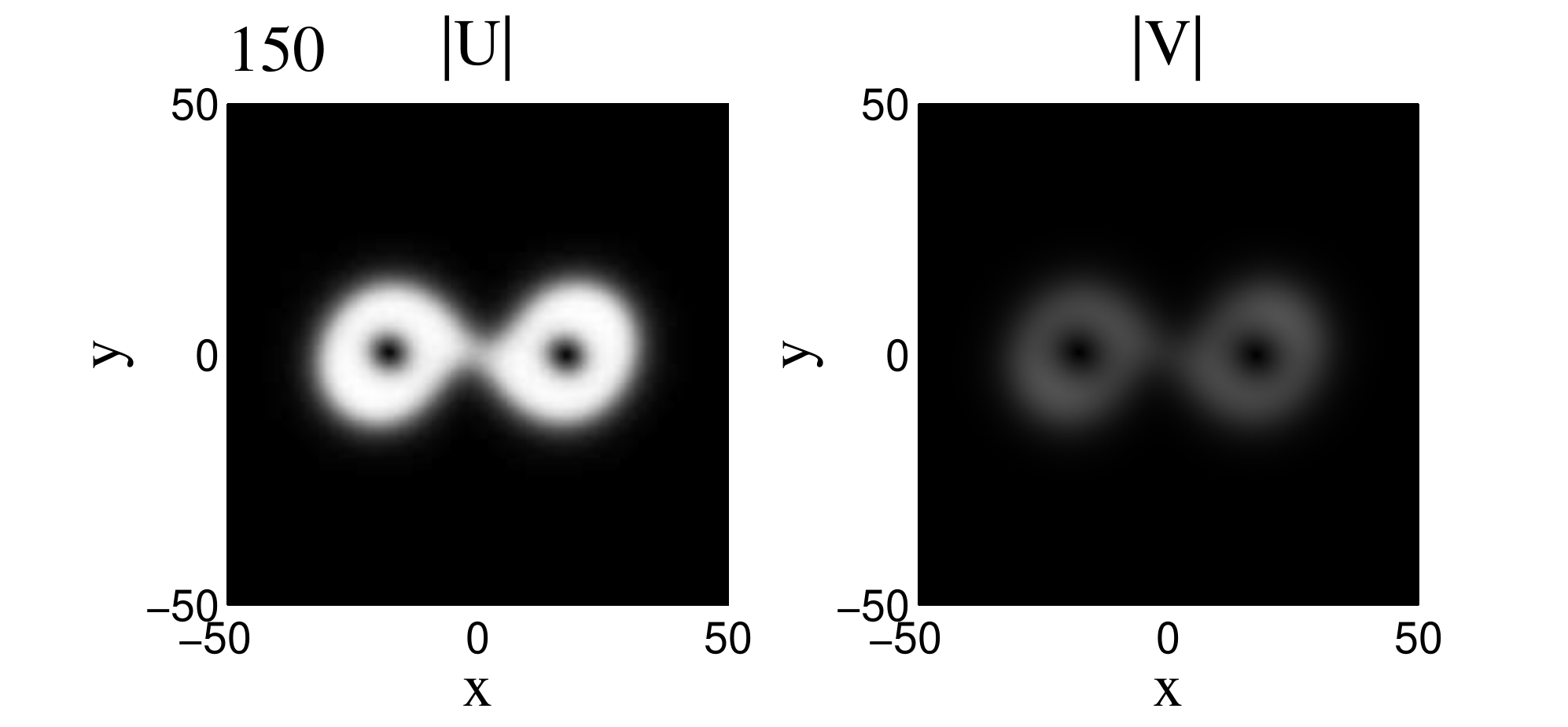}\\
\begin{minipage}{3.3in}
\includegraphics[width=3.2in]{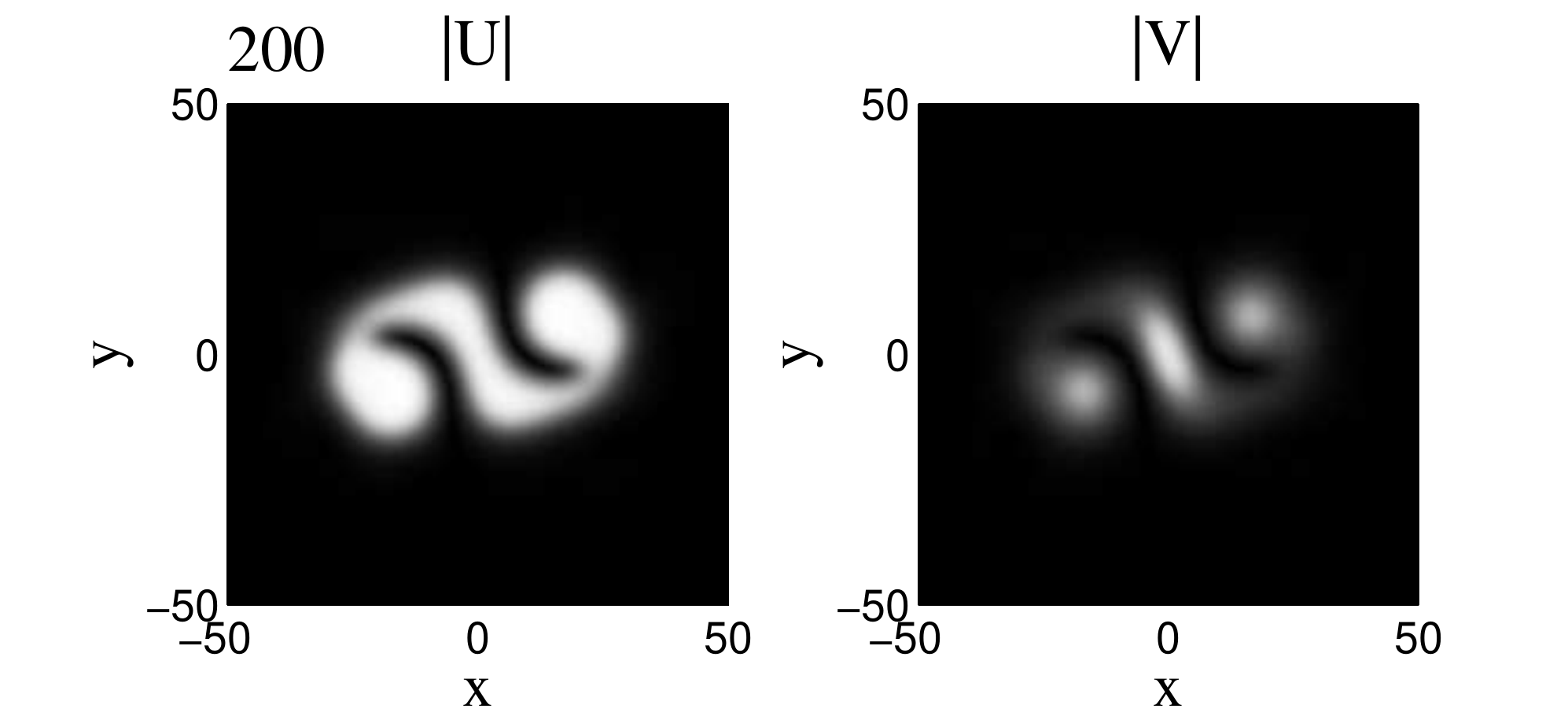}\\
\begin{minipage}{3.3in}
\includegraphics[width=3.2in]{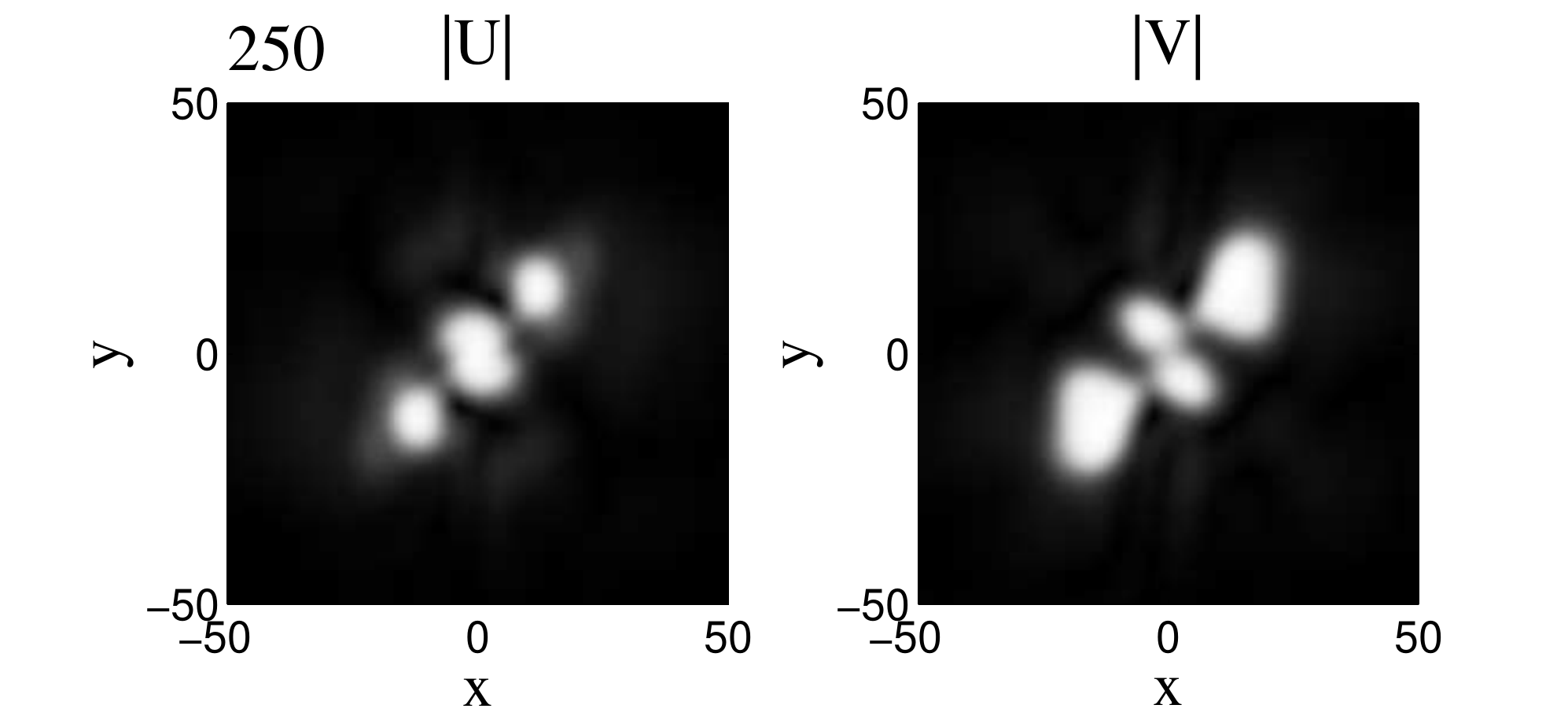}\label{collisions_s1_shift-a}
\end{minipage}
\end{minipage}
\end{minipage}
\end{minipage}
\end{minipage}} \quad
\subfigure[]{
\begin{minipage}{3.3in}
\includegraphics[width=3.2in]{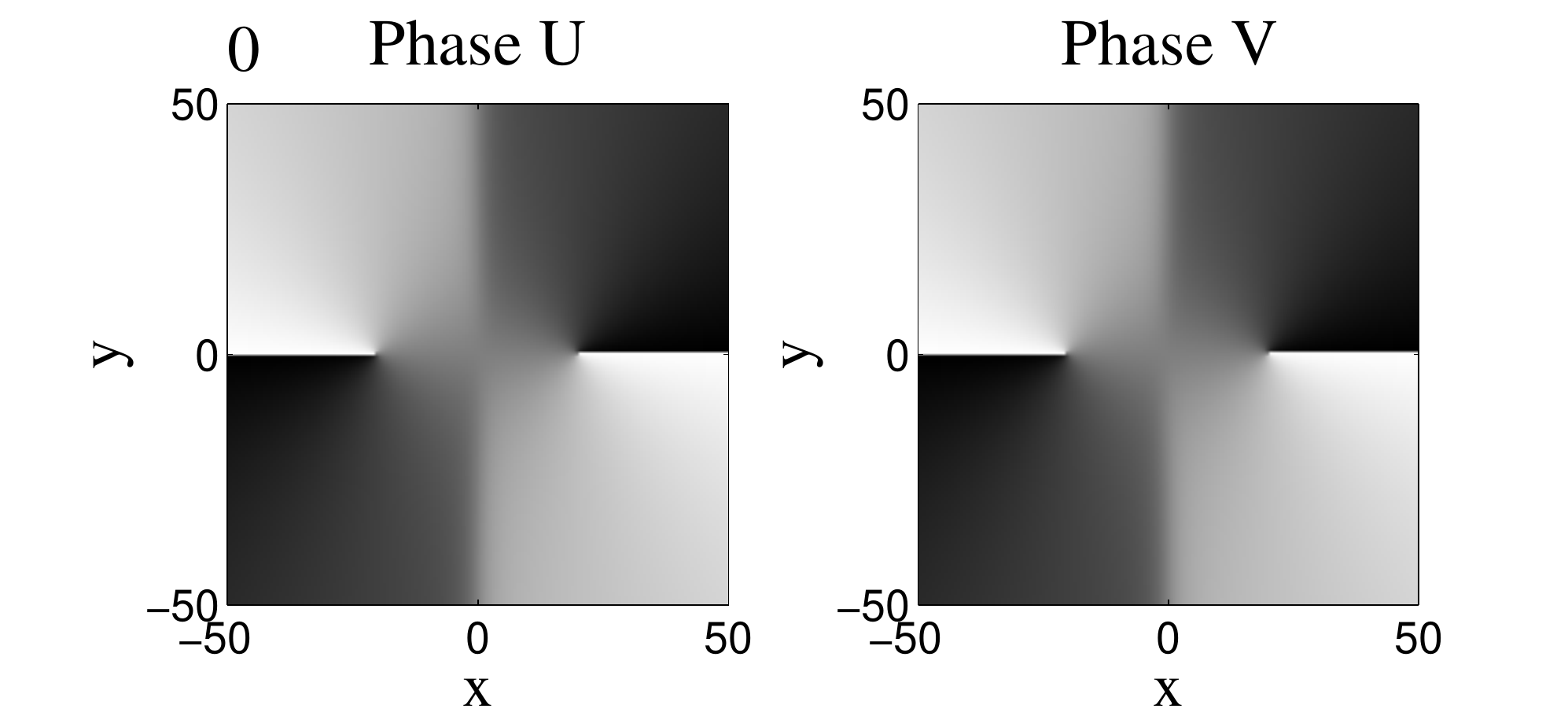} \\
\begin{minipage}{3.3in}
\includegraphics[width=3.2in]{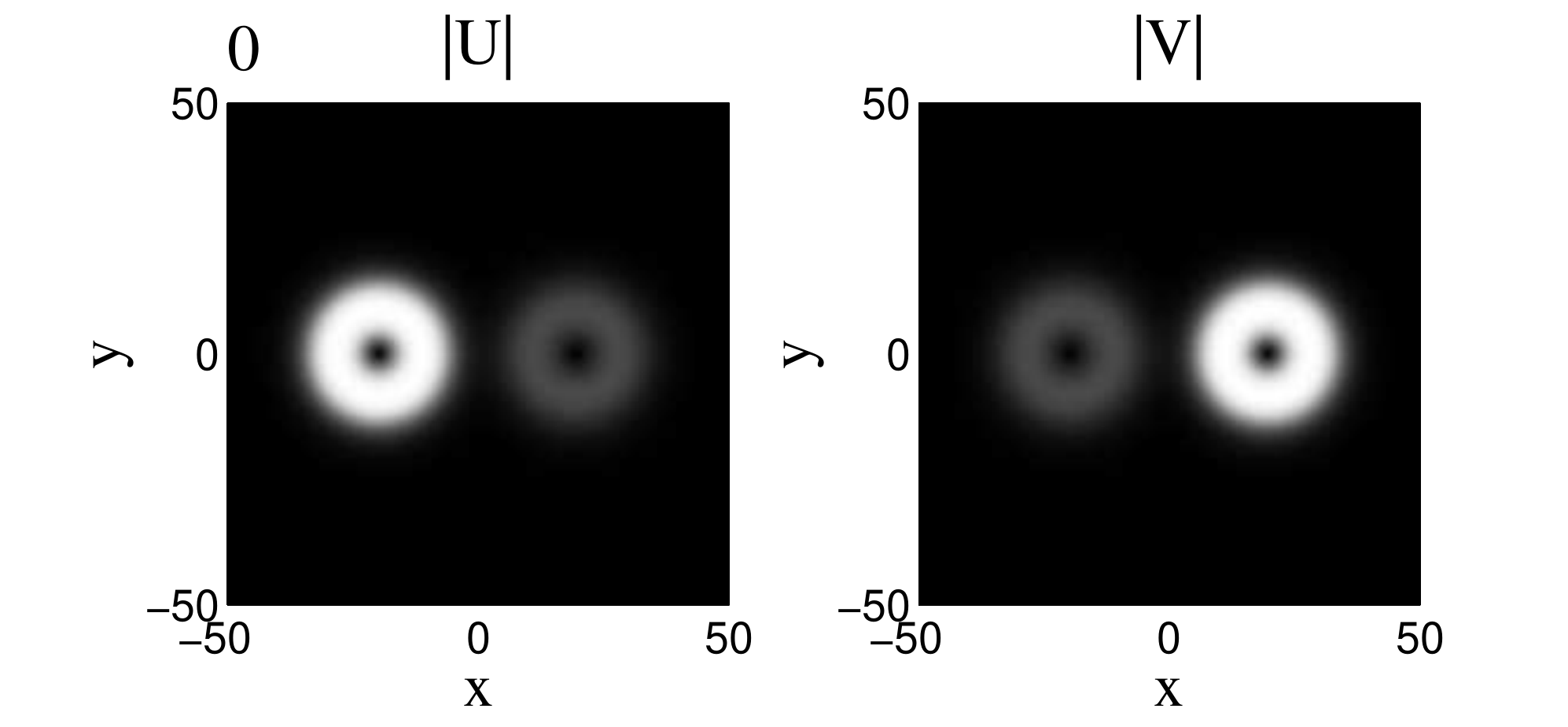}\\
\begin{minipage}{3.3in}
\includegraphics[width=3.2in]{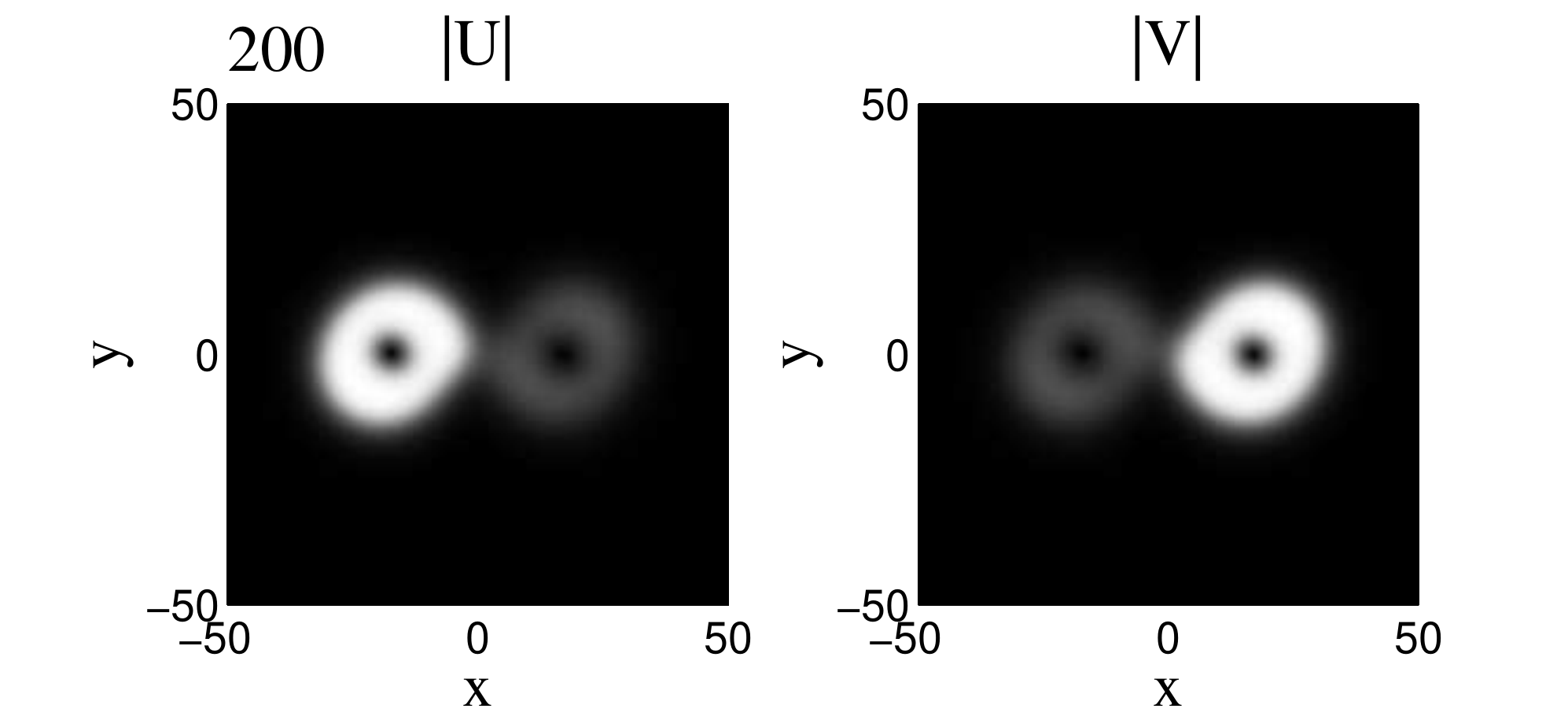}\\
\begin{minipage}{3.3in}
\includegraphics[width=3.2in]{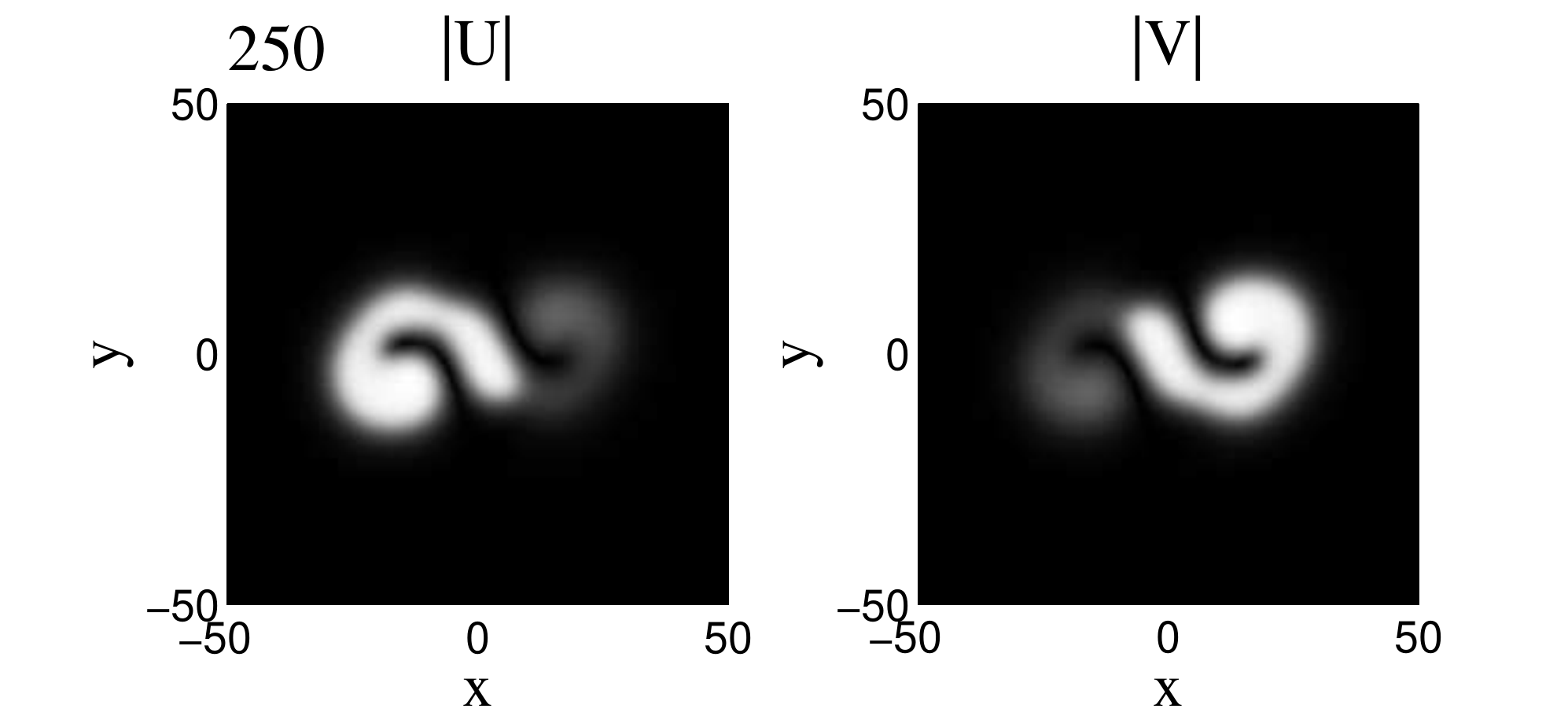}\\
\begin{minipage}{3.3in}
\includegraphics[width=3.2in]{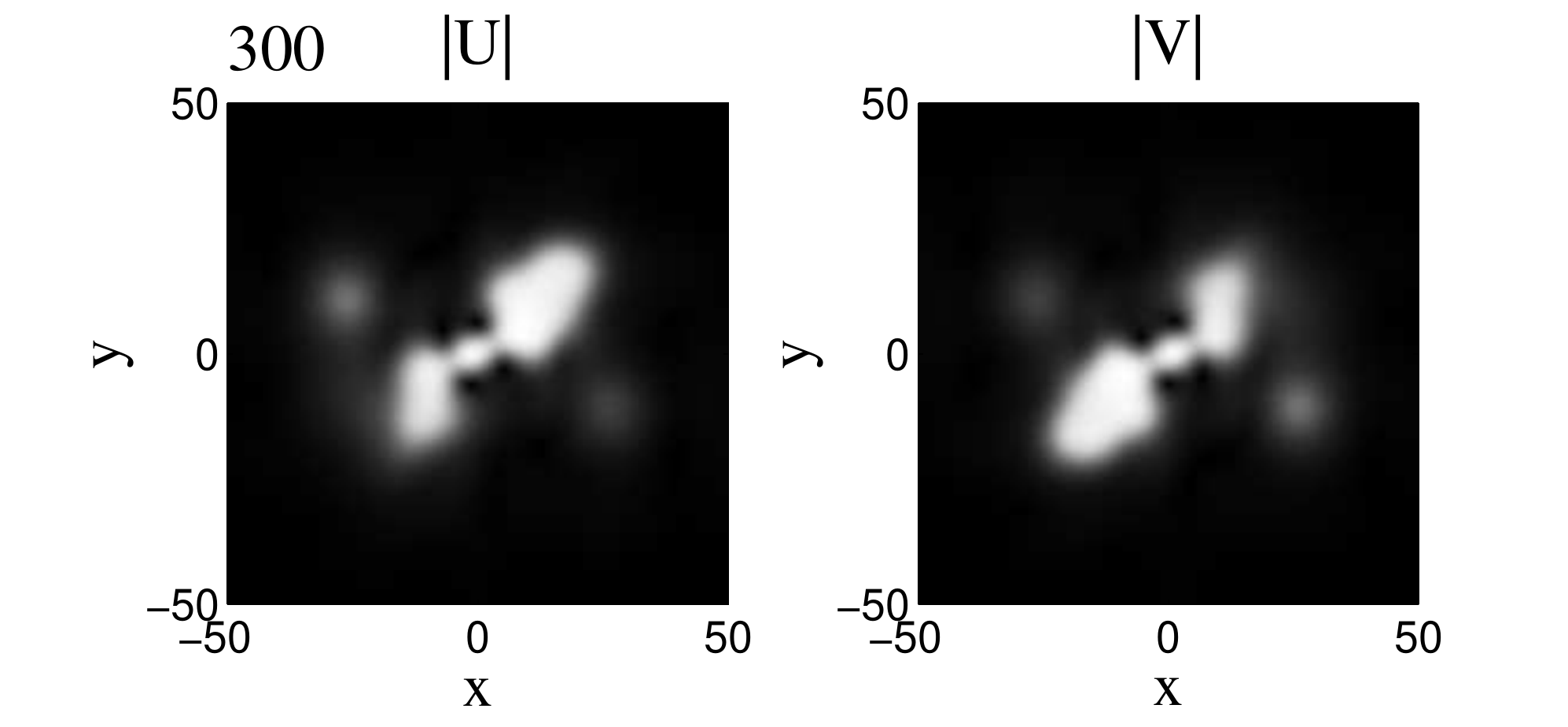}\label{collisions_s1_shift-b}
\end{minipage}
\end{minipage}
\end{minipage}
\end{minipage}
\end{minipage}}
\caption{The interaction between asymmetric vortices with $s=1$, $\Delta
\protect\theta =\protect\pi $, $\protect\lambda =0.05$, $k=0.18$, and $%
\Delta x=40$. Notice the \emph{attraction} between the $\protect\pi $%
-out-of-phase vortex solitons in this case.}
\label{collisions_s1_shift}
\end{figure}

\begin{figure}[tbp]
\subfigure[]{
\begin{minipage}{3.3in}
\includegraphics[width=3.2in]{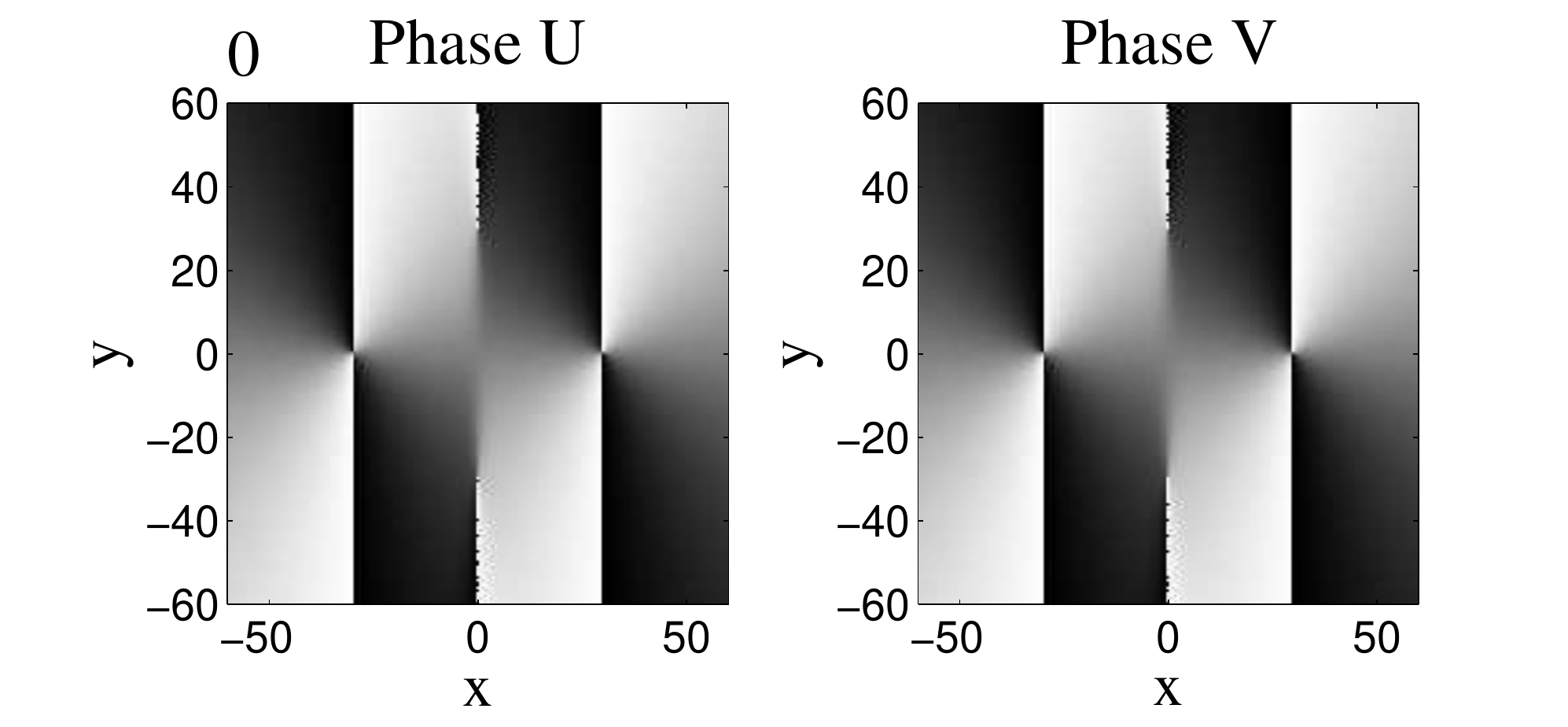} \\
\begin{minipage}{3.3in}
\includegraphics[width=3.2in]{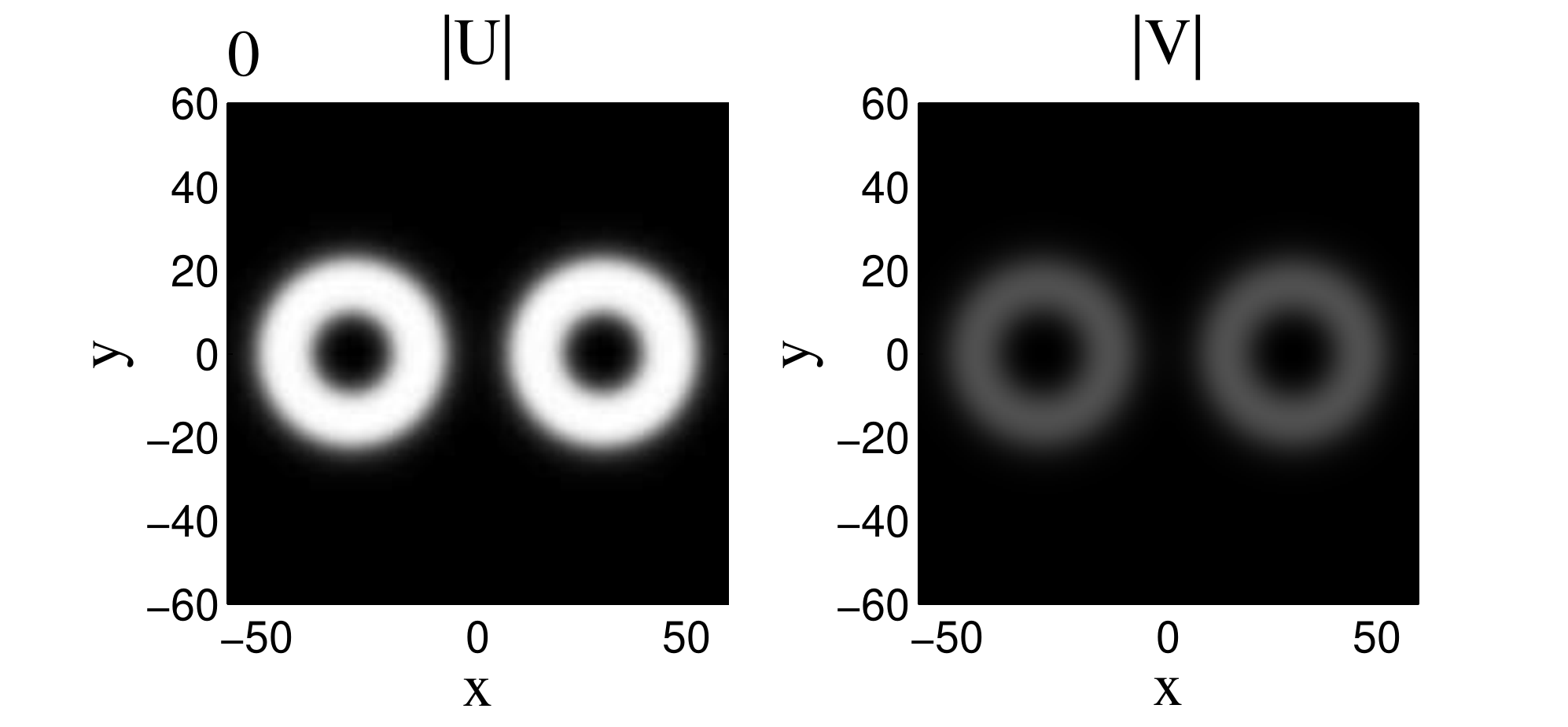}\\
\begin{minipage}{3.3in}
\includegraphics[width=3.2in]{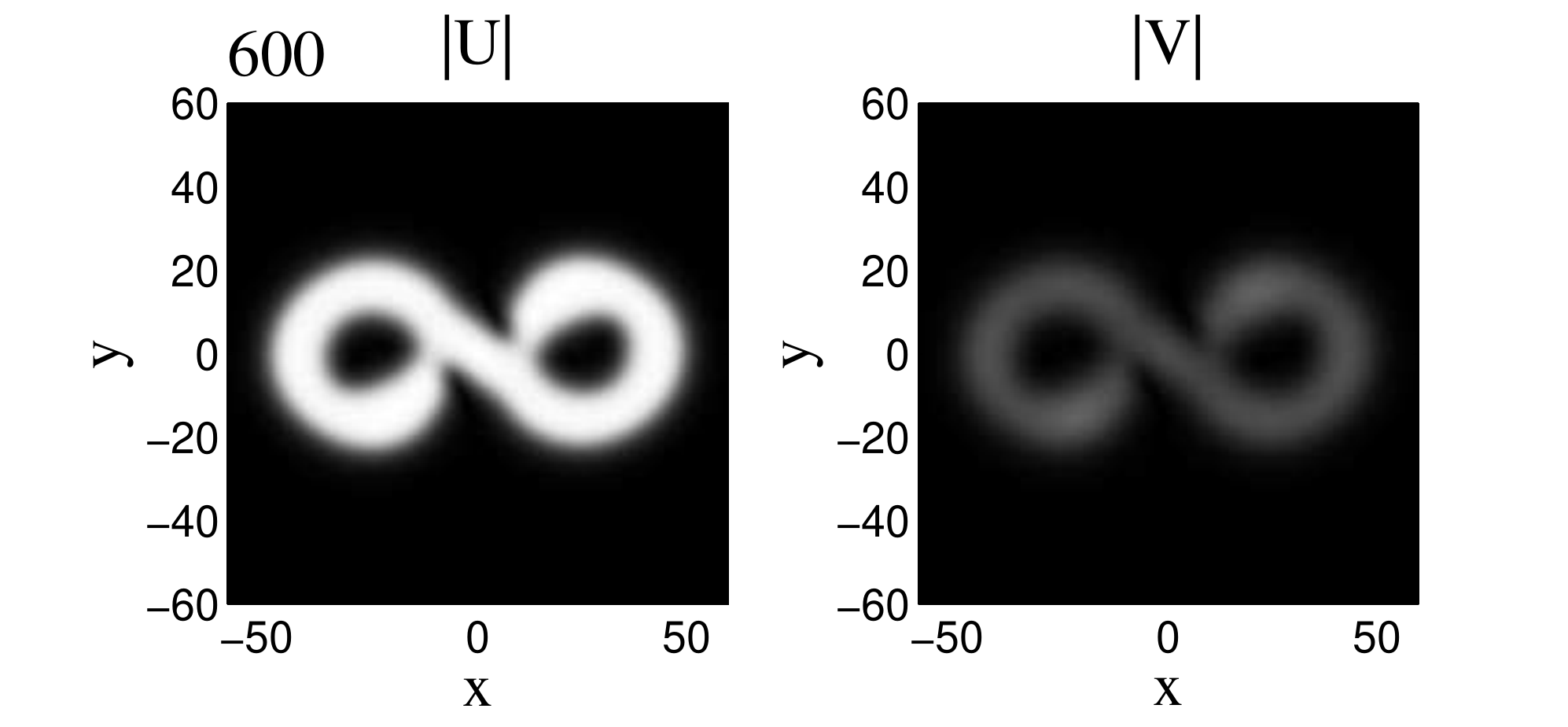}\\
\begin{minipage}{3.3in}
\includegraphics[width=3.2in]{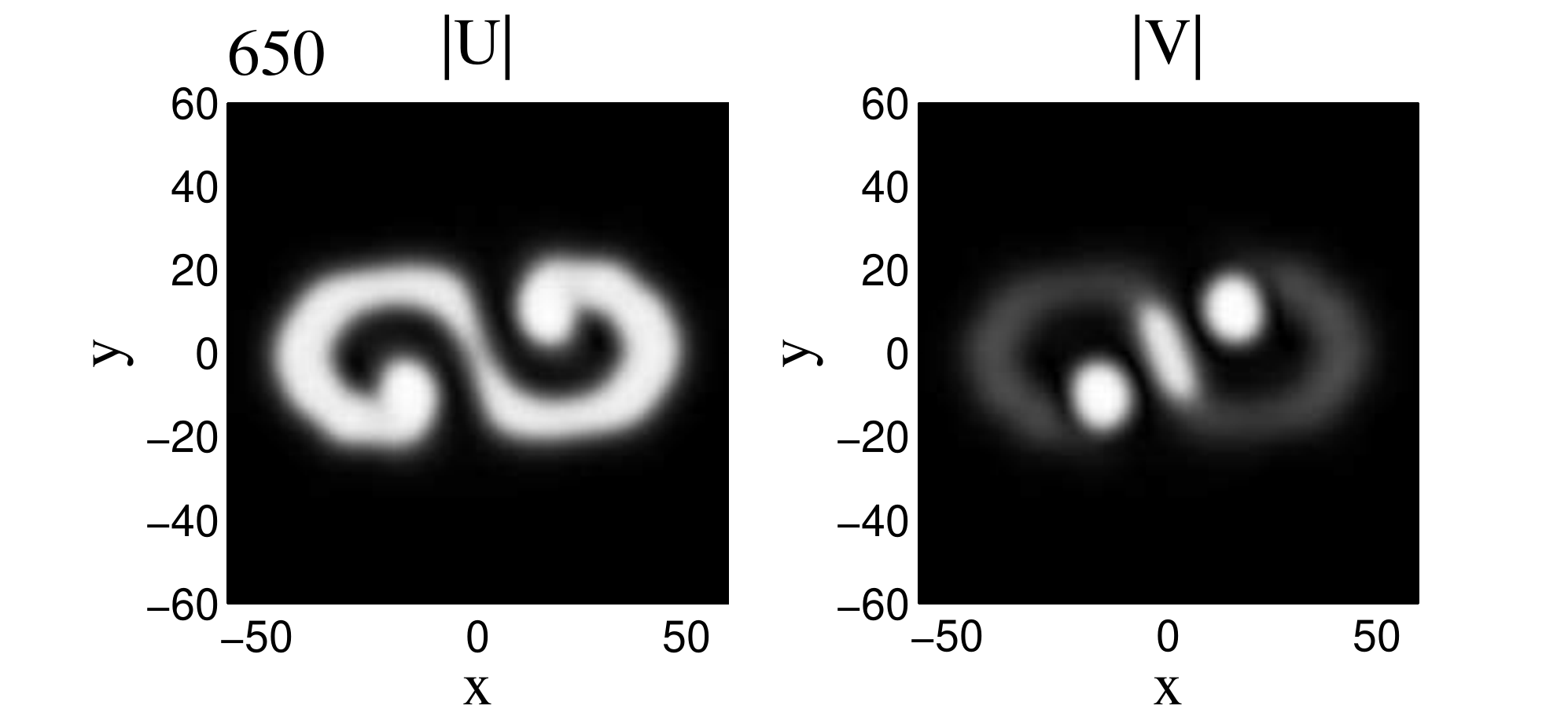}\\
\begin{minipage}{3.3in}
\includegraphics[width=3.2in]{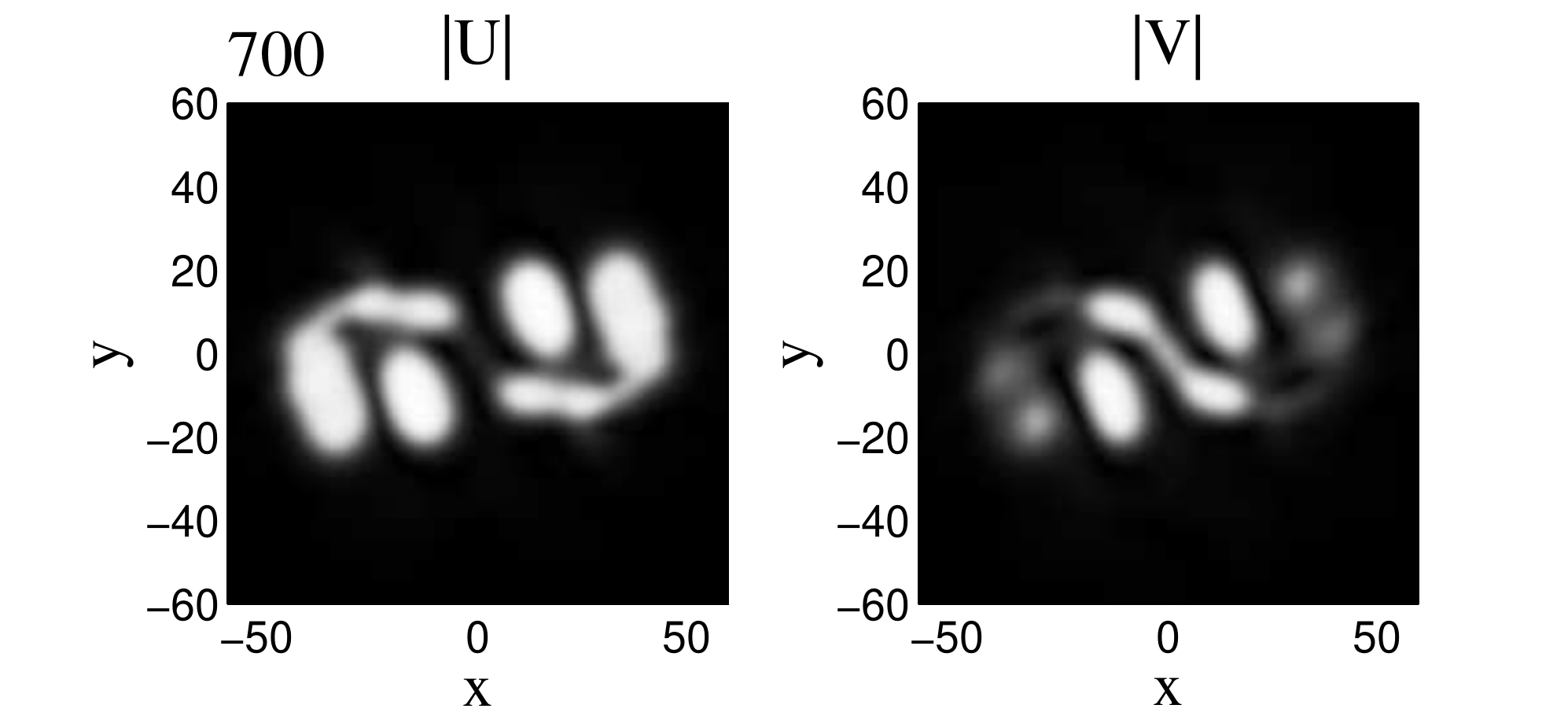}\label{collisions_s2-a}
\end{minipage}
\end{minipage}
\end{minipage}
\end{minipage}
\end{minipage}} \quad
\subfigure[]{
\begin{minipage}{3.3in}
\includegraphics[width=3.2in]{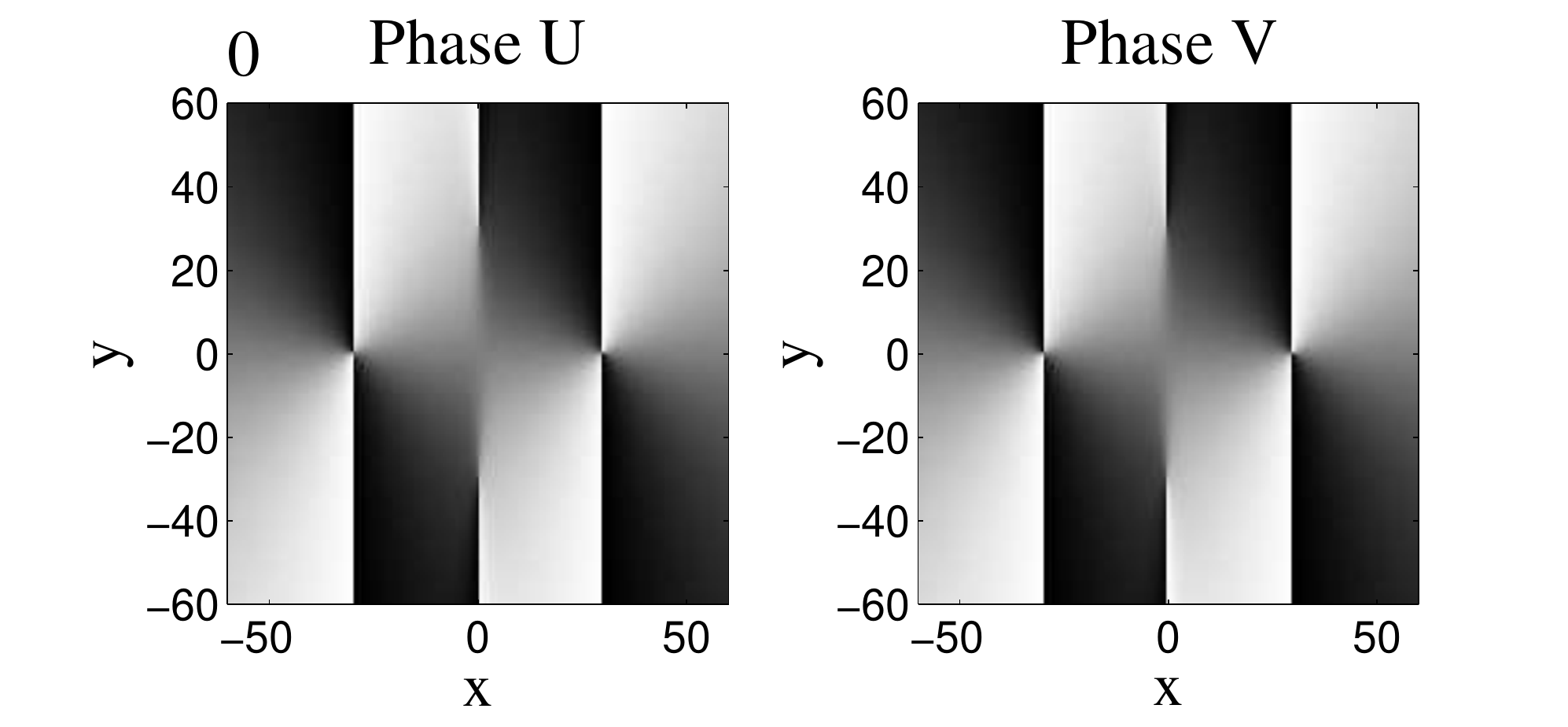} \\
\begin{minipage}{3.3in}
\includegraphics[width=3.2in]{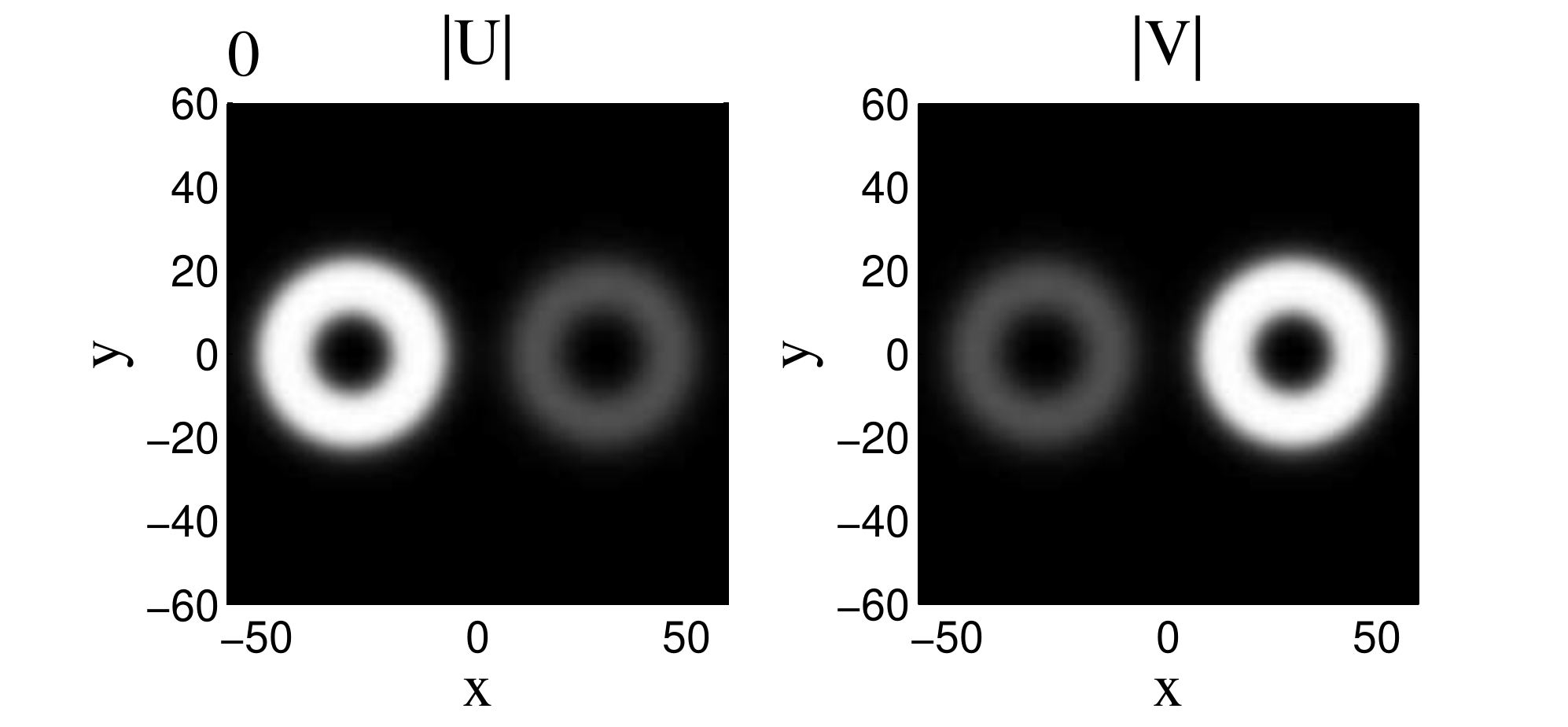}\\
\begin{minipage}{3.3in}
\includegraphics[width=3.2in]{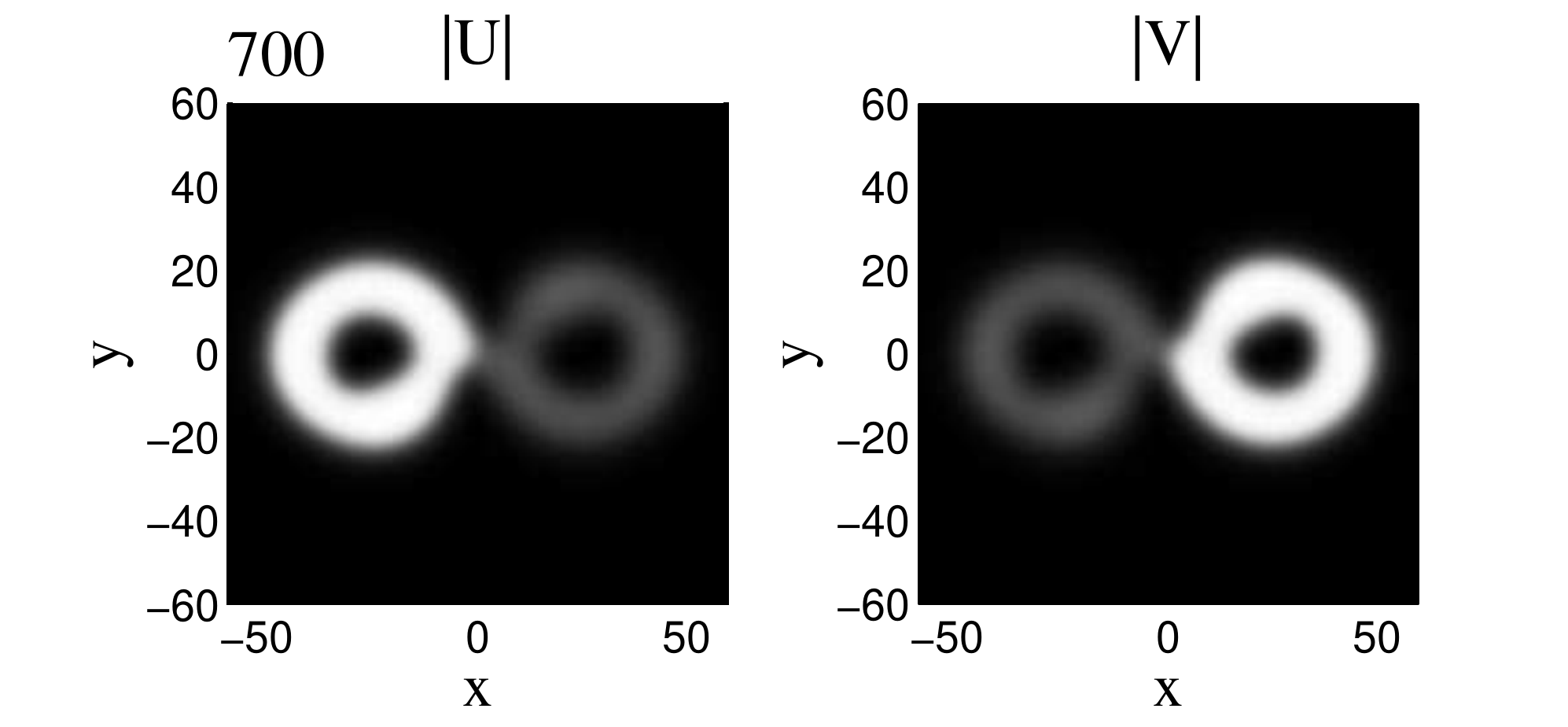}\\
\begin{minipage}{3.3in}
\includegraphics[width=3.2in]{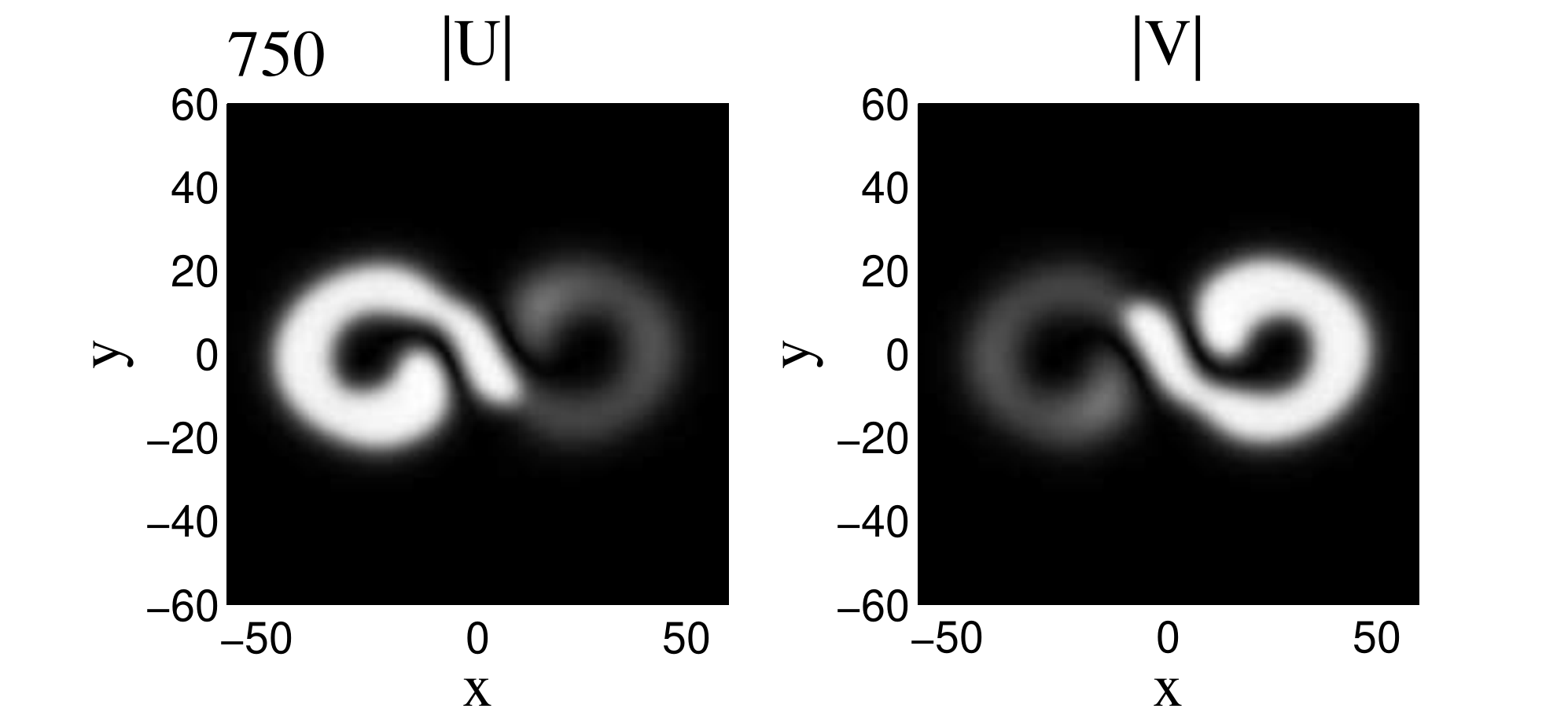}\\
\begin{minipage}{3.3in}
\includegraphics[width=3.2in]{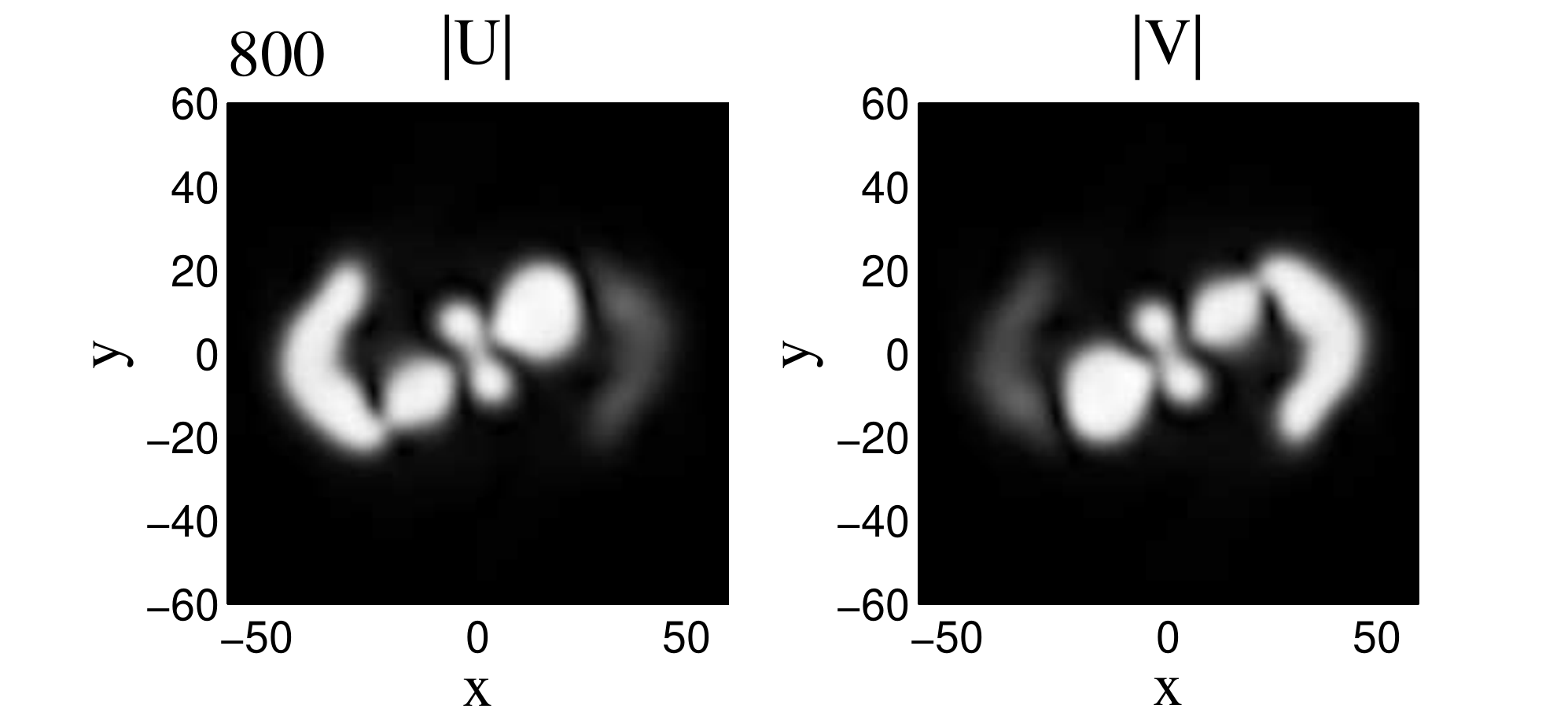}\label{collisions_s2-b}
\end{minipage}
\end{minipage}
\end{minipage}
\end{minipage}
\end{minipage}}
\caption{The interaction between asymmetric in-phase vortices with $s=2$, $%
\protect\lambda =0.05$, $k=0.184$, and $\Delta x=60$.}
\label{collisions_s2}
\end{figure}

The results are presented in Figs.~\ref{collisions_s0}--\ref{collisions_s2}.
In accordance with the known principle \cite{Phase}, we observed that
identical solitons with even values of the spin, $s=0$ or $s=2$, attract
each other in the in-phase configuration, while the vortices with $s=1$
exhibit the attraction when they are \textit{out-of-phase}, with $\Delta
\theta =\pi $. The attraction results in inelastic collisions, as seen in
Figs.~\ref{collisions_s0-a}--\subref{collisions_s0-b}, \ref{collisions_s2-a}%
--\subref{collisions_s2-b}, and~\ref{collisions_s1_shift-a}--%
\subref{collisions_s1_shift-b}. If both solitons are fundamental ones, the
inelastic interaction ends up with their merger into a single pulse (which
is not necessarily another stationary fundamental soliton). If vortex rings
are involved into the collision, the eventual result is destruction of the
ring(s), and formation of a disordered pattern.

On the other hand, repulsion is observed, also in agreement with the
predictions of Ref.~\cite{Phase}, between out-of-phase solitons ($\Delta
\theta =\pi $) with even values of the spin, $s=0$ and $2$ (not shown here
in detail), and between \textit{in-phase} vortices with $s=1$. In the case
of the repulsion, the interactions produce, as a matter of fact, no visible
effect, leading only to a small increase of the separation between the two
solitons, as shown, for $s=1$, in Fig~\ref{collisions_s1-a}--%
\subref{collisions_s1-b}.

\begin{figure}[tbp]
\parbox{3.4in}{
\begin{minipage}{3.3in}
\includegraphics[width=3.2in]{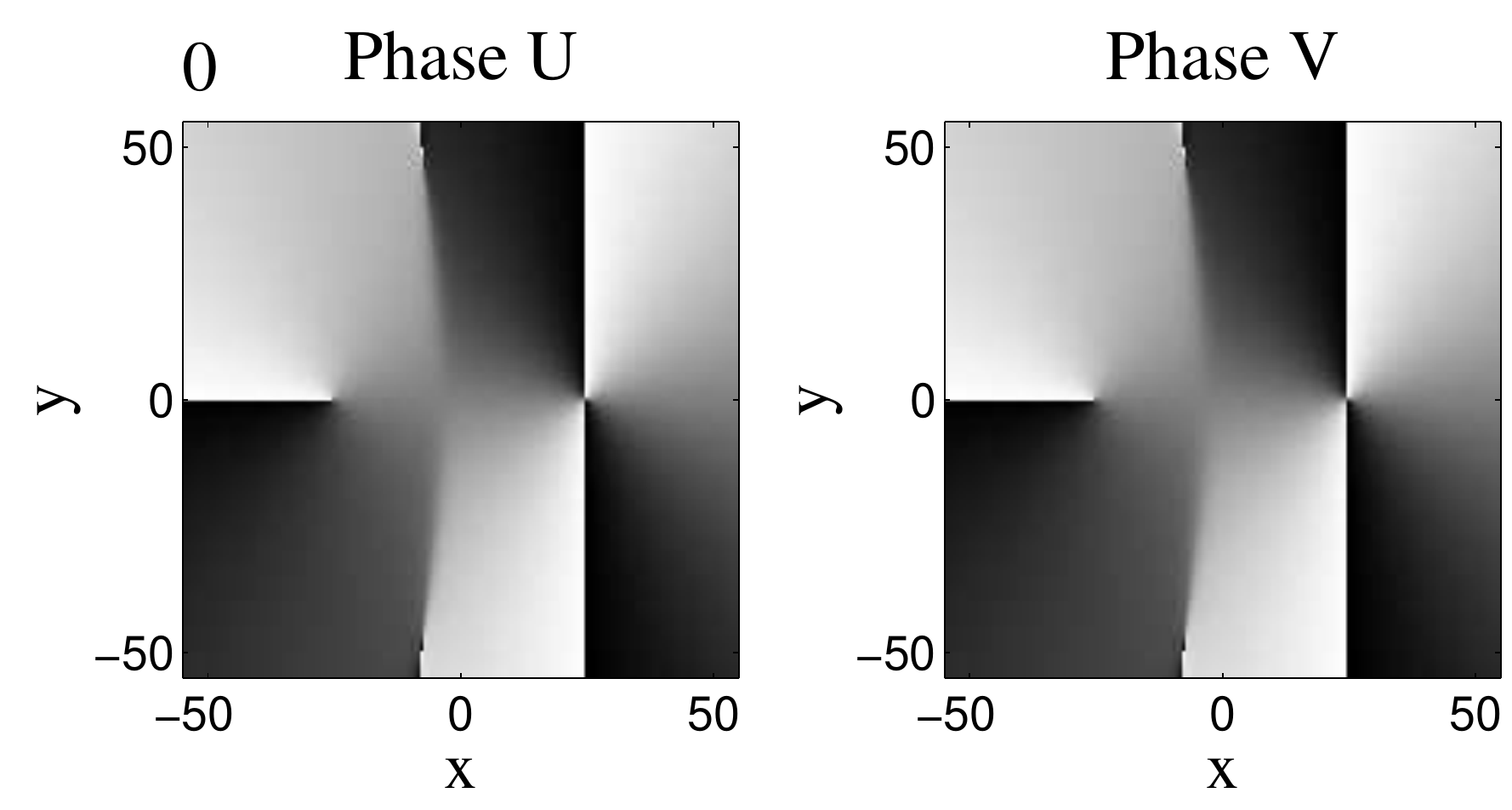} \\
\begin{minipage}{3.3in}
\includegraphics[width=3.2in]{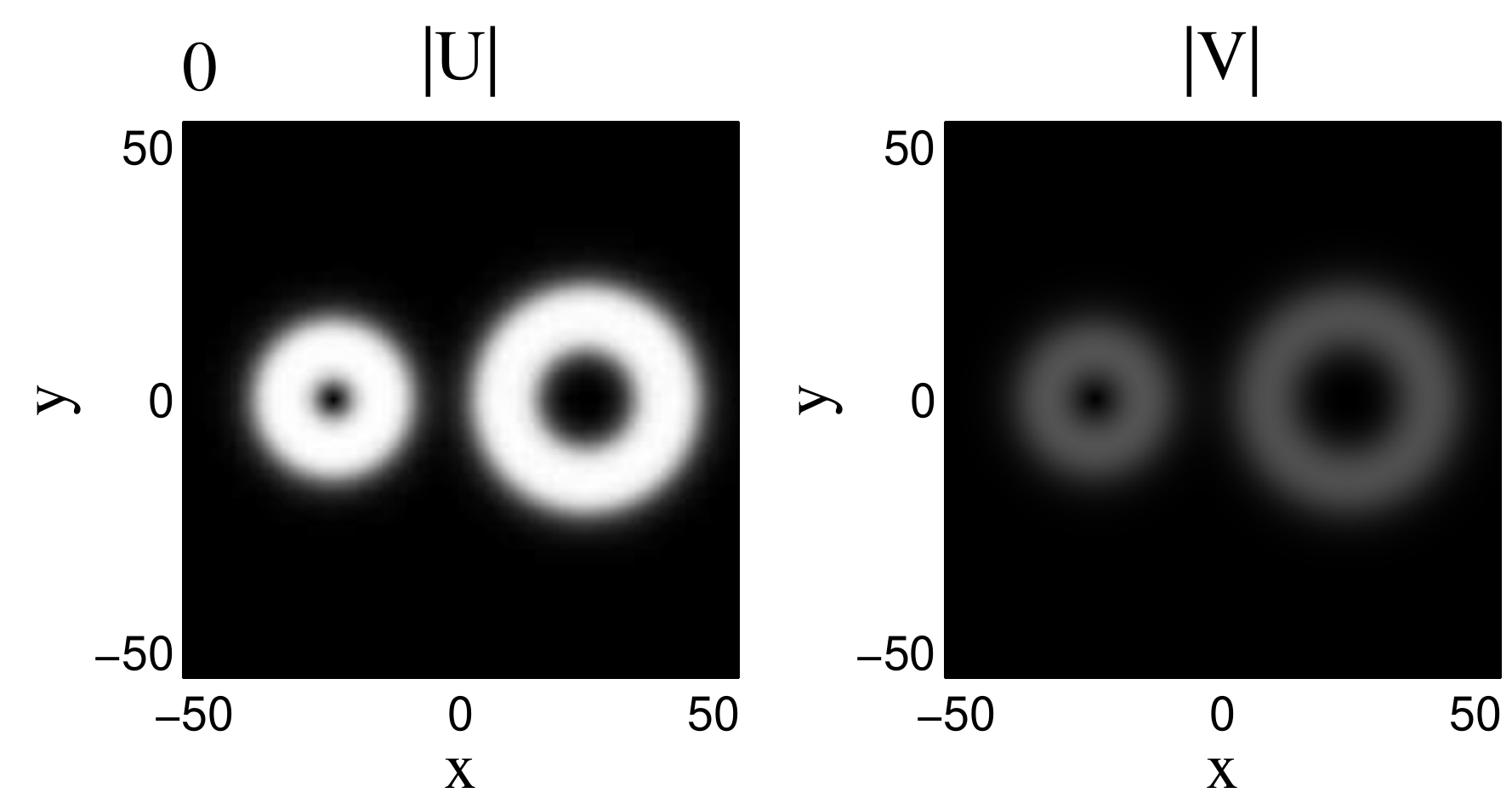}\\
\begin{minipage}{3.3in}
\includegraphics[width=3.2in]{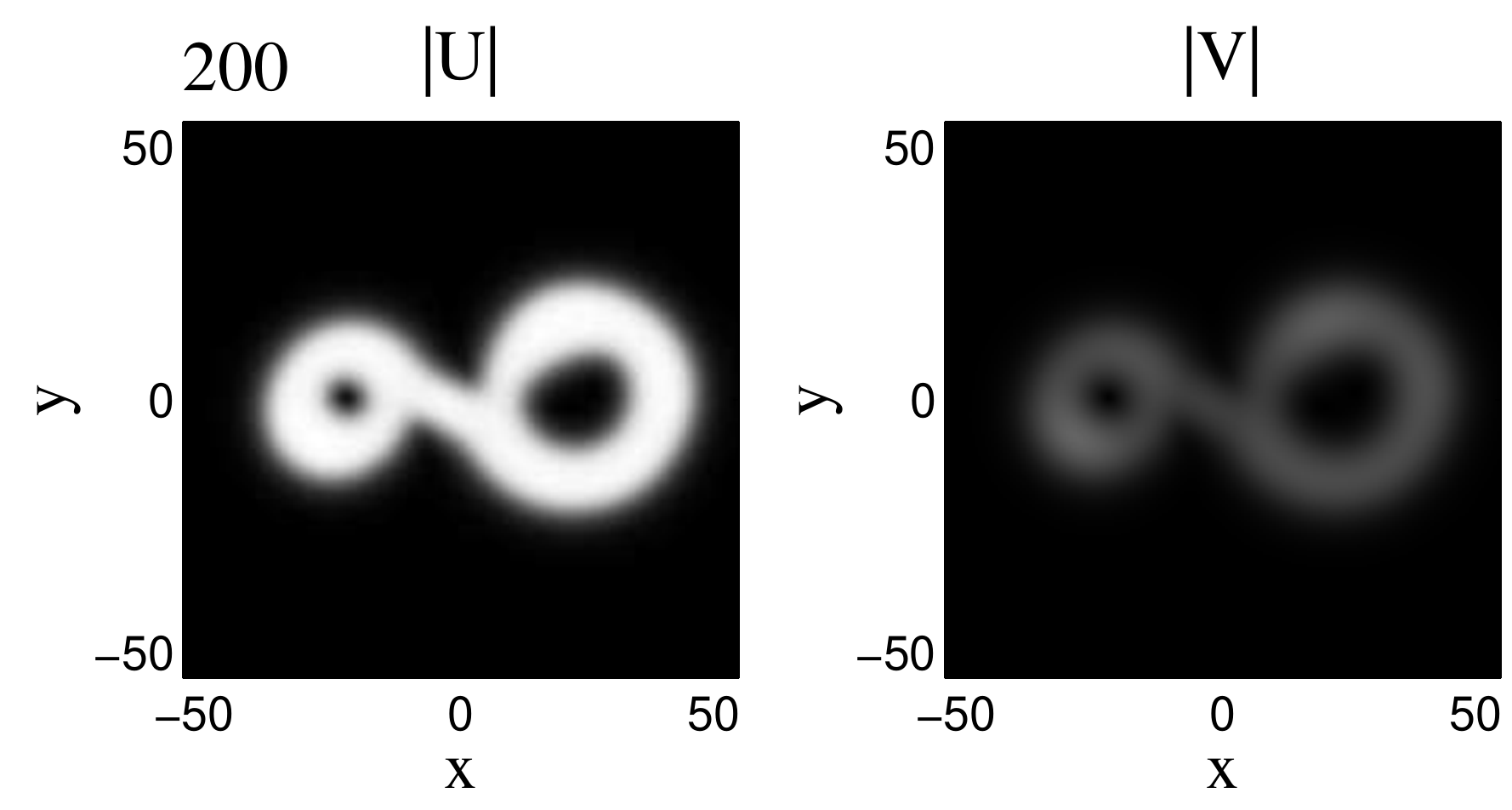}\\
\begin{minipage}{3.3in}
\includegraphics[width=3.2in]{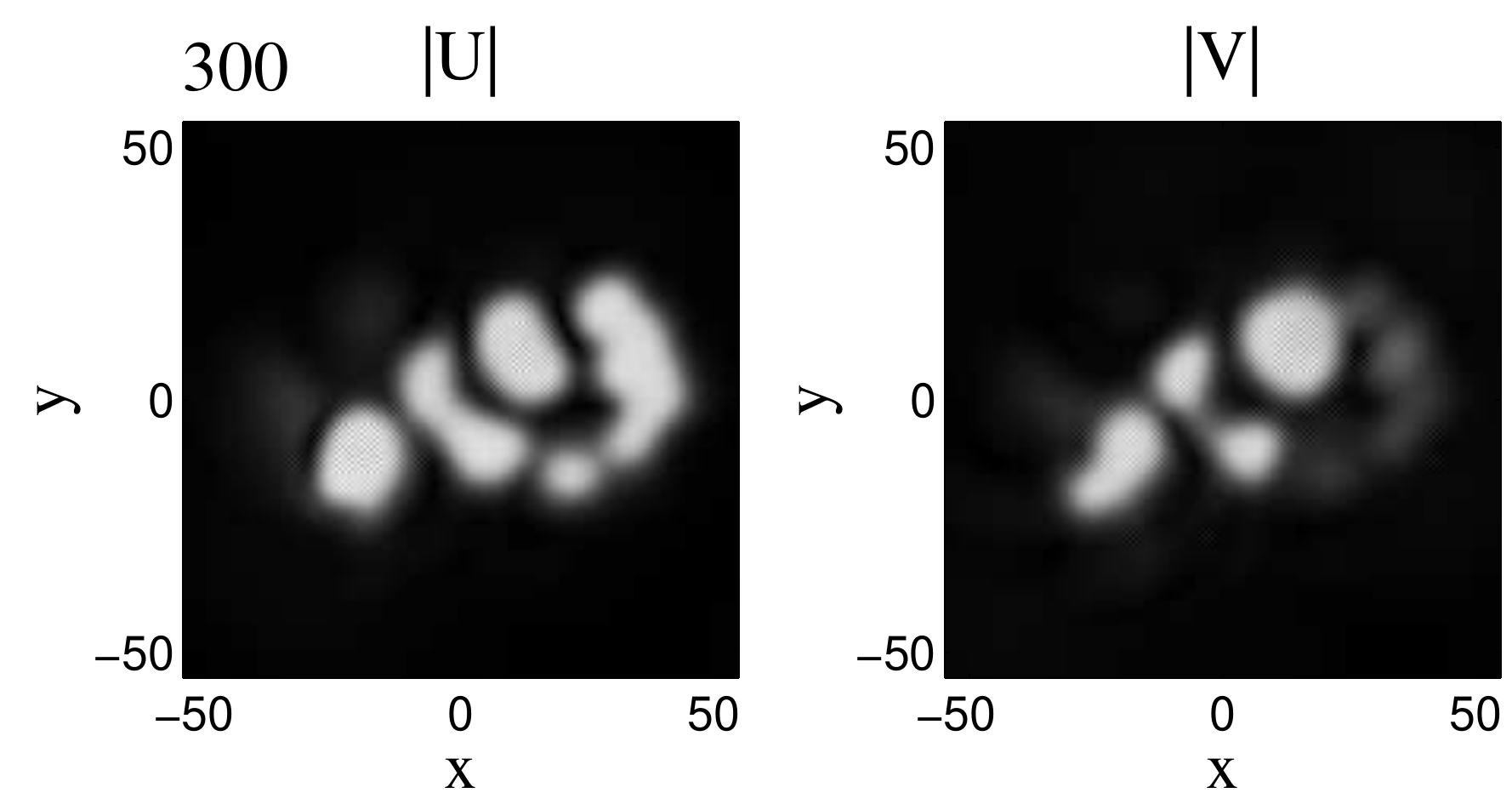}
\end{minipage}
\end{minipage}
\end{minipage}
\end{minipage}
\caption{The interaction between vortices with $s=1$ and $s=2$, for
$\Delta \protect\theta =0$, $\Delta x=50$, and $\protect\lambda
=0.05$, $k=0.184$.} \label{collisions_s1s2}} \qquad
\begin{minipage}{3.3in}
\includegraphics[width=3.2in]{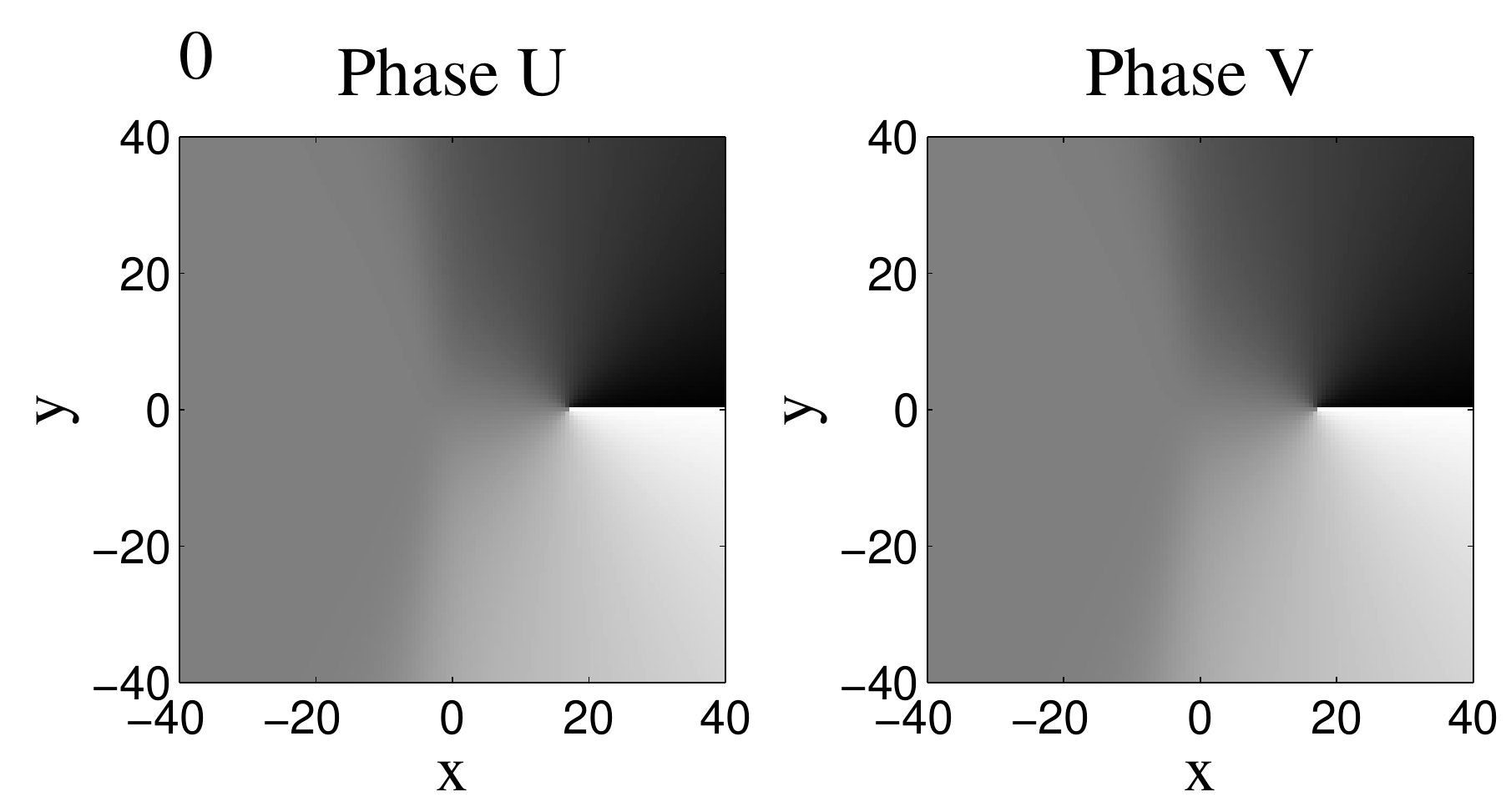} \\
\begin{minipage}{3.3in}
\includegraphics[width=3.2in]{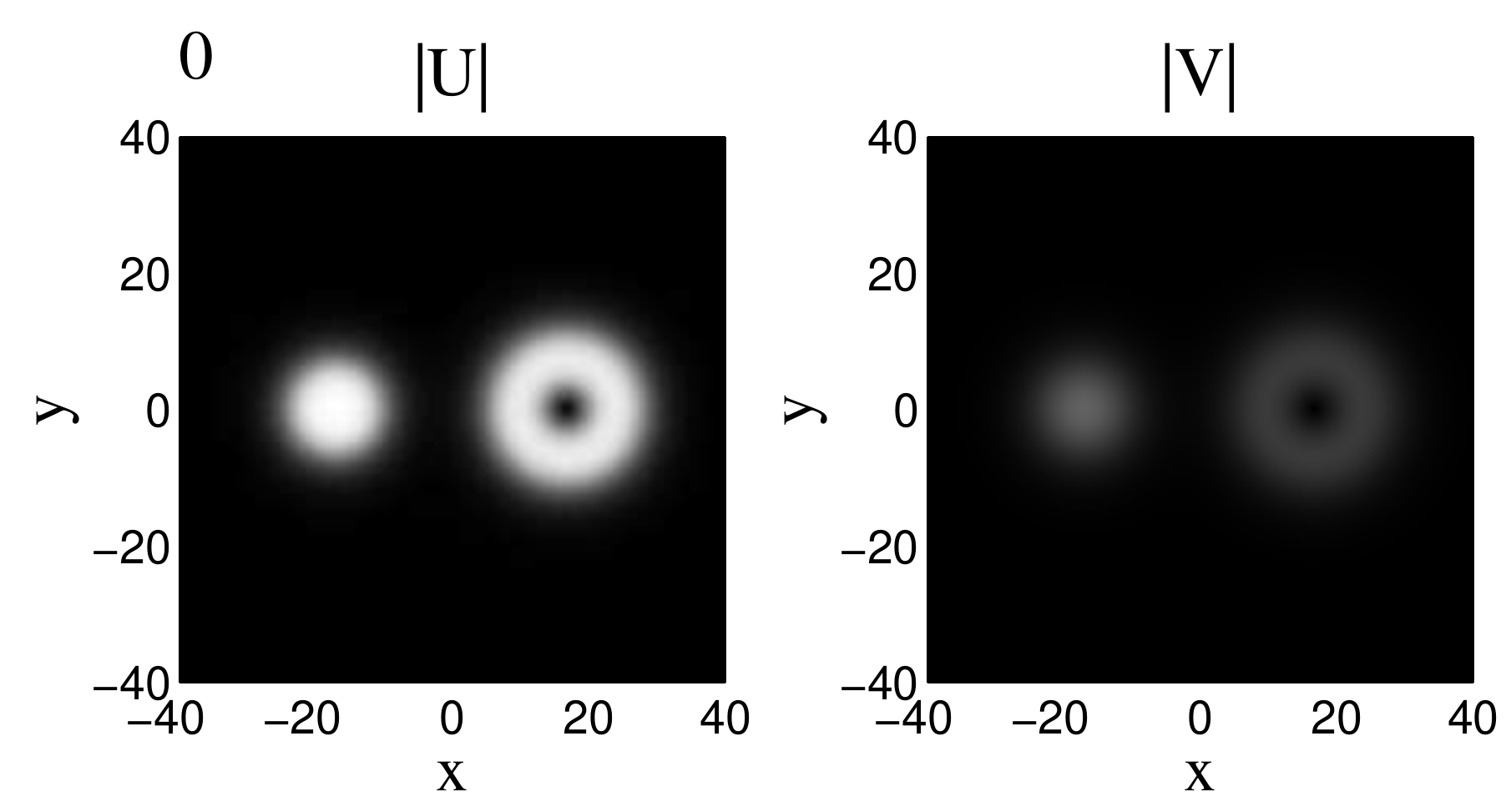}\\
\begin{minipage}{3.3in}
\includegraphics[width=3.2in]{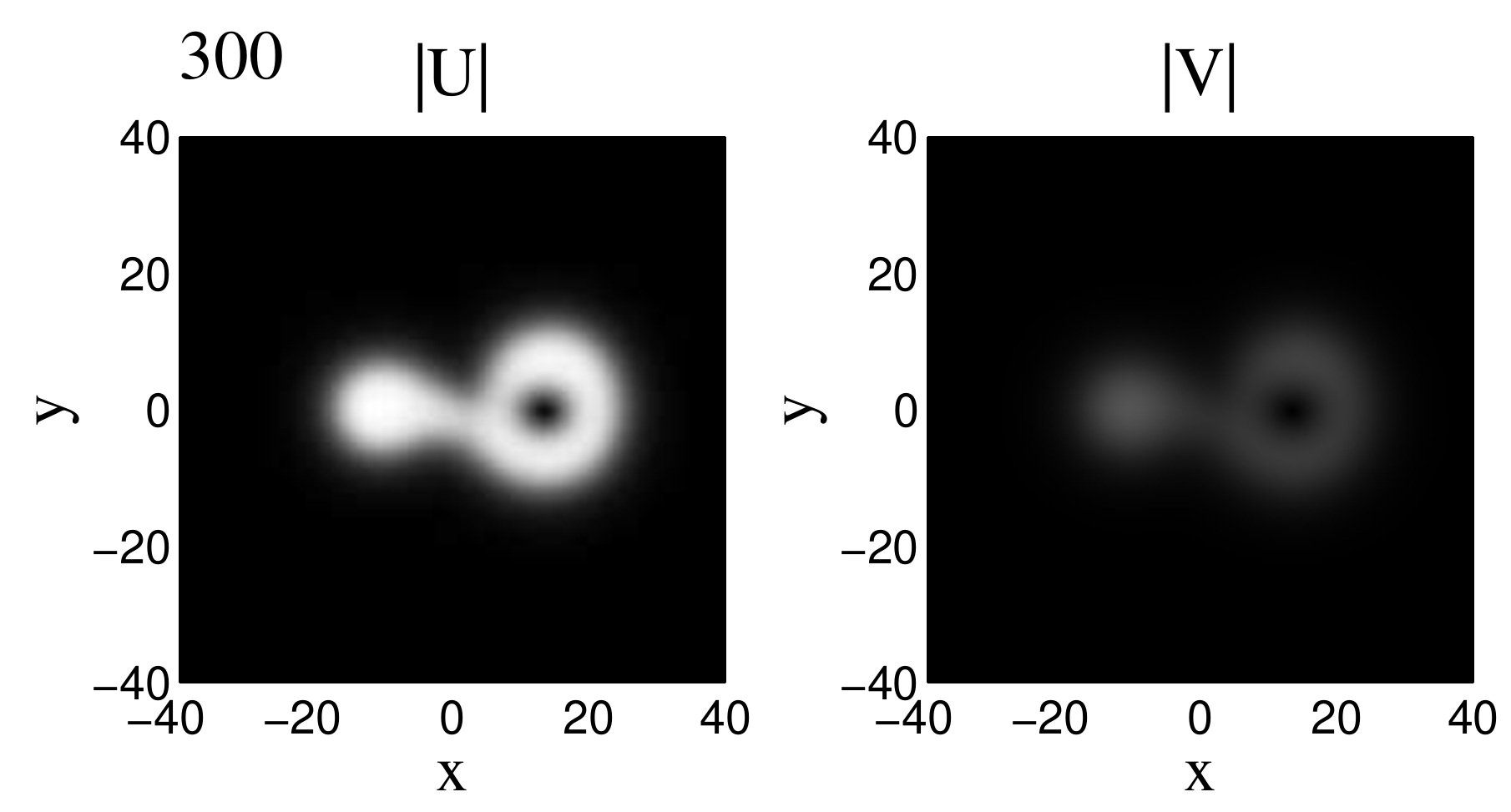}\\
\begin{minipage}{3.3in}
\includegraphics[width=3.2in]{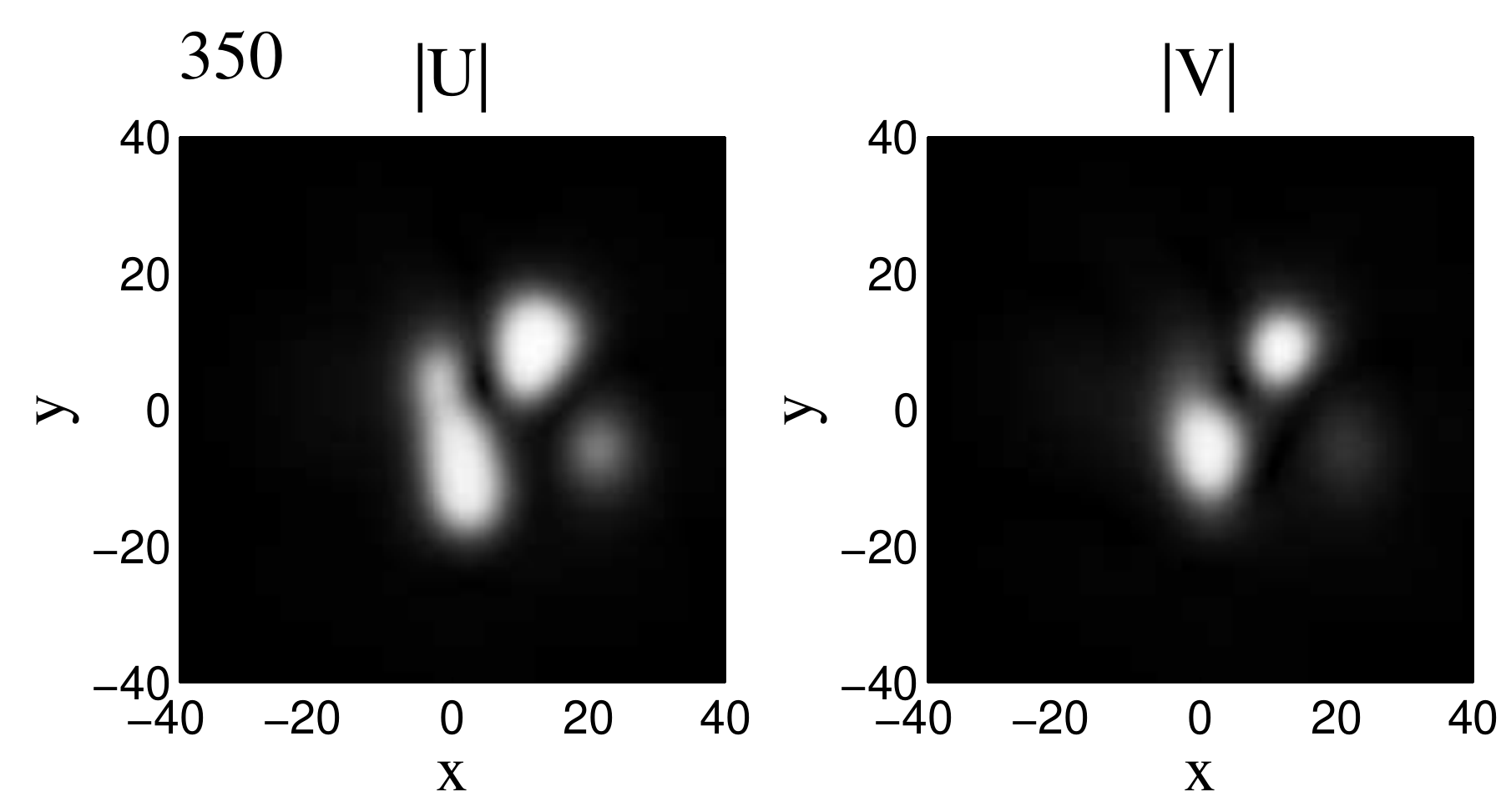}
\caption{The interaction between a fundamental soliton ($s=0$) and a
vortex with $s=1$, for $\Delta x=34,$ $\Delta \protect\theta
=\protect\pi$ and $\protect\lambda =0.05$, $k=0.17$.}
\label{collisions_s0s1}
\end{minipage}
\end{minipage}
\end{minipage}
\end{minipage}
\end{figure}

Additional simulations were carried out for soliton pairs with different
spins. Figure~\ref{collisions_s1s2} shows a typical example for an in-phase
vortex pair with $s=1$ and $s=2$. In this case, the phase varies smoothly
between the centers of the two vortices, similar to the case of the in-phase
pairs with $s=0$ or $s=2$, and the out-of-phase one with $s=1$. Accordingly,
the simulations demonstrate a strong attractive interaction in this case. On
the other hand, for the same pair with $s=1$, $s=2$ and $\Delta \theta =\pi $%
, the change of the phase between the vortices is very steep, similar to the
situations when the solitons with equal spins repel each other, and,
accordingly, the vortices under consideration repel each other too, which
results in slow separation between the vortices (not shown here).

We have also studied the interaction between the vortex with $s=1$ and the
fundamental soliton ($s=0$). In this case, an inelastic collision is
observed in Fig.~\ref{collisions_s0s1} for $\Delta \theta =\pi $, while for $%
\Delta \theta =0$ the interaction is repulsive, producing no conspicuous
effect (not shown here). These outcomes may be explained by the same
character of the phase pattern between the solitons as above, i.e., smooth
in the configuration leading to the attraction and strong interaction, and
steep in the opposite case of the repulsion.

Interactions between stable symmetric solutions have also been
examined, demonstrating results identical to those originally
reported in Ref. \cite{Michinel} (not shown here). In particular, in
this case the collisions maintain the symmetry between the two
components, and do not lead to the appearance of any asymmetric
final states.

\section{Conclusion}

\label{sec:Conclusion} We have introduced a model of a 2D dual-core
waveguide with the CQ (cubic-quintic) nonlinearity inside each core and the
linear coupling between them. Families of fundamental ($s=0$) and vortical
solitons, with spins $s=1$ and $2$, have been constructed, and their
stability has been investigated. The model may be realized as a dual-core
planar optical waveguide, or as a set of two tunnel-coupled parallel
pancake-shaped traps for BEC. In the former case, the solitons are
\textquotedblleft planar light bullets". In particular, vortex solitons may
be interpreted as a new species of the \textquotedblleft bullets", \textit{%
viz}., spatiotemporal vortices.

The main objective of the work was to study symmetry-breaking bifurcations
of the 2D solitons, both fundamental and vortical ones. If the inter-core
coupling constant, $\lambda $, is not too large, the bifurcation diagrams
for the solitons of all the types form closed loops, which connect points of
direct and reverse bifurcations. The stability of all the solutions was
investigated via the calculation of the corresponding eigenvalues for
infinitely small perturbations around the solitons. In particular, it was
found that, at sufficiently small values of $\lambda $, the loop for the
fundamental solitons ($s=0$) includes two bistability regions, which may be
of interest for potential applications, such as all-optical switching. This
is also a new feature of the loop in comparison with its earlier studied
counterpart in the 1D version of the model, which could feature a single
bistability domain. At larger values of $\lambda $, both bistable regions
vanish and the loop's shape becomes plainly convex. With the further
increase of $\lambda $, the loop shrinks to zilch and disappears, leaving
only stable symmetric solitons. The vortical solitons may be easily
destabilized by azimuthal perturbations, but they also have stability
regions, as long as the corresponding bifurcation loop keeps its
double-concave shape. We have also developed a quasi-analytical approach to
the description of the bifurcation diagrams, based on the variational
approximation, which produces reasonably accurate predictions, in comparison
with the numerical results.

In direct simulations, we have demonstrated the splitting of azimuthally
unstable asymmetric solitons into sets of fragments. The number of the
fragments usually corresponds to the azimuthal index of the most unstable
eigenmode of small perturbations. We have also studied interactions between
initially quiescent solitons, and confirmed the earlier prediction \cite%
{Phase}, which states that the usual attractive/repulsive sign of the
interaction between the in-phase/$\pi $-out-of-phase solitons with even
values of the spin ($s=0$ or $2$), is reversed for the odd spin ($s=1$). In
the case of the attraction, the vortex solitons merge into disordered
patterns, losing the initial topological structure.

This work may be naturally extended in other directions. In particular, it
may be interesting to study symmetry-breaking effects in two-component 2D
solitons and vortices in the system of NLSEs coupled by both linear and
nonlinear terms, which may describe the co-propagation of two polarizations
of light in a single nonlinear waveguide \cite{Agrawal}. Moreover, the
latter version of the model is meaningful in the 3D geometry too. Another
relevant generalization still pertains to the dual-core waveguide, with the
linear coupling between the two waves, while the competing nonlinear terms
are quadratic and cubic, rather than cubic and quintic, cf. Ref. \cite{2:3}
and references therein. In that case, one may also expect bifurcation loops
accounting for the breaking and restoration of the symmetry of two-component
solitons.

\end{document}